\documentclass[12pt, letterpaper, hyper, twoside]{JHEP3}  

\usepackage{amsmath,amssymb,amsfonts,amsxtra,makeidx,graphics,graphicx,amsthm,epstopdf,epsfig}
\usepackage[all]{xy}

\newcommand{\be}{\begin{equation}}
\newcommand{\ee}{\end{equation}}
\newcommand{\beq}{\begin{equation}}
\newcommand{\eeq}{\end{equation}}
\newcommand{\ba}{\begin{array}}
\newcommand{\ea}{\end{array}}
\newcommand{\bea}{\begin{eqnarray}}
\newcommand{\eea}{\end{eqnarray}}
\newcommand{\ben}{\begin{enumerate}}
\newcommand{\een}{\end{enumerate}}
\newcommand{\bean}{\begin{eqnarray*}}
\newcommand{\eean}{\end{eqnarray*}}
\newcommand{\eref}[1]{(\ref{#1})}
\newcommand{\sref}[1]{\S\ref{#1}}

\newcommand{\nn}{\nonumber}

\newcommand{\BC}{\mathbb{C}}

\newcommand{\comment}[1]{}

\newcommand{\CM}{{\cal M}}

\newcommand{\CN}{{\cal N}}
\newcommand{\CB}{{\cal B}}
\newcommand{\Li}{\mathrm {Li}}

\newcommand{\ud}{\mathrm{d}}

\newcommand{\PE}{\mathrm{PE}}
\newcommand{\PL}{\mathrm{PL}}

\newtheorem{theorem}{\bf Theorem}

\newtheorem{observation}[theorem]{\bf Observation}
\newtheorem{lemma}[theorem]{\bf Lemma}

\newcommand{\setall}{\setcounter{equation}{0}
        \setcounter{theorem}{0}}

\title{The Hilbert Series of Adjoint SQCD}
\author{Amihay Hanany, Noppadol Mekareeya and Giuseppe Torri\\
Theoretical Physics Group, The Blackett Laboratory \\
Imperial College London, Prince Consort Road\\ 
London,  SW7 2AZ,  UK \\
Email: {\tt a.hanany, n.mekareeya07, giuseppe.torri08@imperial.ac.uk}}

\abstract{We use the plethystic exponential and the Molien--Weyl formula to compute the Hilbert series (generating functions), which count gauge invariant operators in $\CN=1$ supersymmetric $SU(N_c)$, $Sp(N_c)$, $SO(N_c)$ and $G_2$ gauge theories with 1 adjoint chiral superfield, fundamental chiral superfields, and zero classical superpotential.  The structure of the chiral ring through the generators and relations between them is examined using the plethystic logarithm and the character expansion technique.  The palindromic numerator in the Hilbert series implies that the classical moduli space of adjoint SQCD is an affine Calabi--Yau cone over a weighted projective variety.}

\preprint{Imperial/TP/08/AH/11} 

\begin{document}

\section{Introduction and Summary} \setall
The rich structures of $\CN =1$ Supersymmetric Quantum Chromodynamics (SQCD) chiral rings and moduli spaces has been studied in \cite{Gray, Hanany:2008kn} using the Plethystic Programme \cite{BFHH, Noma:2006pe, feng, forcella, Butti:2007jv, Forcella:2007ps, Hanany:2008qc, Hanany:2008cd}, Molien--Weyl formula \cite{pouliot, romelsberger, hanany, Dolan:2007rq, Dolan:2008qi}, and character expansions \cite{Hanany:2008qc, Hanany:2008cd, Forcella:2008bb}.  We can generalise this theory in an interesting way by adding to it a chiral superfield $\phi$ in the adjoint representation, and use the name termed in the literature, $\CN=1$ {\bf adjoint SQCD}.  In this paper, we shall focus on the theory with the $SU(N_c)$, $Sp(N_c)$, $SO(N_c)$ and $G_2$ gauge groups with vanishing classical superpotential.

There have been a series of works \cite{Kutasov:1995ve, Kutasov:1995np, Kutasov:1995ss, Giveon:1998sr, Kutasov:2003iy, Parnachev:2008yt} on the  $SU(N_c)$ adjoint SQCD, as well as \cite{Leigh:1995qp, Luty:1996cg} on the $SO(N_c)$ and $Sp(N_c)$ theories with various classical superpotentials. It is known that the classical moduli space of the $SU(N_c)$ theory does not get quantum corrections \cite{Kutasov:1995ss, Giveon:1998sr, Kutasov:2003iy}.  However, due to technical difficulties, many aspects (e.g., Seiberg duality) of the zero classical superpotential theories have yet to be fully understood\footnote{Regarding this, let us quote the authors of \cite{Kutasov:1995np}: `This interesting model has so far resisted all attempts at a detailed understanding.'}.  The main aim of this paper is to examine the structure of the chiral rings of adjoint SQCD (with zero superpotential) through the generators of the gauge invariant operators (GIOs) and their relations.  

We use the plethystic exponential and the Molien--Weyl formula to obtain Hilbert series\footnote{There are 3 words which are synonymous: partition function, generating function and Hilbert series. The first one is the physics literature name, whereas the second and third ones typically appear in the mathematical literature.}, which count GIOs.  The generators and the relations in the chiral ring can be extracted from the plethystic logarithm of the Hilbert series.  Using the character expansion technique, we can also figure out how these generators and relations transform under the global symmetry.  

Hilbert series also contain information about geometrical properties of the moduli space. We shall see in subsequent sections, for example, that plethystic logarithms of Hilbert series can indicate whether the moduli space is a complete intersection, and that the palindomic property of the Hilbert series implies that the moduli space is Calabi--Yau \cite{Gray, Hanany:2008kn,Forcella:2008bb}.

Below, we collect the main results of our work. 

\paragraph{Outline and key points.}
\begin{itemize}
\item In Section \ref{recap}, we summarise the Plethystic Programme and Molien--Weyl formula.
\item In Section \ref{SU}, Hilbert series of adjoint SQCD with $SU(N_c)$ gauge group are constructed.  We analyse the generators and relations using the plethystic logarithm.  We find that the total number of generators in $SU(N_c)$ theory with $N_f$ flavours is of order $N_f^{N_c}$.  
\begin{itemize}
\item In Subsection \ref{partition}, we count adjoint baryons in the $SU(N_c)$ theory. This leads to an interesting combinatorial problem of partitions. We find an exponentially large number of adjoint baryons when $N_c$ is large.  The asymptotic formula is given by \eref{asympa}.
\item In Subsection \ref{freeenergy}, the canonical free energy of the theory is derived and found to be of order $N_fN_c$.  
\item In Subsection \ref{secCI}, we focus on the complete intersection moduli space of the $SU(N_c)$ theory (\emph{i.e.,} with 1 flavour).  There is exactly one basic relation in this case.  A general expression of such a relation for any $N_c$ is given by \eref{magic}.


\end{itemize}
\item The structures of generators and relations of the $Sp(N_c)$, $SO(N_c)$ and $G_2$ adjoint SQCD are studied using Hilbert series in Sections \ref{secSp}, \ref{secSO}, and \ref{secg2}. It is found that $Sp(N_c)$ theories with 1 flavor, and $SO(N_c)$ theories with 2 flavors have a moduli space which is a complete intersection, while the moduli space of $SO(N_c)$ theories with 1 flavor is freely generated. The number of relations for the complete intersection cases is equal to the rank of the gauge group and not to 1 as in the case of $SU(N_c)$ with 1 flavor.
\item In Section \ref{apercu}, we take a geometric aper\c{c}u of the moduli space of adjoint SQCD.  We establish that the classical moduli space is an irreducible affine Calabi--Yau cone.
\end{itemize}

\paragraph{Notation for representations.} In this paper, we represent an irreducible representation of a group $G$ by its highest weight $[a_1, \ldots, a_r]$, where $r = \mathrm{rank}~G$. Young diagrams are also used in order to avoid cluttered notation.  We also slightly abuse terminology by referring to each character by its corresponding representation.


\section{Plethystic Programme and Molien--Weyl Formula:\\ A Recapitulation} \label{recap} \setall
In order to write down explicit formulae and for performing computations we need to introduce weights for the different elements in the maximal torus of the different groups. We use 
\begin{itemize}
\item  $z_a$ (with $a=1,\ldots, \mathrm{rank}~G$) to denote `colour' weights, \emph{i.e.} coordinates of the maximal torus of the gauge group $G$~; \\
\item  $t_i$  (with $i=1,\ldots,N_f$) to denote `flavour' weights for the fundamental chiral superfield $Q$, \emph{i.e.} coordinates of the maximal torus of  the global $U(N_f)_1$ symmetry~; \\ 
\item  $\tilde{t}_i$  (with $i=1,\ldots,N_f$) to denote `flavour' weights for the antifundamental chiral superfield $\widetilde{Q}$, \emph{i.e.} coordinates of the maximal torus of the global $U(N_f)_2$ symmetry~; \\ 
\item  $s$ to count the chiral superfield $\phi$ in the adjoint representation. 
\end{itemize}
These weights have the interpretation of fugacities for the charges they count and the characters of the representations are functions of these variables.  Henceforth, we shall take $t_i,\: \tilde{t}_i,\: s$ to be complex variables such that their absolute values lie between 0 and 1. 

Below we summarise important facts and conventions of the gauge groups and their Lie algebras which we shall use later \cite{FH}:
\paragraph{The group $SU(N_c)$.} Let us take the weights of the fundamental representation of $SU(N_c)$ to be
\beq
L_1 = (1,0, \ldots, 0) ~, \quad
L_k = (0,0, \ldots, -1,1, \ldots 0) ~, \quad
L_{N_c} = (0, \ldots, -1) ~,
\label{coordsmaxtorus}
\eeq
where all $L$'s are $(N_c-1)$-tuples, and for $L_k$ (with $2 \leq k \leq N_c-1$), we have $-1$ in the $(k-1)$-th position and $1$ in the $k$-th position.
With this choice of weights, we find that the characters of the fundamental and antifundamental representations  of $SU(N_c)$ are
\bea 
[1,0, \ldots,0 ]_{SU(N_c)} (z_a) &=&  z_1 + \sum_{k=2}^{N_c-1} \frac {z_k} {z_{k-1} } + \frac {1} {z_{N_c-1} }~, \nn \\
\left[0, \ldots,0,1 \right]_{SU(N_c)} (z_a) &=& \frac{1} {z_1}  + \sum_{k=2}^{N_c-1} \frac {z_{k-1} } {z_k} + z_{N_c-1} ~.
\eea
The roots of the Lie algebra of $SU(N_c)$ are $\{ L_a -L_b \}_{a \neq b}$. The character of the adjoint representation can be written as
\bea
\left[1,0, \ldots,0,1 \right]_{SU(N_c)} &=&[1,0, \ldots,0 ]_{SU(N_c)} \times \left[0, \ldots,0,1 \right]_{SU(N_c)} -1 \nn \\
&=& (N_c -1) + \sum_{\alpha} \left( \prod_{l =1}^{N_c-1} z_l ^ {\alpha_l} \right)~,
\eea  
where the summation is taken over all roots $\alpha$, and the notation $\alpha_l$ denotes the number in the $l$-th position of the root $\alpha$.  For example,
\beq
[1,1]_{SU(3)} = 2+\frac{z_1}{z_2^2}+\frac{1}{z_1 z_2}+\frac{z_1^2}{z_2}+\frac{z_2}{z_1^2}+z_1 z_2+\frac{z_2^2}{z_1}~.
\eeq


\paragraph{The group $SO(N_c)$.}  We note that a special orthogonal group falls into one of the two categories of the classical groups, namely $B_{n} = SO(2n+1)$ and $D_{n}= SO(2n)$.  The Lie algebras of $B_{n}$ and $D_{n}$ both have the same rank $n$.  The weights of the fundamental (vector) representations of $B_{n}$ and $D_{n}$ are respectively $\{0, \pm L_a \}$ and $\{\pm L_a\}$.  With this choice, we can write down the characters of the fundamental representations of $B_{n}$ and $D_{n}$ respectively as
\bea
 \left[1,0, \ldots,0 \right]_{B_n} (z_a) &=& 1+ \sum_{a=1}^{n} \left( z_a + \frac{1}{z_a} \right) ~, \nn \\
 \left[1,0, \ldots,0 \right]_{D_n} (z_a) &=&  \sum_{a=1}^{n} \left( z_a + \frac{1}{z_a} \right)~. \label{charfunds}
\eea
The adjoint representation of $B_{n}$ and $D_{n}$ is the antisymmetric square of the corresponding fundamental representation: $\mathrm{Adj}_{B_n, D_n} = \Lambda^2 [1,0, \ldots,0]_{B_n, D_n}$, and so its character is given by
\bea
\mathrm{Adj}_{B_n, D_n} (z_a) = \frac{1}{2} \Big( \left( [1,0, \ldots,0]_{B_n, D_n} (z_a) \right)^2 - [1,0, \ldots,0]_{B_n, D_n} (z_a^2) \Big) ~.
\eea
The roots of the Lie algebras of $B_{n}$ and $D_{n}$ are respectively $\{ \pm L_a \pm L_b,\: \pm L_a\}_{a \ne b}$ and $\{ \pm L_a \pm L_b\}_{a \ne b}$.  

\paragraph{The group $Sp(N_c)$.} We shall use the notation such that the rank of the Lie algebra of $Sp(N_c)$ is $N_c$, the fundamental representation of $Sp(N_c)$ is $2N_c$ dimensional, and $Sp(1)$ is isomorphic to $SU(2)$.  We define $L_a = (0, \ldots,0, 1_{a;L},0, \ldots, 0)$, where the length of the tuple is $N_c$. The weights of the fundamental representation are $\{\pm L_m \}$.  With this choice of $L$'s, we find the character of the fundamental representation to be
\bea
[1,0, \ldots,0]_{Sp(N_c)}(z_a) = \sum_{a=1}^{N_c} \left( z_a + \frac{1}{z_a} \right)~.
\eea
The adjoint representation of $Sp(2N_c)$ is the symmetric square of the fundamental representation: $\mathrm{Adj}_{Sp(N_c)} = \mathrm{Sym}^2 [1,0,\ldots,0]_{Sp(N_c)}$, and so its character is given by
\bea
\mathrm{Adj}_{Sp(N_c)}  (z_a) = \frac{1}{2} \Big( \left( [1,0, \ldots,0]_{Sp(N_c)} (z_a) \right)^2 + [1,0, \ldots,0]_{Sp(N_c)} (z_a^2) \Big) ~.
\eea
The roots of the Lie algebra of $Sp(N_c)$ are $\{  \pm L_a \pm L_b \}$, where $1 \leq a,\: b \leq N_c$.  

\paragraph{The group $G_2$.}  The Lie algebra of $G_2$ has rank 2.  We define $L_1 = (2, -1),\: L_2 = (-1,1)$. The weights of the fundamental representation are $\{0,\: \pm L_1,\: \pm L_2,\: \pm (L_1 + L_2) \}$. With this choice of weights, the character of the fundamental representation is
\bea
[1,0]_{G_2} = 1+\frac{1}{z_1}+z_1+\frac{z_1}{z_2}+\frac{z_1^2}{z_2}+\frac{z_2}{z_1^2}+\frac{z_2}{z_1} ~.
\eea
The adjoint representation $[0,1]_{G_2}$ is given by 
\bea
[0,1] &=& \Lambda^2 [1,0]_{G_2} - [1,0]_{G_2} \nn \\
&=& 2+\frac{1}{z_1}+z_1+\frac{z_1^3}{z_2^2}+\frac{1}{z_2}+\frac{z_1}{z_2}+\frac{z_1^2}{z_2}+\frac{z_1^3}{z_2}+z_2+\frac{z_2}{z_1^3}+\frac{z_2}{z_1^2}+\frac{z_2}{z_1}+\frac{z_2^2}{z_1^3}~. \label{G2adj}
\eea

\paragraph{The plethystic exponential.} The chiral GIOs are \emph{symmetric} functions of the fundamental chiral superfields $Q^i_a$, the antifundamental chiral superfields $\widetilde{Q}^a_i$, and the adjoint chiral superfield $\phi$ which transform respectively in the fundamental, the antifundamental, and the adjoint representations of the gauge group $G$.  A convenient combinatorial tool which constructs symmetric products of representations is the {\bf plethystic exponential} \cite{Gray, Hanany:2008kn, BFHH, Noma:2006pe, feng, forcella, Butti:2007jv, Forcella:2007ps, Hanany:2008qc, Hanany:2008cd}, which is a \emph{generator for symmetrisation}. 
To briefly remind the reader, we define the {plethystic exponential} of a  multi-variable function $g (t_1 , \ldots , t_n )$ that vanishes at the origin, $g (0,\ldots, 0) = 0$, to be 
\begin{equation}
\label{PE}
\PE [g (t_1 , \ldots , t_n )] := \exp \left ( \sum_{r=1}^\infty \frac{g(t_1^r,\ldots,t_n^r)}{r}\right )~.
\end{equation}
Using formula \eref{PE} and the expansion $- \log (1-x) = \sum_{k=1}^\infty x^k/k$, we have, for example,
\bea
\mathrm{PE}\: \left[ [1,0, \ldots,0]_{SU(N_c)}  \sum_{i=1}^{N_f} t_i  \right] 
&=&  \frac{1}{ \prod_{i=1}^{N_f} \left[ \left(1-t_i z_1\right) \left(1- \frac{t_i}{ z_{N_c-1}} \right) \prod_{k=2}^{N_c-1} (1-t_i \frac{z_k}{z_{k-1}} ) \right]}~, \nn \\
\mathrm{PE}\: \left[ [1,0, \ldots,0]_{B_n}  \sum_{i=1}^{N_f} t_i  \right] &=& \frac{1}{ \prod_{i=1}^{N_f} \prod_{a=1}^{n} (1-t_i)(1-t_i z_a) \left(1-\frac{t_i}{ z_a} \right)}~, \nn \\
\mathrm{PE}\: \left[ [1,0, \ldots,0]_{D_n} \sum_{i=1}^{N_f} t_i  \right] &=& \frac{1}{ \prod_{i=1}^{N_f} \prod_{a=1}^{n} (1-t_i z_a) \left(1-\frac{t_i}{ z_a} \right)}~.
\eea

\paragraph{The Molien--Weyl formula.} We emphasize that, in order to obtain the generating function that counts \emph{gauge invariant} quantities, we need to project the representations of the gauge group generated by the plethystic exponential onto the trivial subrepresentation, which consists of the quantities \emph{invariant} under the action of the gauge group.
Using knowledge from representation theory, this can be done by integrating over the whole group.  Hence, the generating function for the gauge group $G$ with $N_f$ chiral multiplets in the fundamental representation, $N_f$ chiral multiplets in the anitfundamental representaton, and 1 chiral multiplet in the adjoint representation is given by
\beq \label{genfn}
g^{(N_f, G)} = \int_{G} \ud \mu_{G}\: \mathrm{PE}\: \left[ \chi^{\text{fund}}_G (z_a) \sum_{i=1}^{N_f} t_i + \chi^{\text{antifund}}_G (z_a) \sum_{i=1}^{N_f} \tilde{t}_i  + \chi^{\text{adjoint}}_G (z^a) s \right]~,
\eeq
where the notation $\chi$ signifies the character. This formula is called the {\bf Molien--Weyl formula} \cite{Gray, Hanany:2008kn, pouliot, romelsberger, hanany, Dolan:2007rq, Dolan:2008qi}. 
In the following section, we shall demonstrate in details how to use this formula to count GIOs.
We note that the Haar measure for the gauge group $\mu_{G}$ is given by \cite{DK}
\begin{equation}
\int_{G} \ud \mu_{G} = \frac{1}{(2 \pi i)^{r}} \oint_{|z_1|=1} \ldots \oint_{|z_{r}|=1} \frac{\ud z_1}{z_1}\ldots  \frac{\ud z_{r}}{z_{r}} \prod_{\alpha^+} \left(1- \prod_{l = 1}^{r} z_l^{\alpha^+_l} \right) ~, \label{haar}
\end{equation}
where $\alpha^+$ are positive roots\footnote{We note that the Haar measure we use here is different from those in \cite{Gray, Hanany:2008kn}. The former involves only positive roots and therefore has no Weyl group normalisation.  This proves to extensively reduce the amount of computations.}   of the Lie algebra of the gauge group $G$ and $r = \mathrm{rank}~G$.  For example,
\bea
\int_{SU(2)} \ud \mu_{SU(2)} &=& \frac{1}{2\pi i} \oint_{|z| =1} \frac{dz}{z} (1-z^2)~,\nn \\
\int_{SU(3)}  \ud \mu_{SU(3)} &=& \frac{1}{(2\pi i)^2} \oint_{|z_1| =1} \frac{dz_{1}}{z_{1}} \oint_{|z_2| =1} \frac{dz_{2}}{z_{2}}  \left(1-z_{1}z_{2}\right)\left(1-\frac{z^{2}_{1}}{z_{2}}\right)\left(1-\frac{z^{2}_{2}}{z_{1}}\right)~, \nn \\
\int_{SO(3)} \ud \mu_{SO(3)} &=& \frac{1}{2 \pi i } \oint_{|z|=1}  \frac{ \ud z}{z} \left(1-z\right)~, \nn \\
\int_{SO(4)} \ud \mu_{SO(4)} &=& \frac{1}{(2 \pi i)^2 } \oint_{|z_1|=1} \frac{ \ud z_1}{z_1} \oint_{|z_2|=1} \frac{ \ud z_2}{z_2}  \left(1-\frac{z_1}{z_2}\right)  (1-z_1 z_2)~, \nn \\
\int_{SO(5)} \ud \mu_{SO(5)} &=& \frac{1}{(2 \pi i)^2 } \oint_{|z_1|=1} \frac{ \ud z_1}{z_1} \oint_{|z_2|=1} \frac{ \ud z_2}{z_2}  (1-z_1) (1-z_2)   \left(1-\frac{z_1}{z_2}\right) (1-z_1 z_2)~, \nn \\
\int_{Sp(2)} \ud \mu_{Sp(2)}  &=& \frac{1}{(2 \pi i)^2 } \oint_{|z_1|=1} \frac{ \ud z_1}{z_1} \oint_{|z_2|=1} \frac{ \ud z_2}{z_2} (1-z_1^2)(1-z_2)(1-\frac{z_1^2}{z_2})(1-\frac{z_2^2}{z_1^2})~, \nn \\
\int_{G_2} \ud \mu_{G_2} &=&  \frac{1}{(2 \pi i)^2 } \oint_{|z_1|=1} \frac{ \ud z_1}{z_1} \oint_{|z_2|=1} \frac{ \ud z_2}{z_2} \left(1-z_1\right) \left(1-\frac{z_1^2}{z_2}\right) \left(1-\frac{z_1^3}{z_2}\right)\left(1-z_2\right) \times \nn \\
&&   \left(1-\frac{z_2}{z_1}\right) \left(1-\frac{z_2^2}{z_1^3}\right)~.
\eea

\paragraph{The Hilbert series.}  A Hilbert series is a function that counts chiral gauge invariant operators and has the interpretation of a partition function at zero temperature and non-zero chemical potentials for global conserved $U(1)$ charges.  It can be written as a rational function whose numerator is a polynomial with integer coefficients.  
Importantly, the powers of the denominators are such that \emph{the leading pole captures the dimension of the manifold}. 


\paragraph{The plethystic logarithm.} Information about the generators of the moduli space and the relations they satisfy can be computed by using the {\bf plethystic logarithm} \cite{Gray, Hanany:2008kn, BFHH, feng, forcella, Hanany:2008qc, Forcella:2008bb}, which is the inverse function of the plethystic exponential. Using the M\"obius function $\mu(r)$ we define:
\begin{equation}
\label{PL}
\PL [g (t_1 , \ldots , t_n )] := \sum_{r=1}^\infty \frac{\mu(r) \log g(t_1^r,\ldots,t_n^r)}{r}~. 
\end{equation}
The significance of the series expansion of the plethystic logarithm is that \emph{the first terms with plus sign give the basic generators while the first terms with the minus sign give the constraints between these basic generators.}  If the formula (\ref{PL}) is an infinite series of terms with plus and minus signs, then the moduli space is not a complete intersection\footnote{Mathematicians refer to the dimension of the moduli space (or the order of the pole of the series for $t=1$) as the {\bf Krull dimension}.} 
and the constraints in the chiral ring are not trivially generated by relations between the basic generators, but receive stepwise corrections at higher degree. These are the so-called higher syzygies. 

\section{Dimension of the Moduli Space} \setall
At a generic point of the moduli space, the gauge symmetry $G$ is broken completely, and hence there are $d(G)$ broken generators.  In the Higgs mechanism, a massless vector multiplet `eats' an entire chiral multiplet to form a massive vector multiplet.  Originally, we have $d(G)$ degrees of freedom coming from the chiral superfields in the adjoint representation (which is $d(G)$ dimensional), and $N_{\chi} d(\Box)$ degrees of freedom coming from the $N_{\chi}$ chiral superfields in the fundamental (and antifundamental) representation (which is $d(\Box)$ dimensional).  Therefore, of the original $d(G)+N_{\chi} d(\Box)$ chiral degrees of freedom, 
only $\left[d(G) +N_{\chi} d(\Box) \right] - d(G)  = N_{\chi}d(\Box) $ singlets are left massless.  Therefore, the dimension of the moduli space $\CM$ is
\bea
\dim \CM = N_{\chi}d(\Box)~.  \label{dim}
\eea

For the $SU(N_c)$ adjoint SQCD with $N_f$ chiral superfields in the fundamental representation and $N_f$ chiral superfields in the antifundamental representation, we have $N_{\chi} = 2N_f$ and $d(\tiny{\Box}) = N_c$.  Therefore,
\bea
\dim \CM_{(N_f,SU(N_c))} = 2N_{f} N_c~.  \label{dimSU}
\eea

For the $Sp(N_c)$ adjoint SQCD with $2N_f$ chiral superfields in the fundamental representation, we have $N_{\chi} = 2N_f$ and $d(\tiny{\Box}) = 2N_c$. Therefore,
\bea
\dim \CM_{(N_f,Sp(N_c))} = 4N_{f} N_c~.  \label{dimSp}
\eea

For the $SO(N_c)$ adjoint SQCD with $N_f$ chiral superfields in the fundamental (vector) representation, we have $N_{\chi} = N_f$ and $d(\tiny{\Box}) = N_c$. Therefore,
\bea
\dim \CM_{(N_f,SO(N_c))} = N_{f} N_c~.  \label{dimSO}
\eea

For the $G_2$ adjoint SQCD with $N_f$ chiral superfields in the fundamental representation, we have $N_{\chi} = N_f$ and $d(\tiny{\Box}) = 7$. Therefore,
\bea
\dim \CM_{(N_f,G_2)} = 7N_{f}~.  \label{dimG2}
\eea

\section{The $SU(N_c)$ Gauge Groups} \setall \label{SU}
Let us consider the $SU(N_c)$ theory with $N_{f}$ chiral superfields transforming in the fundamental representation, $N_{f}$ chiral superfields transforming in the antifundamental representation (\emph{i.e.} $N_f$ flavours), and 1 chiral superfield transforming in the adjoint representation. The anomaly-free global symmetry of this theory \cite{Kutasov:1995ss} is $SU(N_f) \times SU(N_f) \times U(1)_B \times U(1)_{R_1} \times U(1)_{R_2}$.

\comment{The field content and global charges of this theory are summarised in Table \ref{TabSU}.
\TABLE{
\begin{tabular}{|c||c|cccc|}
\hline
& \textsc{gauge symmetry}  & \multicolumn{4}{c|}{\textsc{global symmetry}}    \\
& $SU(N_c)$ & $SU(N_f)$ & $SU(N_f)$ & U(1)_B & $U(1)_R$  \\
\hline \hline
$Q$ & $\Box$ & $\Box$ &  $1$ &9  \\
$\Phi$ & {\bf Adj} & $\mathbf{1}$ & $0$  \\
\hline
\end{tabular}
\caption{The field content and global charges of the $SO(N_c)$ gauge theory.  We note that $\Box$ represents the fundamental representation, ${\bf Adj}$ represents the adjoint representation, and $\mathbf{1}$ represents the trivial representation.}  \label{TabSU}} }

\subsection{Examples of Hilbert Series}
Below we shall derive Hilbert series for various cases.

\subsubsection{The $SU(2)$ Gauge Group} \label{su2gaugegroup}
We start the analysis by the simplest case of the $SU(2)$ gauge theory with $2N_f$ chiral superfields transforming in the fundamental representation ($N_f$ flavours) \footnote{Note that the number of fundamental chiral superfields must be even due to the global $\mathbb{Z}_2$ anomaly.} and 1 chiral multiplet transforming in the adjoint representation.   The Molien--Weyl formula can be written explicitly as:
\bea
g^{(N_{f},SU(2))}(s,t) &=& \frac{1}{2\pi i} \oint_{|z| =1} \frac{dz}{z}  (1-z^2)~\PE \: \left[ 2N_f [1]t + [2] s \right] \nn \\
&=& \frac{1}{2\pi i} \oint_{|z| =1} \frac{dz}{z} \frac{1-z^2}{(1-tz)^{2N_{f}}(1-\frac{t}{z})^{2N_{f}}(1-s)(1-sz^{2})(1-\frac{s}{z^{2}})}~. \nn \\
\eea
Noting that $0< |t| , |s| <1$, we use the residue theorem with the poles $z =t,\: \sqrt{s}, -\sqrt{s}$ and find that
{\small
\bea
g^{(1,SU(2))}(s,t) &=& \frac{1+st^{2}}{(1-s^{2})(1-t^{2})(1-st^{2})^{2}} \nn \\
&=& 1+s^2+s^4+t^2+3 s t^2+s^2 t^2+3 s^3 t^2+s^4 t^2+3 s^5 t^2+t^4+3 s t^4+6 s^2 t^4+ \nn \\
&& 3 s^3 t^4+6 s^4 t^4+3 s^5 t^4+ O(s^6) O(t^6) ~, \nn \\
g^{(2,SU(2))}(s,t) &=& \frac{1+t^{2}+6st^{2}-9st^{4}+s^{2}t^{4}+st^{6}-9s^{2}t^{6}+6s^{2}t^{8}+s^{3}t^{8}+s^{3}t^{10}}{(1-s^{2})(1-t^{2})^{5}(1-st^{2})^{4}} \nn \\
&=& 1+s^2+s^4+6 t^2+10 s t^2+6 s^2 t^2+10 s^3 t^2+6 s^4 t^2+10 s^5 t^2+20 t^4+ \nn \\
&& 45 s t^4+55 s^2 t^4+45 s^3 t^4+55 s^4 t^4+45 s^5 t^4+  O(s^6) O(t^6)~, \nn \\
g^{(3,SU(2))}(s,t) &=& 1+s^2+s^4+15 t^2+21 s t^2+15 s^2 t^2+21 s^3 t^2+15 s^4 t^2+21 s^5 t^2+105 t^4+ \nn\\ 
&& 210 s t^4+231 s^2 t^4+210 s^3 t^4+231 s^4 t^4+210 s^5 t^4+ O(s^6) O(t^6)~, \nn \\
g^{(4,SU(2))}(s,t) &=& 1+s^2+s^4+28 t^2+36 s t^2+28 s^2 t^2+36 s^3 t^2+28 s^4 t^2+36 s^5 t^2+336 t^4+ \nn \\
&& 630 s t^4+666 s^2 t^4+630 s^3 t^4+666 s^4 t^4+630 s^5 t^4+ O(s^6) O(t^6)~, \nn \\
g^{(5,SU(2))}(s,t) &=& 1+s^2+s^4+45 t^2+55 s t^2+45 s^2 t^2+55 s^3 t^2+45 s^4 t^2+55 s^5 t^2+825 t^4+ \nn \\
&& 1485 s t^4+1540 s^2 t^4+1485 s^3 t^4+1540 s^4 t^4+1485 s^5 t^4+ O(s^6) O(t^6)~. \label{gensu2}
\eea}

Looking at these generating functions, it is possible to predict the order of the numerator and the terms in the denominator of the generating function for a case with $N_{f}$ fundamental quarks:
\bea
g^{(N_{f}, SU(2))} &=& \frac{P_{(2N_{f}-1),(8N_{f}-6)}(s,t)}{(1-s^2)(1-t^2)^{4N_{f}-3}(1-st^2)^{2N_{f}}},
\eea
where $P_{a,b}(s,t)$ is a polynomial of degree $a$ in $s$ and of degree $b$ in $t$.
The calculation is simpler when we make a further identification $s=t$. In which case, we can write down a general form of the generating function:
\bea
g^{(N_{f}, SU(2))} (t) &=& \frac{P_{8N_{f}-6}(t)}{(1+t)^{2N_{f}-3}(1-t^2)^{2N_{f}}(1-t^3)^{2N_{f}}}~, \label{HSsu2}
\eea
where $P_{8N_{f}-6}(t)$ is a palindromic polynomial of degree $8N_{f}-6$ in $t$ with $P_{8N_{f}-6}(1) \neq 0$ for all $N_{f}$. 
Observe that the order of the pole at $t=1$ of $g^{(N_{f}, SU(2))} (t)$ is $4N_f$.  Therefore, the dimension of the moduli space $\CM_{(N_f, SU(2))}$ is $4N_f$, in agreement with \eref{dimSU} and \eref{dimSp}.

\paragraph{Character expansion.} We can write down the generating function for an \emph{arbitary} number of flavous $N_f$ in terms of representations of the global symmetry $SU(2N_f)$ as follows:
\bea
g^{(N_{f},SU(2))} &=&  \sum^{\infty}_{n_1=0} \sum^{\infty}_{n_2=0} \sum_{m=0}^\infty [2n_1,n_2,0, \ldots, 0] {s^{n_1+2m}t^{2n_1+2n_2}} \nn \\
&=& \frac{1}{1-s^{2}}\sum^{\infty}_{n_1=0} \sum^{\infty}_{n_2=0} [2n_1,n_2,0, \ldots, 0] {s^{n_1}t^{2n_1+2n_2}}~.
\eea
We emphasise that $\frac{1}{1-s^2}$ does factor out from the character expansion.


\paragraph{Plethystic logarithms.} We shall calculate the plethystic logarithms of the generating functions in \eref{gensu2} using formula \eref{PL}:
\bea \label{PLsu2}
\PL[g^{(1, SU(2))}(s,t) ] &=& s^{2}+t^{2}+3st^{2} - s^{2}t^{4}~, \nn \\
\PL[g^{(2, SU(2))}(s,t)] &=& s^2+6 t^2+10 s t^2-t^4-15 s t^4-20 s^2 t^4 + O(s^6) O(t^6)~, \nn \\
\PL[g^{(3, SU(2))}(s,t)] &=& s^2+15 t^2+21 s t^2-15 t^4-105 s t^4-105 s^2 t^4 + O(s^6) O(t^6) \nn \\
\PL[g^{(4, SU(2))}(s,t)] &=& s^{2}+28t^{2}+ 36st^{2} - 70t^{4} - 378st^{4} - 336s^{2}t^{4} + O(s^6) O(t^6) \nn \\
\PL[g^{(5, SU(2))}(s,t)] &=& s^{2}+45t^{2}+ 55st^{2} - 210t^{4} - 990st^{4} - 825s^{2}t^{4} +  O(s^6) O(t^6)~. \nn \\ 
\eea
Observe that only plethystic logarithm of $g^{(1, SU(2))}$ is a polynomial.  Therefore, the moduli space of the $SU(2)$ gauge theory with 1 flavour and 1 adjoint matter is a \emph{complete intersection}.  

\paragraph{Generators of the GIOs.} According to \eref{PLsu2}, we see that there are only 3 types of generators of the GIOs in the $SU(2)$ theory, namely
\beq
\ba{llllll}
\mbox{\bf Casimir invariants}: & s^{2}  &\rightarrow & u \equiv {\rm Tr} (\phi^{2})  &\hbox{:} \quad  [0,\ldots,  0] & 1~\text{dimensional}~, \nn  \\
\mbox{\bf Mesons}: & t^{2}  &\rightarrow &  M^{i j} \equiv \epsilon^{a b} Q^{i}_{a} \: {Q}^{j}_{b} &\hbox{:}\quad [0,1,\ldots,0] &  {2N_f \choose 2}~\text{dimensional}~,\nn \\
\mbox{\bf Adjoint mesons}: & st^{2} &\rightarrow &  A^{i j} \equiv \epsilon^{ab} \epsilon^{cd} Q^{i}_{a} \phi_{bc} \:{Q}^{j}_{d} &\hbox{:}\quad [2,0,\ldots,0] &  N_f(2N_f+1) ~\text{dimensional}~.
\ea
\eeq
Note that the total number of generators is quadratic in $N_{f}$,
\bea
1+ {2N_f \choose 2} + N_f(2N_f+1) = 4N_f^2+1~. \label{totgensu2}
\eea

\paragraph{Relations between the generators.} From plethystic logarithms \eref{PLsu2},  we see that there are 3 types of basic relations:
\begin{itemize}
\item {\bf Order $t^4$:} The relations are known from the theory without adjoint:
\bea
\mathrm{Pf}~M = \epsilon_{i_1  \ldots i_{2N_f}} M^{i_1 i_2} M^{i_3 i_4} = 0~.
\eea
They transform in the $SU(2N_f)$ representation $[0,0,0,1,0,\ldots,0]$.  We note that this is contained in the decomposition of the symmetric square of the representation $[0,1,0,\ldots,0]$ at order $t^2$.
\item {\bf Order $st^4$:} The relations transform in the representation $[1,0,1,0,\ldots,0]$, which is contained in the decomposition of the antisymmetric product of the representation $[2,0,\ldots,0]$ at order $st^2$ and the representation $[0,1,0,\ldots,0]$ at order $t^2$. 
\item {\bf Order $s^2t^4$:} The relations transform in the representation $[0,2,0,\ldots,0]$, which is contained in the decomposition of the symmetric square of the representation $[2,0,\ldots,0]$ at order $st^2$.  In the case of 1 flavour, there is only 1 basic relation which can be written out explicitly as 
\bea
A^{11} A^{22} - \left( A^{12} \right)^2 + \frac{1}{2} u \left( M^{12} \right)^2 = 0~. \label{relsu2}
\eea
\end{itemize}
In summary, for the $SU(2)$ theory, we have the basic relations which transform in the $SU(2N_f)$ representations $[0,0,0,1,0,\ldots,0]$, $[1,0,1,0,\ldots,0]$, and $[0,2,0,\ldots,0]$.  

Therefore, we may write down a general expression of the plethystic logarithm in terms of $SU(2N_f)$ representations as
\bea
\PL[g^{(N_f, SU(2))}(s,t)] &=& [0, \ldots,0] s^{2}+ [0,1,\ldots,0] t^{2}+ [2,0,\ldots,0] st^{2} \nn \\
&& - [0,0,0,1,0,\ldots,0]t^{4} - [1,0,1,0,\ldots,0]st^{4} - [0,2,0,\ldots,0]s^{2}t^{4} +  O(s^6) O(t^6)~. \nn \\
\eea





\subsubsection{The $SU(3)$ Gauge Group} \label{trick}
Now let us turn to the $SU(3)$ theory with $N_f$ flavours and 1 adjoint matter.  We have $N_{f}$ chiral superfields transforming in the fundamental representation, $N_{f}$ chiral superfields transforming in the antifundamental representation, and 1 chiral superfield transforming in the adjoint representation.  Therefore, we can apply the Molien--Weyl formula to our theory as follows:
{\small
\bea
 g^{(N_{f},SU(3))} &=& \frac{1}{(2 \pi i)^2} \oint \limits_{|z_1|=1} \frac{dz_{1}}{z_{1}} \oint\limits_{|z_2|=1} \frac{dz_{2}}{z_{2}}  \left(1-z_{1}z_{2}\right)\left(1-\frac{z^{2}_{1}}{z_{2}}\right)\left(1-\frac{z^{2}_{2}}{z_{1}}\right) \PE\left[ N_{f}[1,0]t  +  N_{f}[0,1]\tilde{t}  +  [1,1]s \right]  \nn \\
 &=&   \frac{1}{(2 \pi i)^2} \oint \limits_{|z_1|=1} \frac{dz_{1}}{z_{1}} \oint \limits_{|z_2|=1} \frac{dz_{2}}{z_{2}}\frac{\left(1-z_{1}z_{2}\right)\left(1-\frac{z^{2}_{1}}{z_{2}}\right)\left(1-\frac{z^{2}_{2}}{z_{1}}\right)}{\left((1-tz_{1})(1-t\frac{z_{2}}{z_{1}})(1-\frac{t}{z_{2}}) \right)^{N_{f}} \left( (1-\tilde{t}z_{2})(1-\tilde{t}\frac{z_{1}}{z_{2}})(1-\frac{\tilde{t}}{z_{1}}) \right)^{N_{f}}} \times \nn \\ 
&& \frac{1}{(1-sz_{1}z_{2})(1-s\frac{z^{2}_{1}}{z_{2}})(1-s\frac{z^{2}_{2}}{z_{1}})(1-s\frac{z_{1}}{z^{2}_{2}})(1-s\frac{z_{2}}{z^{2}_{1}})(1-s\frac{1}{z_{1}z_{2}})(1-s)^{2}}~.
\eea}
Applying the residue theorem, we have
{\small
\bea
 g^{(1,SU(3))} (s,t, \tilde{t}) &=& \frac{1-s^6 t^3 \tilde{t}^3}{\left(1-s^2\right) \left(1-s^3\right) \left(1-t \tilde{t}\right) \left(1-s t \tilde{t}\right) \left(1-s^2 t \tilde{t}\right) \left(1-s^3 t^3\right)\left(1-s^3 \tilde{t}^3\right)}  \nn \\ 
&=& 1+s^2+s^3+s^4+s^5+2 s^6+s^3 t^3+s^5 t^3+s^6 t^3+t \tilde{t}+s t \tilde{t}+2 s^2 t \tilde{t}+\nn \\
&& 2 s^3 t \tilde{t}+3 s^4 t \tilde{t}+3 s^5 t \tilde{t}+4 s^6 t \tilde{t}+t^2 \tilde{t}^2+s t^2 \tilde{t}^2+3 s^2 t^2 \tilde{t}^2+3 s^3 t^2 \tilde{t}^2+ \nn \\
&& 5 s^4 t^2 \tilde{t}^2+5 s^5 t^2 \tilde{t}^2+7 s^6 t^2 \tilde{t}^2+s^3 \tilde{t}^3+s^5 \tilde{t}^3+s^6 \tilde{t}^3+t^3 \tilde{t}^3+s t^3 \tilde{t}^3+ \nn \\
&& 3 s^2 t^3 \tilde{t}^3+4 s^3 t^3 \tilde{t}^3+6 s^4 t^3 \tilde{t}^3+7 s^5 t^3 \tilde{t}^3+10 s^6 t^3 \tilde{t}^3+O(s^7)O(t^4)O(\tilde{t}^4)~,  \nn \\
g^{(2,SU(3))} (s,t, \tilde{t}) 
&=& 1+s^2+s^3+s^4+s^5+2 s^6+2 s t^3+2 s^2 t^3+6 s^3 t^3+4 s^4 t^3+8 s^5 t^3+8 s^6 t^3+ \nn \\
&& 4 t \tilde{t}+4 s t \tilde{t}+8 s^2 t \tilde{t}+8 s^3 t \tilde{t}+12 s^4 t \tilde{t}+12 s^5 t \tilde{t}+16 s^6 t \tilde{t}+10 t^2 \tilde{t}^2+ 16 s t^2 \tilde{t}^2+\nn \\
&& 35 s^2 t^2 \tilde{t}^2+41 s^3 t^2 \tilde{t}^2+60 s^4 t^2 \tilde{t}^2+66 s^5 t^2 \tilde{t}^2+ 85 s^6 t^2 \tilde{t}^2+2 s \tilde{t}^3+2 s^2 \tilde{t}^3+\nn \\
&& 6 s^3 \tilde{t}^3+4 s^4 \tilde{t}^3+8 s^5 \tilde{t}^3+8 s^6 \tilde{t}^3+20 t^3 \tilde{t}^3+ 40 s t^3 \tilde{t}^3+96 s^2 t^3 \tilde{t}^3+136 s^3 t^3 \tilde{t}^3+\nn \\
&& 204 s^4 t^3 \tilde{t}^3+ 244 s^5 t^3 \tilde{t}^3+316 s^6 t^3 \tilde{t}^3 +O(s^7)O(t^4)O(\tilde{t}^4)~, \nn \\
g^{(3,SU(3))} (s,t, \tilde{t}) 
&=& 1+s^2+s^3+s^4+s^5+2 s^6+t^3+8 s t^3+9 s^2 t^3+19 s^3 t^3+17 s^4 t^3+27 s^5 t^3+ \nn \\
&& 28 s^6 t^3+9 t \tilde{t}+9 s t \tilde{t}+18 s^2 t \tilde{t}+18 s^3 t \tilde{t}+27 s^4 t \tilde{t}+27 s^5 t \tilde{t}+36 s^6 t \tilde{t}+ \nn \\
&& 45 t^2 \tilde{t}^2+81 s t^2 \tilde{t}^2+162 s^2 t^2 \tilde{t}^2+198 s^3 t^2 \tilde{t}^2+279 s^4 t^2 \tilde{t}^2+ 315 s^5 t^2 \tilde{t}^2+ \nn \\
&& 396 s^6 t^2 \tilde{t}^2+\tilde{t}^3+8 s \tilde{t}^3+9 s^2 \tilde{t}^3+19 s^3 \tilde{t}^3+ 17 s^4 \tilde{t}^3+27 s^5 \tilde{t}^3+ 28 s^6 \tilde{t}^3+\nn \\
&& 165 t^3 \tilde{t}^3+404 s t^3 \tilde{t}^3+893 s^2 t^3 \tilde{t}^3+1301 s^3 t^3 \tilde{t}^3+1881 s^4 t^3 \tilde{t}^3+2289 s^5 t^3 \tilde{t}^3+ \nn \\
&& 2878 s^6 t^3 \tilde{t}^3 +O(s^7)O(t^4)O(\tilde{t}^4)~, \nn\\
g^{(4,SU(3))} (s,t, \tilde{t}) 
&=& 1+s^2+s^3+s^4+s^5+2 s^6+4 t^3+20 s t^3+24 s^2 t^3+44 s^3 t^3+44 s^4 t^3+64 s^5 t^3+ \nn \\
&& 68 s^6 t^3+16 t \tilde{t}+16 s t \tilde{t}+32 s^2 t \tilde{t}+32 s^3 t \tilde{t}+48 s^4 t \tilde{t}+48 s^5 t \tilde{t}+64 s^6 t \tilde{t}+ \nn \\
&& 136 t^2 \tilde{t}^2+256 s t^2 \tilde{t}^2+492 s^2 t^2 \tilde{t}^2+612 s^3 t^2 \tilde{t}^2+848 s^4 t^2 \tilde{t}^2+ 968 s^5 t^2 \tilde{t}^2+ \nn \\
&& 1204 s^6 t^2 \tilde{t}^2+4 \tilde{t}^3+20 s \tilde{t}^3+24 s^2 \tilde{t}^3+44 s^3 \tilde{t}^3+ 44 s^4 \tilde{t}^3+64 s^5 \tilde{t}^3+ 68 s^6 \tilde{t}^3+\nn \\
&& 816 t^3 \tilde{t}^3+2160 s t^3 \tilde{t}^3+4576 s^2 t^3 \tilde{t}^3+6736 s^3 t^3 \tilde{t}^3+9536 s^4 t^3 \tilde{t}^3+11696 s^5 t^3 \tilde{t}^3+ \nn \\
&& 14512 s^6 t^3 \tilde{t}^3 +O(s^7)O(t^4)O(\tilde{t}^4)~, \nn \\
g^{(5,SU(3))} (s,t, \tilde{t}) 
&=& 1+s^2+s^3+s^4+s^5+2 s^6+10 t^3+40 s t^3+50 s^2 t^3+85 s^3 t^3+90 s^4 t^3+125 s^5 t^3+ \nn \\
&& 135 s^6 t^3+25 t \tilde{t}+25 s t \tilde{t}+50 s^2 t \tilde{t}+50 s^3 t \tilde{t}+75 s^4 t \tilde{t}+75 s^5 t \tilde{t}+100 s^6 t \tilde{t}+ \nn \\
&& 325 t^2 \tilde{t}^2+625 s t^2 \tilde{t}^2+1175 s^2 t^2 \tilde{t}^2+1475 s^3 t^2 \tilde{t}^2+2025 s^4 t^2 \tilde{t}^2+ 2325 s^5 t^2 \tilde{t}^2+ \nn \\
&& 2875 s^6 t^2 \tilde{t}^2+10 \tilde{t}^3+40 s \tilde{t}^3+50 s^2 \tilde{t}^3+85 s^3 \tilde{t}^3+ 90 s^4 \tilde{t}^3+125 s^5 \tilde{t}^3+ 135 s^6 \tilde{t}^3+\nn \\
&& 2925 t^3 \tilde{t}^3+8025 s t^3 \tilde{t}^3+16575 s^2 t^3 \tilde{t}^3+24500 s^3 t^3 \tilde{t}^3+34250 s^4 t^3 \tilde{t}^3+42175 s^5 t^3 \tilde{t}^3+ \nn \\
&& 51950 s^6 t^3 \tilde{t}^3 +O(s^7)O(t^4)O(\tilde{t}^4)~, \nn \\
g^{(6,SU(3))} (s,t, \tilde{t}) 
&=& 1+s^2+s^3+s^4+s^5+2 s^6+20 t^3+70 s t^3+90 s^2 t^3+146 s^3 t^3+160 s^4 t^3+216 s^5 t^3+ \nn \\
&& 236 s^6 t^3+36 t \tilde{t}+36 s t \tilde{t}+72 s^2 t \tilde{t}+72 s^3 t \tilde{t}+108 s^4 t \tilde{t}+108 s^5 t \tilde{t}+144 s^6 t \tilde{t}+666 t^2 \tilde{t}^2+ \nn \\
&& 1296 s t^2 \tilde{t}^2+2403 s^2 t^2 \tilde{t}^2+3033 s^3 t^2 \tilde{t}^2+4140 s^4 t^2 \tilde{t}^2+ 4770 s^5 t^2 \tilde{t}^2+  5877 s^6 t^2 \tilde{t}^2+\nn \\
&&20 \tilde{t}^3+70 s \tilde{t}^3+90 s^2 \tilde{t}^3+146 s^3 \tilde{t}^3+ 160 s^4 \tilde{t}^3+216 s^5 \tilde{t}^3+ 236 s^6 \tilde{t}^3+ 8436 t^3 \tilde{t}^3+\nn \\
&& 23576 s t^3 \tilde{t}^3+47888 s^2 t^3 \tilde{t}^3+70904 s^3 t^3 \tilde{t}^3+98316 s^4 t^3 \tilde{t}^3+121332 s^5 t^3 \tilde{t}^3+ \nn \\
&& 148780 s^6 t^3 \tilde{t}^3 +O(s^7)O(t^4)O(\tilde{t}^4)~. 
\eea}
We remark that, although these results seem to be rather lengthy, they contain information which proves to be extremely useful for analyses of the chiral ring.  As we shall see from plethystic logarithms, $s^6 t^3 \tilde{t}^3$ is the minimum order up to which Hilbert series contain all necessary information about the generators and their basic relations.

The calculation is significantly simpler when we make an identification $s=t=\tilde{t}$. In which case, we can write down a general form of the generating function:
\bea
g^{(N_{f},SU(3))}(t)=  \frac{P_{14N_{f}-10}(t)}{(1-t)^{N_{f}-1}(1-t^{2})^{N_{f}+2}(1-t^{3})^{N_{f}+1}(1-t^{4})^{2N_{f}-2}(1-t^{6})^{N_{f}}} ~, \label{HSsu3}
\eea 
where $P_{14N_{f}-10}(t)$ is a palindromic polynomial of degree $14N_f -10$ with $P_{14N_{f}-10}(1)$ being a non-zero number for any number of flavour.
Observe that the order of the pole at $t=1$ of $g^{(N_{f}, SU(3))} (t)$ is $6N_f$. Therefore, the dimension of the moduli space $\CM_{(N_f, SU(3))}$ is $6N_f$ in agreement with \eref{dimSU}.

\paragraph{Plethystic logarithms.} We shall calculate plethystic logarithms of generating functions  using formula \eref{PL}:
{\small
\bea
\PL [g^{(1, SU(3))}(s,t, \tilde{t})] &=& s^2+s^3+t \tilde{t} +s t \tilde{t} +s^2 t \tilde{t}+ s^3 t^3 + s^3\tilde{t}^3 -s^6 t^3 \tilde{t}^3~, \nn \\  
\PL [g^{(2, SU(3))}(s,t, \tilde{t})] &=& s^2+s^3+2 s t^3+2 s^2 t^3+4 s^3 t^3+4 t \tilde{t}+4 s t \tilde{t}+4 s^2 t \tilde{t}-s^2 t^2 \tilde{t}^2-s^3 t^2 \tilde{t}^2-s^4 t^2 \tilde{t}^2 \nn \\
&& +2 s \tilde{t}^3+2 s^2 \tilde{t}^3+4 s^3 \tilde{t}^3-4 s^2 t^3 \tilde{t}^3-8 s^3 t^3 \tilde{t}^3-20 s^4 t^3 \tilde{t}^3-16 s^5 t^3 \tilde{t}^3-16 s^6 t^3 \tilde{t}^3\nn\\
&& + O(s^7)O(t^4)O(\tilde{t}^4) ~, \nn \\
\PL [g^{(3, SU(3))}(s,t, \tilde{t})] &=& s^2+s^3+t^3+8 s t^3+8 s^2 t^3+10 s^3 t^3+9 t \tilde{t}+9 s t \tilde{t}+9 s^2 t \tilde{t}-9 s^2 t^2 \tilde{t}^2-9 s^3 t^2 \tilde{t}^2 \nn \\
&&-9 s^4 t^2 \tilde{t}^2+\tilde{t}^3+8 s \tilde{t}^3+8 s^2 \tilde{t}^3+10 s^3 \tilde{t}^3-t^3 \tilde{t}^3-17 s t^3 \tilde{t}^3 -81 s^2 t^3 \tilde{t}^3-148 s^3 t^3 \tilde{t}^3 \nn \\ 
&&-207 s^4 t^3 \tilde{t}^3-143 s^5 t^3 \tilde{t}^3 - 84 s^6 t^3 \tilde{t}^3 + O(s^7)O(t^4)O(\tilde{t}^4) ~, \nn \\
\PL [g^{(4, SU(3))}(s,t, \tilde{t})] &=& s^2+s^3+4 t^3+20 s t^3+20 s^2 t^3+20 s^3 t^3+16 t \tilde{t}+16 s t \tilde{t}+16 s^2 t \tilde{t}-36 s^2 t^2 \tilde{t}^2 \nn \\
&& -36 s^3 t^2 \tilde{t}^2- 36s^4 t^2 \tilde{t}^2+4 \tilde{t}^3+20 s \tilde{t}^3+20 s^2 \tilde{t}^3+20 s^3 \tilde{t}^3-16 t^3 \tilde{t}^3-176 s t^3 \tilde{t}^3 \nn \\
&& -576 s^2 t^3 \tilde{t}^3-960 s^3 t^3 \tilde{t}^3-1024 s^4 t^3 \tilde{t}^3-624 s^5 t^3 \tilde{t}^3-240 s^6 t^3 \tilde{t}^3 \nn \\
&& +O(s^7)O(t^4)O(\tilde{t}^4) ~, \nn \\
\PL [g^{(5, SU(3))}(s,t, \tilde{t})] &=& s^2+s^3+10 t^3+40 s t^3+40 s^2 t^3+35 s^3 t^3+25 t \tilde{t}+25 s t \tilde{t}+25 s^2 t \tilde{t}-100 s^2 t^2 \tilde{t}^2 \nn \\
&& -100 s^3 t^2 \tilde{t}^2- 100s^4 t^2 \tilde{t}^2+10 \tilde{t}^3+40 s \tilde{t}^3+40 s^2 \tilde{t}^3+35 s^3 \tilde{t}^3-100 t^3 \tilde{t}^3-900 s t^3 \tilde{t}^3 \nn\\
&& -2500 s^2 t^3 \tilde{t}^3-3900 s^3 t^3 \tilde{t}^3-3500 s^4 t^3 \tilde{t}^3-1900 s^5 t^3 \tilde{t}^3-425 s^6 t^3 \tilde{t}^3 \nn\\
&& + O(s^7)O(t^4)O(\tilde{t}^4) ~, \nn \\
\PL [g^{(6, SU(3))}(s,t, \tilde{t})] &=& s^2+s^3+20 t^3+70 s t^3+70 s^2 t^3+56 s^3 t^3+36 t \tilde{t}+36 s t \tilde{t}+36 s^2 t \tilde{t}-225 s^2 t^2 \tilde{t}^2 \nn \\
&& -225 s^3 t^2 \tilde{t}^2-225 s^4 t^2 \tilde{t}^2+20 \tilde{t}^3+70 s \tilde{t}^3+70 s^2 \tilde{t}^3+56 s^3 \tilde{t}^3-400 t^3 \tilde{t}^3-3200 s t^3 \tilde{t}^3 \nn \\
&& -8100 s^2 t^3 \tilde{t}^3-12040 s^3 t^3 \tilde{t}^3-9540 s^4 t^3 \tilde{t}^3-4640 s^5 t^3 \tilde{t}^3-336 s^6 t^3 \tilde{t}^3 \nn \\
&& +O(s^7)O(t^4)O(\tilde{t}^4) ~. \label{plsu3}
\eea}
As for the case of $N_c =2$, the moduli space of the $SU(3)$ theory with 1 flavour and 1 adjoint matter is a complete intersection. 

\paragraph{Generators of the GIOs.} Armed with plethystic logarithms, we can write down the generators of the GIOs. 
\beq
\ba{llll}
\mbox{\bf Casimir invariants}: & s^{k} &\rightarrow & u_k ={\rm Tr} (\phi^{k}),~k=2,\: 3  \nn \\
&&& [0,\ldots,  0; 0, \ldots, 0] \quad 1~\text{dimensional}~, \nn  \\
\mbox{\bf Mesons}: & t \tilde{t}  &\rightarrow &  M^i_j=Q^{i}_{a} \widetilde{Q}^{a}_{j}   \nn \\
&&& [1,0,\ldots,0; 0,\ldots, 0, 1]  \quad N_f^2~\text{dimensional}~, \nn \\
\mbox{\bf Adjoint mesons}: & s^l t \tilde{t}  &\rightarrow &  (A_l)^i_j = Q^{i}_{a} (\phi^l)^a_b   \widetilde{Q}^{b}_{j} ,~l = 1,\:2 \nn \\
&&& [1,0,\ldots,0; 0,\ldots, 0, 1] \quad  N_f^2~\text{dimensional}~,\nn \\
\mbox{\bf Baryons}: & t^{3}  &\rightarrow & B^{i_1 i_2 i_3} = \epsilon^{a_1a_2 a_{3}}  Q^{i_1}_{a_1} {Q}^{i_2}_{a_2} Q^{i_{3}}_{a_{3}} \nn \\
&&& [0,0,1,0, \ldots, 0; 0,\ldots, 0]  \quad  {N_f \choose 3}~\text{dimensional}~,\nn \\
\mbox{\bf Antibaryons}: & \tilde{t}^{3}  &\rightarrow & \widetilde{B}_{i_1 i_2 i_3} = \epsilon_{a_1 a_2 a_{3}}  \widetilde{Q}^{a_1}_{i_1} \widetilde{Q}^{a_2}_{i_2} \widetilde{Q}^ {a_{3}}_{i_{3}}  \nn \\
&&&  [0, \ldots,0 ; 0, \ldots, 1,0, 0] \quad  {N_f \choose 3}~\text{dimensional}~.\nn \\
\ea
\eeq
In addition, we have {\bf adjoint baryons}:
\beq
\label{su3adjbar}
\ba{lll}
s t^{3} &\rightarrow & \CB_{0,0,1}^{i_1 i_2 j_1} = \epsilon^{a_1 a_2 b_1}  Q^{i_1}_{a_1} Q^{i_2}_{a_2}  (P_1)^{j_1}_{b_1}  \nn \\
&& [1,1,0, \ldots, 0; 0,\ldots, 0]^*  \qquad  \frac{1}{3}(N_f-1)N_f(N_f+1) ~\text{dimensional}~, \nn \\
s^2 t^{3} &\rightarrow &  \CB_{0,1,1}^{i_1 j_1 j_2} = \epsilon^{a_1  b_1 b_2}  Q^{i_1}_{a_1} (P_1)^{j_1}_{b_1} (P_1)^{j_2}_{b_2} \; , \; \CB_{0,0,2}^{i_1 i_2 j_1} = \epsilon^{a_1  a_2 b_1}  Q^{i_1}_{a_1} Q^{i_2}_{a_2} (P_2)^{j_1}_{b_1} 
\nn  \\
&& [1,1,0, \ldots, 0; 0,\ldots, 0]^{**} \qquad  \frac{1}{3}(N_f-1)N_f(N_f+1) ~\text{dimensional}~, \nn \\
s^3 t^{3} &\rightarrow & \CB_{1,1,1}^{i j k} = \epsilon^{a  b c}  (P_1)^{i}_{a}  (P_1)^{j}_{b} (P_1)^{k}_{c} \; , \; \CB_{0,1,2}^{i_1 j_1 k_1} = \epsilon^{a_1  b_1 c_1}  Q^{i_1}_{a_1}  (P_1)^{j_1}_{b_1} (P_2)^{k_2}_{c_2} \; , \;  \CB_{0,0,3}^{i_1 i_2 j_1} = \epsilon^{a_1 a_2 b_1}  Q^{i_1}_{a_1} Q^{i_2}_{a_2}  (P_3)^{j_1}_{b_1}  \nn \\
&& [3,0, \ldots, 0; 0,\ldots, 0]^{***}  \qquad \quad   \frac{1}{3!}N_f(N_f+1)(N_f +2) ~\text{dimensional}~, \nn \\
\ea
\eeq
where $(P_{m})^i_a = \phi^{b_1}_{a} \phi^{b_2}_{b_1} \ldots \phi^{b_m}_{b_{m-1}}  Q^i_{b_m}$, and the subscript of $\CB$ indicates the partition of the power of $s$ in the adjoint baryon.  Moreover, in the same spirit as antibaryons, we also have {\bf adjoint antibaryons} which transform in the conjugate representations of adjoint baryons. 

\paragraph{${}^{*}$The generator at order $s t^3$.}  Note that the generator  $\CB_{0,0,1}^{i_1 i_2 j_1}$ is subject to a relation:
\bea
\CB_{0,0,1}^{[i_1 i_2 j_1]} = 0~, \label{rel13}
\eea 
where the square bracket denotes an antisymmetrisation without a normalisation factor. This means that the completely antisymmetric part, which transforms in the $SU(N_f)$ representation $[0,0,1,0,\ldots,0]$, vanishes. Note that we can construct the generator by considering the following $SU(N_f)$ tensor product:
\bea
[0,1,0,\ldots,0] \times [1,0,\ldots,0]= [1,1,0, \ldots, 0] + [0,0,1,0, \ldots, 0]~. \nn
\eea
Therefore, after taking \eref{rel13} into account, we conclude that $\CB_{0,0,1}^{i_1 i_2 j_1}$ transforms in the $SU(N_f) \times SU(N_f)$ representation [1,1,0, \ldots, 0; 0,\ldots, 0], as stated in the list above. 

\paragraph{${}^{**}$Two generators at order $s^2 t^3$.}  We can construct \emph{each} of the generators $\CB_{0,1,1}$ and $\CB_{0,0,2}$ by considering the following $SU(N_f)$ tensor product:
\bea
[0,1,0,\ldots,0] \times [1,0,\ldots,0]= [1,1,0, \ldots, 0] + [0,0,1,0, \ldots, 0]~. \nn 
\eea
Therefore, if there were no relations, we would say that the generators transform in $2[1,1,0, \ldots, 0] + 2[0,0,1,0, \ldots, 0]$.
However, $\CB_{0,1,1}$ and $\CB_{0,0,2}$ are subject to the relations:  
\bea
\CB_{0,0,2}^{i j  k} = -\CB_{0,1,1}^{[i j] k}~, \qquad
\CB_{0,0,2}^{[i j  k]} = - 2\CB_{0,1,1}^{[i j  k]} ~,
\eea
which transforms respectively in the $SU(N_f)$ representation $[1,1,0, \ldots, 0] + [0,0,1,0, \ldots, 0]$ and $[0,0,1,0, \ldots, 0]$.  Therefore, we are left with the global $SU(N_f) \times SU(N_f)$ representation $[1,1,0,\ldots,0; 0, \ldots,0]$, as stated in the above list.  

\paragraph{${}^{***}$Three generators at order $s^3 t^3$.} We can construct the generators $\CB_{1,1,1}$, $\CB_{0,1,2}$, $\CB_{0,0,3}$ from the following $SU(N_f)$ tensor products: 
\bea
\Lambda^3 [1,0,\ldots,0] &=&  [0,0,1,0,\ldots,0]~,  \nn \\
\left[1,0,\ldots,0 \right]^3 &=& [3,0,\ldots,0] + 2[1,1,0,\ldots,0] + [0,0,1,0,\ldots,0]~, \nn \\
\left[0,1,0,\ldots,0 \right] \times [1,0,\ldots,0] &=& [1,1,0, \ldots, 0] + [0,0,1,0, \ldots, 0]  ~. \nn
\eea
Therefore, if there were no relations, we would say that the generators transform in
\bea
[3,0,\ldots,0] + 3[1,1,0,\ldots,0] +3[0,0,1,0,\ldots,0]~.\nn 
\eea
However, these generators are subject to the relations:
\bea
\CB_{0,0,3}^{i j  k} &=& -\CB_{0,1,2}^{[i j] k}~, \\
\CB_{1,1,1}^{i j k} &=& - \CB_{0,1,2}^{i [j k]}~, \\
2\CB_{1,1,1}^{i j k}+ \CB_{0,0,3}^{i k j} &=& - \CB_{0,1,2}^{[i |j| k]} \equiv - \left( \CB_{0,1,2}^{ijk}- \CB_{0,1,2}^{k j i} \right) ~.
\eea
These relations transform in the $SU(N_f)$ reperesentation $3[1,1,0,\ldots,0] +3[0,0,1,0,\ldots,0]$. Therefore, we are left with the global $SU(N_f) \times SU(N_f)$ representation $[3,0,\ldots,0; 0, \ldots,0]$, as stated in the above list. 

\paragraph{Total number of generators.}The total number of generators is cubic in $N_{f}$:
\bea
2+  3 N_f^2+ 2 N_f^3 ~.  \label{totgensu3}
\eea

\paragraph{Dimensions from plethystic logarithms: A trick.}  In the above, we computed dimensions of representations for relations using the following technique.  For definiteness, let us consider the relations at order $s^2 t^3 \tilde{t}^3$.  We know that the dimension $D(N_f)$ of the $SU(N_f) \times SU(N_f)$ representation $[0, 0,1,0, \ldots,0; 0,\ldots,0,1,0,0] + [1,1,0, \ldots,0; 0,\ldots,0,1,0,0]+ [0,0,1,0 \ldots,0; 0,\ldots,0,1,1] + [1,1,0,\ldots,0; 0,\ldots,0,1,1]$ must be a polynomial of order 6 in $N_f$:
\beq
D(N_f)= \sum_{k=0}^6 a_k N_f^k~. \label{D}
\eeq
Observe that we can determine the unknowns $a_0, \ldots, a_6$ from the 7 data points which come from the coefficients of $s^2 t^3 \tilde{t}^3$ in \eref{plsu3} for $N_f = 0, \ldots, 6$ (where the Hilbert series for $N_f=0$ can be obtained by setting $t=\tilde{t} =0$).  Solving the following 7 equations simultaneously
\bea
D(0) &=& 0, \quad D(1) = 0, \quad D(2)=4,\quad D(3) = 81~, \nn \\
D(4) &=& 576, \qquad D(5) = 2500, \qquad D(6) = 8100~,
\eea
we find that
\bea
a_0 &=& 0,\quad a_1 = 0,\quad a_2 =0, \quad a_3 =0, \nn \\
a_4 &=& \frac{1}{4}, \qquad a_5 = -\frac{1}{2}, \qquad a_6=\frac{1}{4}~.
\eea  
Substituting back to \eref{D}, we arrive at
\bea
D(N_f) = \frac{1}{4} N_f^4 (N_f-1)^2~.
\eea

\subsubsection{The $SU(4)$ Gauge Group}
Let us examine the $SU(4)$ theory with $N_f$ flavours and 1 adjoint matter.  We have $N_{f}$ chiral superfields transforming in the fundamental representation, $N_{f}$ chiral superfields transforming in the antifundamental representation, and 1 chiral superfield transforming in the adjoint representation. Therefore, we can apply the Molien--Weyl formula to our theory as follows:
{\small
\bea
 g^{(N_{f},SU(4))} &=& \frac{1}{(2 \pi i)^3} \oint \limits_{|z_1|=1} \frac{dz_{1}}{z_{1}} \oint\limits_{|z_2|=1} \frac{dz_{2}}{z_{2}} \oint\limits_{|z_3|=1} \frac{dz_{3}}{z_{3}}  \left(1-\frac{z_1^2}{z_2}\right) \left(1-\frac{z_1 z_2}{z_3}\right) \left(1-z_1 z_3\right) \left(1-\frac{z_2^2}{z_1 z_3}\right) 
 \times \nn \\ 
 &&  \left(1-\frac{z_2 z_3}{z_1}\right) \left(1-\frac{z_3^2}{z_2}\right)  \PE\left[ N_{f}[1,0,0]t  +  N_{f}[0,0,1]\tilde{t}  +  [1,0,1]s \right]  \nn \\
 &=&  \frac{1}{(2 \pi i)^3} \oint \limits_{|z_1|=1} \frac{dz_{1}}{z_{1}} \oint\limits_{|z_2|=1} \frac{dz_{2}}{z_{2}} \oint\limits_{|z_3|=1} \frac{dz_{3}}{z_{3}}
  \left(1-\frac{z_1^2}{z_2}\right) \left(1-\frac{z_1 z_2}{z_3}\right) \left(1-z_1 z_3\right) \left(1-\frac{z_2^2}{z_1 z_3}\right) 
 \times \nn \\ 
 &&  \left(1-\frac{z_2 z_3}{z_1}\right) \left(1-\frac{z_3^2}{z_2}\right) \times  \frac{1}{\left( \left(1-t z_1\right) \left(1- t \frac{z_2}{z_1}\right) \left(1-t \frac{z_3}{z_2} \right) \left(1- \frac{t}{z_3}\right) \right)^{N_f}} \times \nn \\
&& \frac{1}{\left( \left(1- \frac{\tilde{t}}{z_1}\right)  \left(1- \tilde{t} \frac{z_1}{z_2}\right) \left(1-\tilde{t} \frac{z_2}{z_3} \right) \left(1-\tilde{t} z_3\right) \right)^{N_f}} \times \nn \\
&& \frac{1}{(1-s)^3 \left(1-\frac{s z_1^2}{z_2}\right) \left(1-\frac{s z_2}{z_1^2}\right) \left(1-\frac{s z_2}{z_3^2}\right) \left(1-\frac{s}{z_1 z_3}\right) \left(1-\frac{s z_1}{z_2 z_3}\right) \left(1-\frac{s z_1 z_2}{z_3}\right) \left(1-\frac{s z_2^2}{z_1 z_3}\right)} \times \nn \\
&& \frac{1}{\left(1-s z_1 z_3\right) \left(1-\frac{s z_1 z_3}{z_2^2}\right) \left(1-\frac{s z_3}{z_1 z_2}\right) \left(1-\frac{s z_2 z_3}{z_1}\right) \left(1-\frac{s z_3^2}{z_2}\right)}~. 
\eea}
Applying the residue theorem, we find that
{\small
\bea
 g^{(1,SU(4))}(s,t, \tilde{t}) &=& \frac{1-s^{12} t^4 \tilde{t}^4}{\left(1-s^2\right) \left(1-s^3\right) \left(1-s^4\right) \left(1-t \tilde{t}\right) \left(1-s t \tilde{t}\right) \left(1-s^2 t \tilde{t}\right) \left(1-s^3 t \tilde{t}\right) \left(1-s^6 t^4\right)\left(1-s^6 \tilde{t}^4\right)} \nn\\
&=& 1+s^2+s^3+2 s^4+s^5+3 s^6+2 s^7+4 s^8+3 s^9+5 s^{10}+4 s^{11}+7 s^{12}+s^6 t^4+\nn \\
&& s^8 t^4+ s^9 t^4+2 s^{10} t^4+s^{11} t^4+ 3 s^{12} t^4+ t \tilde{t}+s t \tilde{t}+2 s^2 t \tilde{t}+3 s^3 t \tilde{t}+ 4 s^4 t \tilde{t}+\nn \\
&& 5 s^5 t \tilde{t}+7 s^6 t \tilde{t}+8 s^7 t \tilde{t}+10 s^8 t \tilde{t}+12 s^9 t \tilde{t}+14 s^{10} t \tilde{t}+16 s^{11} t \tilde{t}+19 s^{12} t \tilde{t}+ \nn\\ 
&& t^2 \tilde{t}^2+s t^2 \tilde{t}^2+3 s^2 t^2 \tilde{t}^2+4 s^3 t^2 \tilde{t}^2+7 s^4 t^2 \tilde{t}^2+8 s^5 t^2 \tilde{t}^2+13 s^6 t^2 \tilde{t}^2+ 14 s^7 t^2 \tilde{t}^2+\nn \\
&& 20 s^8 t^2 \tilde{t}^2+22 s^9 t^2 \tilde{t}^2+29 s^{10} t^2 \tilde{t}^2+31 s^{11} t^2 \tilde{t}^2+40 s^{12} t^2 \tilde{t}^2+t^3 \tilde{t}^3+ s t^3 \tilde{t}^3+3 s^2 t^3 \tilde{t}^3+\nn \\
&& 5 s^3 t^3 \tilde{t}^3+8 s^4 t^3 \tilde{t}^3+11 s^5 t^3 \tilde{t}^3+17 s^6 t^3 \tilde{t}^3+21 s^7 t^3 \tilde{t}^3+28 s^8 t^3 \tilde{t}^3+35 s^9 t^3 \tilde{t}^3+ \nn \\
&& 43 s^{10} t^3 \tilde{t}^3+51 s^{11} t^3 \tilde{t}^3+62 s^{12} t^3 \tilde{t}^3+s^6 \tilde{t}^4+s^8 \tilde{t}^4+ s^9 \tilde{t}^4+ 2 s^{10} \tilde{t}^4+s^{11} \tilde{t}^4+3 s^{12} \tilde{t}^4+\nn \\
&&t^4 \tilde{t}^4+s t^4 \tilde{t}^4+3 s^2 t^4 \tilde{t}^4+5 s^3 t^4 \tilde{t}^4+ 9 s^4 t^4 \tilde{t}^4+12 s^5 t^4 \tilde{t}^4+20 s^6 t^4 \tilde{t}^4+25 s^7 t^4 \tilde{t}^4+36 s^8 t^4 \tilde{t}^4+\nn \\
&& 44 s^9 t^4 \tilde{t}^4+58 s^{10} t^4 \tilde{t}^4+ 68 s^{11} t^4 \tilde{t}^4+87 s^{12} t^4 \tilde{t}^4 + O(s^{13})O(t^5)O(\tilde{t}^5)~, \nn \\
 g^{(2,SU(4))}(s,t, \tilde{t}) 
 &=&  1+s^2+s^3+2 s^4+s^5+3 s^6+2 s^7+4 s^8+s^2 t^4+3 s^3 t^4+5 s^4 t^4+7 s^5 t^4+14 s^6 t^4+ \nn \\
 &&14 s^7 t^4+22 s^8 t^4+4 t \tilde{t}+4 s t \tilde{t}+8 s^2 t \tilde{t}+12 s^3 t \tilde{t}+16 s^4 t \tilde{t}+20 s^5 t \tilde{t}+28 s^6 t \tilde{t}+32 s^7 t \tilde{t}+ \nn \\
 && 40 s^8 t \tilde{t}+10 t^2 \tilde{t}^2+16 s t^2 \tilde{t}^2+36 s^2 t^2 \tilde{t}^2+57 s^3 t^2 \tilde{t}^2+87 s^4 t^2 \tilde{t}^2+114 s^5 t^2 \tilde{t}^2+ \nn \\
 && 163 s^6 t^2 \tilde{t}^2+196 s^7 t^2 \tilde{t}^2+255 s^8 t^2 \tilde{t}^2+20 t^3 \tilde{t}^3+40 s t^3 \tilde{t}^3+100 s^2 t^3 \tilde{t}^3+180 s^3 t^3 \tilde{t}^3+\nn \\
 && 296 s^4 t^3 \tilde{t}^3+432 s^5 t^3 \tilde{t}^3+624 s^6 t^3 \tilde{t}^3+816 s^7 t^3 \tilde{t}^3+1064 s^8 t^3 \tilde{t}^3+s^2 \tilde{t}^4+ 3 s^3 \tilde{t}^4+\nn \\
 && 5 s^4 \tilde{t}^4+7 s^5 \tilde{t}^4+14 s^6 \tilde{t}^4+14 s^7 \tilde{t}^4+22 s^8 \tilde{t}^4+35 t^4 \tilde{t}^4+80 s t^4 \tilde{t}^4+215 s^2 t^4 \tilde{t}^4+ \nn \\
 && 425 s^3 t^4 \tilde{t}^4+759 s^4 t^4 \tilde{t}^4+1193 s^5 t^4 \tilde{t}^4+ 1816 s^6 t^4 \tilde{t}^4+2508 s^7 t^4 \tilde{t}^4+3404 s^8 t^4 \tilde{t}^4  +  \nn \\ 
 && O(s^{9})O(t^5)O(\tilde{t}^5)~,\nn \\
 g^{(3,SU(4))}(s,t, \tilde{t}) 
 &=& 1+s^2+s^3+2 s^4+s^5+3 s^6+2 s^7+4 s^8+3 s t^4+9 s^2 t^4+21 s^3 t^4+33 s^4 t^4+48 s^5 t^4+ \nn \\
 && 75 s^6 t^4+90 s^7 t^4+123 s^8 t^4+9 t \tilde{t}+9 s t \tilde{t}+18 s^2 t \tilde{t}+27 s^3 t \tilde{t}+36 s^4 t \tilde{t}+45 s^5 t \tilde{t}+63 s^6 t \tilde{t}+\nn \\
 && 72 s^7 t \tilde{t}+90 s^8 t \tilde{t}+45 t^2 \tilde{t}^2+81 s t^2 \tilde{t}^2+171 s^2 t^2 \tilde{t}^2+279 s^3 t^2 \tilde{t}^2+414 s^4 t^2 \tilde{t}^2+ \nn \\
 && 558 s^5 t^2 \tilde{t}^2+774 s^6 t^2 \tilde{t}^2+954 s^7 t^2 \tilde{t}^2+1215 s^8 t^2 \tilde{t}^2+165 t^3 \tilde{t}^3+405 s t^3 \tilde{t}^3+974 s^2 t^3 \tilde{t}^3+\nn\\
 && 1787 s^3 t^3 \tilde{t}^3+2920 s^4 t^3 \tilde{t}^3+4297 s^5 t^3 \tilde{t}^3+6103 s^6 t^3 \tilde{t}^3+8044 s^7 t^3 \tilde{t}^3+10414 s^8 t^3 \tilde{t}^3+\nn \\
 && 3 s \tilde{t}^4+9 s^2 \tilde{t}^4+21 s^3 \tilde{t}^4+33 s^4 \tilde{t}^4+48 s^5 \tilde{t}^4+75 s^6 \tilde{t}^4+90 s^7 \tilde{t}^4+123 s^8 \tilde{t}^4+495 t^4 \tilde{t}^4+\nn \\
 && 1485 s t^4 \tilde{t}^4+3996 s^2 t^4 \tilde{t}^4+8172 s^3 t^4 \tilde{t}^4+14625 s^4 t^4 \tilde{t}^4+23292 s^5 t^4 \tilde{t}^4+34848 s^6 t^4 \tilde{t}^4+\nn \\
 && 48465 s^7 t^4 \tilde{t}^4+65043 s^8 t^4 \tilde{t}^4+O(s^{9})O(t^5)O(\tilde{t}^5)~, \nn\\
 g^{(4,SU(4))}(s,t, \tilde{t}) 
&=&1+s^2+s^3+2 s^4+s^5+3 s^6+2 s^7+4 s^8+t^4+15 s t^4+36 s^2 t^4+76 s^3 t^4+117 s^4 t^4+ \nn \\
&& 171 s^5 t^4+248 s^6 t^4+312 s^7 t^4+409 s^8 t^4+16 t \tilde{t}+16 s t \tilde{t}+32 s^2 t \tilde{t}+48 s^3 t \tilde{t}+64 s^4 t \tilde{t}+\nn \\
&& 80 s^5 t \tilde{t}+112 s^6 t \tilde{t}+128 s^7 t \tilde{t}+160 s^8 t \tilde{t}+136 t^2 \tilde{t}^2+256 s t^2 \tilde{t}^2+528 s^2 t^2 \tilde{t}^2+868 s^3 t^2 \tilde{t}^2+ \nn\\
&& 1276 s^4 t^2 \tilde{t}^2+1736 s^5 t^2 \tilde{t}^2+2380 s^6 t^2 \tilde{t}^2+2960 s^7 t^2 \tilde{t}^2+3740 s^8 t^2 \tilde{t}^2+816 t^3 \tilde{t}^3+ \nn\\
&& 2176 s t^3 \tilde{t}^3+5152 s^2 t^3 \tilde{t}^3+9488 s^3 t^3 \tilde{t}^3+15424 s^4 t^3 \tilde{t}^3+22720 s^5 t^3 \tilde{t}^3+32032 s^6 t^3 \tilde{t}^3+ \nn \\
&& 42288 s^7 t^3 \tilde{t}^3+54560 s^8 t^3 \tilde{t}^3+\tilde{t}^4+15 s \tilde{t}^4+36 s^2 \tilde{t}^4+76 s^3 \tilde{t}^4+117 s^4 \tilde{t}^4+171 s^5 \tilde{t}^4+\nn \\
&& 248 s^6 \tilde{t}^4+312 s^7 \tilde{t}^4+409 s^8 \tilde{t}^4+3876 t^4 \tilde{t}^4+ 13055 s t^4 \tilde{t}^4+35171 s^2 t^4 \tilde{t}^4+72451 s^3 t^4 \tilde{t}^4+\nn \\ 
&& 129257 s^4 t^4 \tilde{t}^4+205641 s^5 t^4 \tilde{t}^4+305193 s^6 t^4 \tilde{t}^4+423847 s^7 t^4 \tilde{t}^4+565889 s^8 t^4 \tilde{t}^4+\nn \\
&& O(s^{9})O(t^5)O(\tilde{t}^5)~.
\eea}
We emphasise that these results with explicit expansion up to order $s^8 t^4 \tilde{t}^4$, albeit rather lengthy, turn out to be essential for analyses of the generators and their basic relations in the chiral ring.

\paragraph{Plethystic logarithms.} We shall calculate plethystic logarithms of generating functions  using formula \eref{PL}:
\beq \ba{rcl}
\PL \left[ g^{(1,SU(4))} (s, t, \tilde{t}~) \right] &=& s^2+s^3+s^4+t \tilde{t}+s t \tilde{t}+s^2 t \tilde{t}+s^3 t \tilde{t}+s^6 t^4+s^6 \tilde{t}^4-s^{12} t^4 \tilde{t}^4~, \nn \\
\PL \left[ g^{(2,SU(4))} (s, t, \tilde{t}~) \right] &=& s^2 + s^3 + s^4 + s^2 t^4 + 3 s^3 t^4 + 4 s^4 t^4 + 3 s^5 t^4 + 5 s^6 t^4 + 4 t {\tilde{t}} + 4 s t {\tilde{t}} \nn \\
&&  +4 s^2 t {\tilde{t}} + 4 s^3 t {\tilde{t}} -s^3 t^2 {\tilde{t}}^2 - s^4 t^2 {\tilde{t}}^2 - s^5 t^2 {\tilde{t}}^2 - s^6 t^2 {\tilde{t}}^2 + s^2 {\tilde{t}}^4 +3 s^3 {\tilde{t}}^4   \nn\\
&& + 4 s^4 {\tilde{t}}^4 + 3 s^5 {\tilde{t}}^4 + 5 s^6 {\tilde{t}}^4 - s^4 t^4 {\tilde{t}}^4 -6 s^5 t^4 {\tilde{t}}^4 - 17 s^6 t^4 {\tilde{t}}^4 - 30 s^7 t^4 {\tilde{t}}^4 \nn \\
&&  - 44 s^8 t^4 {\tilde{t}}^4 + O(s^9) O(t^5) O({\tilde{t}}^5)~, \nn \\
\PL \left[ g^{(3,SU(4))} (s, t, \tilde{t}~) \right] 
&=& s^2+s^3+s^4+3 s t^4+9 s^2 t^4+18 s^3 t^4+21 s^4 t^4+15 s^5 t^4+15 s^6 t^4\nn \\
&& +9 t \tilde{t}+9 s t \tilde{t}+9 s^2 t \tilde{t}+9 s^3 t \tilde{t}-9 s^3 t^2 \tilde{t}^2-9 s^4 t^2 \tilde{t}^2-9 s^5 t^2 \tilde{t}^2 -9 s^6 t^2 \tilde{t}^2 \nn \\
&& -s^2 t^3 \tilde{t}^3-s^3 t^3 \tilde{t}^3  -s^4 t^3 \tilde{t}^3+17 s^6 t^3 \tilde{t}^3 +17 s^7 t^3 \tilde{t}^3 +17 s^8 t^3 \tilde{t}^3+3 s \tilde{t}^4  \nn \\
&& +9 s^2 \tilde{t}^4+18 s^3 \tilde{t}^4 +21 s^4 \tilde{t}^4+15 s^5 \tilde{t}^4 +15 s^6 \tilde{t}^4 -9 s^2 t^4 \tilde{t}^4 -54 s^3 t^4 \tilde{t}^4 \nn \\
&& -189 s^4 t^4 \tilde{t}^4-441 s^5 t^4 \tilde{t}^4 -783 s^6 t^4 \tilde{t}^4 -1107 s^7 t^4 \tilde{t}^4-1251 s^8 t^4 \tilde{t}^4 \nn \\
&& + O(s^9) O(t^5) O({\tilde{t}}^5)~, \nn \\
\PL \left[ g^{(4,SU(4))} (s, t, \tilde{t}~) \right] 
&=& s^2+s^3+s^4+t^4+15 s t^4+35 s^2 t^4+60 s^3 t^4+65 s^4 t^4+45 s^5 t^4+35 s^6 t^4 \nn \\
&& +16 t \tilde{t}+16 s t \tilde{t}+16 s^2 t \tilde{t}+16 s^3 t \tilde{t}-36 s^3 t^2 \tilde{t}^2-36 s^4 t^2 \tilde{t}^2-36 s^5 t^2 \tilde{t}^2 \nn \\
&& -36 s^6 t^2 \tilde{t}^2-16 s^2 t^3 \tilde{t}^3-16 s^3 t^3 \tilde{t}^3 -16 s^4 t^3 \tilde{t}^3 +176 s^6 t^3 \tilde{t}^3+176 s^7 t^3 \tilde{t}^3\nn \\ 
&& +176 s^8 t^3 \tilde{t}^3 +\tilde{t}^4+15 s \tilde{t}^4+35 s^2 \tilde{t}^4+60 s^3 \tilde{t}^4+65 s^4 \tilde{t}^4 +45 s^5 \tilde{t}^4+35 s^6 \tilde{t}^4 \nn \\
&& -t^4 \tilde{t}^4-31 s t^4 \tilde{t}^4-296 s^2 t^4 \tilde{t}^4-1170 s^3 t^4 \tilde{t}^4-3124 s^4 t^4 \tilde{t}^4 -5912 s^5 t^4 \tilde{t}^4\nn \\
&& -9243 s^6 t^4 \tilde{t}^4-11704 s^7 t^4 \tilde{t}^4-12106 s^8 t^4 \tilde{t}^4 + O(s^9) O(t^5) O({\tilde{t}}^5)~.
\ea \eeq

\paragraph{Generators of the GIOs.} Armed with plethystic logarithms, we can write down the generators of the GIOs.  
\beq
\ba{llll}
\mbox{\bf Casimir invariants}: & s^{k} &\rightarrow & u_k = {\rm Tr} (\phi^{k}),~k=2,\: 3,\: 4  \nn \\
&&& [0,\ldots,  0; 0, \ldots, 0] \quad 1~\text{dimensional}~, \nn  \\
\mbox{\bf Mesons}: & t \tilde{t}  &\rightarrow & M^i_j = Q^{i}_{a} \widetilde{Q}^{a}_{j}   \nn \\
&&& [1,0,\ldots,0; 0,\ldots, 0, 1]  \quad N_f^2~\text{dimensional}~, \nn \\
\mbox{\bf Adjoint mesons}: & s^l t \tilde{t}  & \rightarrow & (A_l)^i_j =  Q^{i}_{a} \: (\phi^l)^a_b \: \widetilde{Q}^{b}_{j},~l = 1,\:2,\:3 \nn \\
&&& [1,0,\ldots,0; 0,\ldots, 0, 1] \quad  N_f^2~\text{dimensional}~, \nn \\
\mbox{\bf Baryons}: & t^{4}  &\rightarrow & B^{i_1 i_2 i_3 i_4} = \epsilon^{a_1a_2 a_{3} a_{4}}  Q^{i_1}_{a_1}  {Q}^{i_2}_{a_2} Q^{i_{3}}_{a_{3}} Q^{i_{4}}_{a_{4}}  \nn \\
&&& [0,0,0,1,0, \ldots, 0; 0,\ldots, 0] \quad  {N_f \choose 4}~\text{dimensional}~,\nn \\
\mbox{\bf Antibaryons}: & \tilde{t}^{4}  &\rightarrow & \text{similar expression to baryons with $Q \rightarrow \widetilde{Q}$} \nn \\
&&& [0, \ldots,0 ; 0, \ldots,0, 1,0,0, 0] \quad  {N_f \choose 4}~\text{dimensional}~.\nn \\
\ea
\eeq
In addition,  we have {\bf adjoint baryons}  (in a similar fashion to the case of $N_c=3$):
\beq
\label{su4adjbar}
\ba{lll}
 s t^{4}  &\rightarrow & \CB_{0,0,0,1}= \epsilon Q {Q}Q P_1 \nn \\
 & & [1,0,1,0\ldots,0; 0, \ldots,0]^{\dagger} \nn \\
 
 s^2 t^{4} &\rightarrow &  \CB_{0,0,0,2} = \epsilon QQQP_2~,~\CB_{0,0,1,1}= \epsilon  Q{Q} P_1P_1 \nn \\ 
 & & [0,2,0,\ldots,0; 0, \ldots,0] + [1,0,1,0,\ldots,0; 0, \ldots,0]^{\ddagger}  \nn \\
 
   s^3 t^{4}  &\rightarrow & \CB_{0,0,0,3}= \epsilon QQQP_3 ~,~ \CB_{0,0,1,2} = \epsilon QQP_1P_2 ~,~ \CB_{0,1,1,1} = \epsilon QP_1P_1P_1\nn \\
   & & [2,1,0,\ldots,0; 0, \ldots,0] + [1,0,1,0,\ldots,0; 0, \ldots,0]  \nn \\
   
   s^4 t^{4}  &\rightarrow &  \CB_{0,0,0,4}= \epsilon QQQP_4 ~,~ \CB_{0,0,1,3}= \epsilon QQP_1P_3 ~,~\CB_{0,0,2,2}= \epsilon QQP_2P_2  \nn \\
   && \CB_{0,1,1,2}= \epsilon QP_1P_1P_2~,~\CB_{1,1,1,1} = \epsilon P_1 P_1P_1P_1 \nn \\
& & [2,1,0,\ldots,0; 0, \ldots,0] + [0,2,0,\ldots,0; 0, \ldots,0] \nn \\ 

 s^5 t^{4} &\rightarrow &  \CB_{0,0,0,5}= \epsilon QQQP_5 ~,~  \CB_{0,0,1,4}= \epsilon QQP_1P_4 ~,~  \CB_{0,0,2,3}= \epsilon QQP_2P_2 \nn \\
 &&  \CB_{0,1,1,3}= \epsilon QP_1P_1P_3 ~,~  \CB_{0,1,2,2}= \epsilon QP_1P_2P_2  ~,~  \CB_{1,1,1,2}= \epsilon P_1 P_1P_1P_2 \nn \\
   & & [2,1,0,\ldots,0; 0, \ldots,0] \nn \\ 

 s^6 t^{4} &\rightarrow &   \CB_{0,0,0,6}= \epsilon QQQP_6 ~,~ \CB_{0,0,1,5}= \epsilon QQP_1P_5 ~,~ \CB_{0,0,3,3}= \epsilon QQP_3P_3 \nn \\
 && \CB_{0,1,1,4}= \epsilon Q P_1 P_1P_4 ~,~ \CB_{0,1,2,3}= \epsilon Q P_1 P_2P_3 ~,~ \CB_{0,2,2,2}= \epsilon Q P_2 P_2P_2\nn \\
 && \CB_{1,1,1,3}= \epsilon P_1 P_1 P_1P_3 ~,~  \CB_{1,1,2,2}= \epsilon P_1 P_1 P_2P_2 ~,~  \CB_{0,0,2,4}= \epsilon Q Q P_2P_4 \nn\\
   & &  [4,0,\ldots,0;0,\ldots,0]
\ea
\eeq
\noindent where $(P_{m})^i_a = \phi^{b_1}_{a} \phi^{b_2}_{b_1} \ldots \phi^{b_m}_{b_{m-1}}  Q^i_{b_m}$, and the subscript of $\CB$ indicates the partition of the power of $s$ in the adjoint baryon.  In the above, we suppressed the indices with the understanding that each epsilon tensor is contracted over all colour indices.  Moreover, we have {\bf adjoint antibaryons} which transform in the conjugate representations of adjoint baryons.  

As for the case of $SU(3)$ gauge group, we emphasise that the representations written above are \emph{not} the ones in which  the generators transform; however, they are the ones in which the relations have already been taken into account.  For example,

\paragraph{${}^{\dagger}$The generator at order $s t^4$.}  Note that the generator {\footnotesize $\CB_{0,0,0,1}^{i_1 i_2 i_3 j_1} = \epsilon^{a_1a_2 a_{3} b_{1}}  Q^{i_1}_{a_1} {Q}^{i_2}_{a_2} Q^{i_{3}}_{a_{3}} (P_1)^{j_{1}}_{b_{1}}$} satisfies a relation:
\bea
 \CB_{0,0,0,1}^{[i_1 i_2 i_3 j_1]} = 0~, \label{rel14}
\eea 
where the square bracket denotes an antisymmetrisation without a normalisation factor. This means that the completely antisymmetric part, which transforms in the $SU(N_f)$ representation [0,0,0,1,0,\ldots,0], vanishes. Note that we can construct the generator by considering the following $SU(N_f)$ tensor product:
\bea
[0,0,1,0,\ldots,0] \times [1,0,\ldots,0]= [1,0,1,0, \ldots, 0] + [0,0,0,1,0, \ldots, 0]~. \nn
\eea
Therefore, after taking \eref{rel14} into account, we conclude that $\CB_{0,0,0,1}^{i_1 i_2 i_{3} j_1}$ transforms in the $SU(N_f) \times SU(N_f)$ representation [1,0,1,0, \ldots, 0; 0,\ldots, 0], as stated in the list above. 

\paragraph{${}^{\ddagger}$Two generators at order $s^2 t^4$.}  We can construct $\CB_{0,0,1,1}, \CB_{0,0,0,2}$ by considering the $SU(N_f)$ tensor products:
\bea
[0,1,0,\ldots,0]^2 &=& [0,2,0,\ldots,0] +[1,0,1,0,\ldots,0] +[0,0,0,1,0,\ldots,0]~, \nn \\
\left[0,0,1,0, \ldots,0 \right] \times [1,0,\ldots,0] &=&  \left[1,0,1,0, \ldots,0 \right]+ \left[0,0,0,1,0, \ldots,0 \right] ~. \nn
\eea
They are however subject to the relations: 
\bea
  \CB_{0,0,0,2}^{[i j k m]} &=& 6\left(\CB_{0,0,0,2}^{i[jkm]} + \CB_{0,0,1,1}^{i[ j k m]} \right)~,  \\
 \CB_{0,0,0,2}^{[i j k m]} &=& -3\CB_{0,0,1,1}^{[i j k m]} ~, 
\eea
which transform respectively in the $SU(N_f)$ representations $ \left[1,0,1,0, \ldots,0 \right]+ \left[0,0,0,1,0, \ldots,0 \right]$,  $\left[0,0,0,1,0, \ldots,0 \right]$.  Therefore, we are left with the global $SU(N_f) \times SU(N_f)$ representation  $[0,2,0,\ldots,0; 0, \ldots,0] + [1,0,1,0,\ldots,0; 0, \ldots,0]$, as stated in the above list.

\paragraph{Total number of generators.} Using the trick mentioned in the previous subsection, we find that the total number of generators is  
\bea
3 +  4 N_f^2+ 2 N_f^4 ~.
\label{totgensu4}
\eea
From \eref{totgensu2}, \eref{totgensu3} and \eref{totgensu4}, we establish the following observation\footnote{From now on, we use the word {\bf Observation} to refer to a strong conjecture which can be deduced, in a consistent manner, from a number of non-trivial results presented earlier.}:
\begin{observation}
The total number of generators in the $SU(N_c)$ theory with $N_f$ fundamental chiral superfields and 1 adjoint chiral superfield is of order $N_f^{N_c}$.  
\end{observation}
\noindent Note that this is substantially higher that the theory with no adjoints.

\paragraph{A comment on representations.}  From a number of examples in the cases of $SU(3)$ and $SU(4)$ gauge groups, we establish the following observations:
\begin{observation} \label{anythingallowed}
Any adjoint baryon of the form $\CB_{\alpha_1, \ldots, \alpha_{N_c}}$ (with $0 \leq \alpha_1 \leq \ldots \leq \alpha_{N_c} \leq N_c$) exists in the theory as a generator.  Note that $N_A \equiv \alpha_1 + \ldots + \alpha_{N_c}$ is the total number of adjoint fields appearing in this particular adjoint baryon.  It satisfies the bounds: $0 \leq N_A \leq \frac{1}{2}N_c(N_c-1)$.
\end{observation}
\begin{observation}
Whenever there is more than one way in partitioning adjoint fields into an adjoint baryon, there exists a relation between those options.  The relation must transform in such a way that it cancels some representations associated with the generators, so that the leftover agrees with plethystic logarithms.
\end{observation}

\subsection{Adjoint Baryons: A Combinatorial Problem of Partitions} \label{partition}
From a combinatorial point of view, Observation \ref{anythingallowed} suggests that an adjoint baryon is simply a partition of the $N_A$ objects into $N_c$ slots (without distinction between the slots).  This leads to an interesting problem:  For given $N_A$ and $N_c$, how many adjoint baryons can be constructed?

\paragraph{The partition function.} This problem can be elegantly solved using a partition function (Hilbert series).  Suppose that the number of slots $N_c$ is held fixed. Let $t$ be a fugacity conjugate to the number of adjoint fields $N_A$.
The required partition function is
\bea
\mathcal{Z}_{N_c}(t) =  \sum_{N_A = 0}^{\infty} a_{N_A, N_c} t^{N_A} ~, \label{z1}
\eea
where $a_{N_A, N_c}$ is the number of adjoint baryons which can be constructed for given $N_A$ and $N_c$.  We can write the partition $\mathcal{Z}$ in another way as follows.  Let $n_i$ be the number of slots which contain $i$ adjoint fields.  It is then easy to see that  $n_1 + 2n_2 +\ldots + N_c n_{N_c}$ is the total number of adjoint fields.  We can therefore write
\bea
\mathcal{Z}_{N_c}(t) &=&  \sum_{\{n_i\}} t^{n_1 + 2n_2 +\ldots + N_c n_{N_c}}  = \frac{1}{(1-t)(1-t^2) \ldots (1-t^{N_c})}  = \prod_{j =1}^{N_c} \frac{1}{1-t^j}  ~.\quad \label{z2}
\eea
This formula is also known as a partition function of $N_c$ bosonic one-dimensional harmonic oscillators \cite{BFHH}. 
Equating \eref{z1} and \eref{z2}, we find that
\bea
 \sum_{N_A=0}^{\infty} a_{N_A, N_c} t^{N_A}=  \prod_{j =1}^{N_c} \frac{1}{1-t^j}~. \label{product}
\eea
Thus, the number of adjoint baryons $a_{N_A, N_c}$ is given by the coefficient of $t^{N_A}$ in the power series of the product $\prod_{j =1}^{N_c} \frac{1}{1-t^j}$~. In other words,
\bea
a_{N_A, N_c} = \frac{1}{2 \pi i} \oint_{|t| =1} \frac{\ud t}{t^{N_A+1}} \prod_{j =1}^{N_c} \frac{1}{1-t^j}~. \label{a}
\eea
We note that this result is correct for any $N_A \geq 0$ but we are particularly interested in the case of $ 0 \leq N_A \leq {Nc \choose 2}$.

\paragraph{Example: $N_c=3$.}  The power series of the last expression in \eref{product} is given by
\bea
\prod_{j =1}^{3} \frac{1}{1-t^j} = 1+t+2 t^2+3 t^3+ \ldots~.
\eea
Therefore, for 0, 1, 2 and 3 adjoint fields, we can construct 1, 1, 2 and 3 adjoint baryons, respectively.  This agrees with the earlier results above Equation \eref{su3adjbar} for the $SU(3)$ gauge group.

\paragraph{Example: $N_c=4$.} The power series of the last expression in \eref{product} is given by
\bea
\prod_{j =1}^{4} \frac{1}{1-t^j} = 1+t+2 t^2+3 t^3+5 t^4+6 t^5+9 t^6+\ldots~.
\eea
Therefore, for 0, 1, 2, 3, 4, 5 and 6 adjoint fields, we can construct 1, 1, 2, 3, 5, 6 and 9 adjoint baryons, respectively.  This agrees with the earlier results above Equation \eref{su4adjbar} for the $SU(4)$ gauge group.

\subsubsection{An Asymptotic Formula}
In this subsection, we derive an asymptotic formula for \eref{a}.  Since the upper bound of the number of adjoint fields $N_A$ is of order $N_c^2$, we consider the limit of large $N_c$ and fixed ratio 
\bea
x \equiv \frac{N_A}{N_c^2}~, \label{x}
\eea
where $x$ is of order one\footnote{For adjoint baryons, we are interested in the range $0 \leq x \lesssim \frac{1}{2}$.}.
Recall from \eref{a} that
\bea
a_{N_A, N_c} =  \frac{1}{2 \pi i} \oint_{|t| =1} \frac{\ud t}{t^{N_A +1}} \mathcal{Z}_{N_c}(t) ~. \label{aa}
\eea 
The main contribution to the integral comes from $t \lesssim 1$, and so we set 
\bea
t = 1- \frac{a}{N_c}~,
\eea
where $a$ is a number of order 1. The behavior of $t$ with $N_c$ is determined by observing that this is a 2 dimensional partition problem in which the scaling of $1-t$ goes like $N_A^{-1/2}$.

\paragraph{Asymptotic formula for $\mathcal{Z}_{N_c}(t)$.} We approximate $\mathcal{Z}_{N_c}(t)$ using the Euler--Maclaurin formula\footnote{This formula states that $\sum_{n=a}^b f(n) \sim \int_{a}^b f(s) \ud s + \frac{1}{2} \left(f(a) + f(b) \right)$~.} as follows: 
\bea
\log \mathcal{Z}_{N_c}(t) &=&   \log \left( \prod_{j =1}^{N_c} \frac{1}{1-t^j} \right) 
= - \sum_{j=1}^{N_c} \log \left( 1-t^j \right) \nn \\
&\sim& -  \int_{1}^{N_c} \log \left( 1-t^s\right) \ud s - \frac{1}{2}  \log \left[ (1-t)\left(1-t^{N_c} \right)\right] \nn \\
&=& -\frac{ \Li_2 (t) - \Li_2 (t^{N_c}) }{ \log t} - \frac{1}{2} \log \left[ (1-t)(1-t^{N_c}) \right] \nn \\
&\sim&  N_c \left(\frac{\pi ^2}{6 a}-\frac{\Li_2 (e^{-a})}{a} \right) - \frac{1}{2} \log N_c + \left[ \frac{1}{2} \log \left( \frac{a}{(e^a-1)(1-e^{-a})^a} \right) - \frac{\pi^2}{12} +\frac {a}{2} -1 \right] ~. \nn \label{preZ}
\eea
Thus, we find that
\bea
\mathcal{Z}_{N_c}(t) \sim \frac{C(a)}{\sqrt{N_c}} \exp \left[ N_c \left(\frac{\pi ^2}{6 a}-\frac{\Li_2 (e^{-a})}{a} \right) \right]~,  \label{asympZ} 
\eea
where $C(a)$ is given by
\bea
C(a) = \sqrt{ \frac{a}{(e^a-1)(1-e^{-a})^a} } \exp \left(- \frac{\pi^2}{12} +\frac {a}{2} -1 \right)~.
\eea

\begin{figure}[htbp]
\begin{center}
\epsfig{file=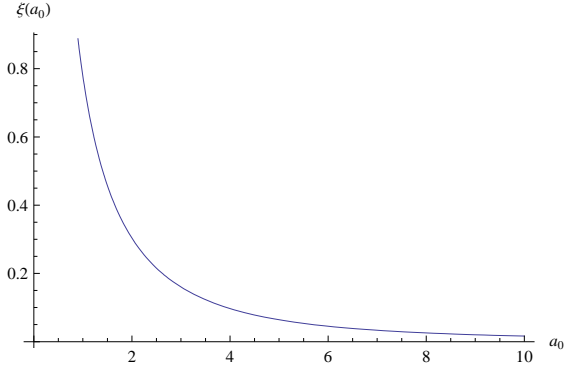,scale=0.7}
\caption{{\sf The graph of $\xi(a_0)$ against $a_0$.}}
\label{f:xi}
\end{center}
\end{figure}

\paragraph{The saddle point method.} Substituting \eref{asympZ} into \eref{a} and writing 
\bea
\frac{1}{t^{N_A+1}} \sim \exp ( -N_A \log t ) \sim  \exp \left[ {N_A(1-t)} \right] =  \exp \left( \frac{N_A}{N_c} a \right) = \exp \left( a x N_c \right) ~,  \nn
\eea
we find that
\bea
a_{N_A, N_c} \sim  \frac{1}{2 \pi i} \oint_{\mathcal{C}}  f(a, N_c) e^{N_c \phi(a)} \ud a~,  \label{asa}
\eea
where the contour $\mathcal{C}$ is taken to be a circle (with a small radius) enclosing the origin $a=0$ in the anticlockwise direction and we define
\bea
f(a, N_c) =  \frac{C(a)}{N_c^{3/2}} \quad, \quad \phi(a) = ax + \frac{1}{a} \left( \frac{\pi^2}{6} -\Li_2 (e^{-a}) \right)  ~. \label{fphi}
\eea
We deform the contour $\mathcal{C}$ to a new contour passing through the location $a_0$ of the saddle point in the direction of steepest descent.  We calculate $a_0$ from the relation $\phi' (a_0) =0$~:
\bea
x = \frac{1}{a_0^2}\left(\frac{\pi ^2}{6} - \Li_2 (e^{-a_0})\right) + \frac{1}{a_0}{\log \left(1-e^{-a_0}\right)} \equiv \xi(a_0)~, \label{sad}
\eea
Since $\xi(a_0)$ is transcendental, it is difficult to obtain an analytical expression of $a_0$ in terms of $x$.  However, given a numerical value of $x$, it is possible to determine the numerical value of $a_0$ (Table \ref{t:a0}).  
\begin{table}[htdp]
\begin{center}
\begin{tabular}{|c|c|}
\hline 
$x$ & $a_0$  \\
\hline
1/2 & 1.405 \\
1/4 & 2.273 \\
1/8 & 3.468 \\
1/10 & 3.934 \\
\hline
\end{tabular}
\end{center}
\caption{\sf \small Numerical values of $a_0$ for various values of $x$.}
\label{t:a0}
\end{table}
The graph of $\xi(a_0)$ against $a_0$ is given in Figure \ref{f:xi}. 

\noindent We note that the second derivative $\phi''(a_0) >0$, and so the steepest descent direction is parallel to the imaginary axis.  
We find that
\bea
a_{N_A, N_c} &\sim& \frac{1}{2 \pi i} \int_{a_0 - i \epsilon}^{a_0+i \epsilon} \ud a ~ f(a_0, N_c) \exp \left(N_c \phi(a_0) + \frac{1}{2}N_c \phi''(a_0)( a - a_0)^2 \right) \nn \\
&=& \frac{1}{2 \pi} e^{N_c \phi(a_0)}  f(a_0, N_c) \int_{- \epsilon}^{ \epsilon}  \ud \alpha~ e^{- \frac{1}{2}N_c \phi''(a_0) \alpha^2}  \qquad  \text{($a= a_0 + i \alpha$)} \nn \\
&\sim&  \frac{1}{\pi} \frac{e^{N_c \phi(a_0)}  f(a_0, N_c)}{\sqrt{2N_c \phi''(a_0)}}  \int_{-\infty}^{\infty}  \ud s ~ e^{-s^2} \qquad \qquad \quad \; \text{$\left( s = \alpha  \sqrt{\frac{1}{2} N_c \phi''(a_0)} \right)$} \nn \\
&\sim& \frac{ e^{N_c \phi(a_0)}  f(a_0, N_c)}{\sqrt{2 \pi N_c \phi''(a_0)}}~. \label{prea}
\eea
Therefore, substituting \eref{fphi} and \eref{sad} into \eref{prea}, we have
\bea
a_{N_A, N_c} = \frac{F(a_0)}{N_c^2} \exp \left(  N_c \left[ \frac{2}{a_0} \left( \frac{\pi^2}{6} -\Li_2 (e^{-a_0}) \right) + \log \left(1-e^{-a_0} \right)  \right] \right)~, \label{asympa}
\eea
where 
\bea
F(a_0) &=& \frac{C(a_0)}{\sqrt{2 \pi \phi''(a_0)}} \ ,  \nn \\
\phi''(a_0) &=& \frac{2}{a_0^3}  \left( \frac{\pi^2}{6} -\Li_2 (e^{-a_0}) \right) + \frac{2}{a_0^2} \log(1- e^{-a_0}) - \frac{1}{a_0(e^{a_0} -1)} \ .
\eea

\paragraph{Numerical values.}  We compare numerical values of the logarithm of \eref{asympa} with the logarithm of the corresponding exact value in Table \ref{t:comp}.
\TABULAR{|c|c|c|c|}{
\hline 
$N_c$ & $x$ &  Logarithm of \eref{asympa} & Logarithm of the exact $a_{N_A, N_c}$ \\
\hline
50 & 1/2 & 74.640 &  75.333 \\
100 & 1/2 & 157.58 & 158.28\\
50 & 1/4 & 53.180 &  53.945 \\
100 & 1/4 & 114.06 & 114.84 \\
100 & 1/8 &  80.008 & 80.837 \\
1000 & 1/10 &  70.975 & 71.828 \\
\hline
}{ {\sf \small Comparison of numerical values of the logarithm of \eref{asympa} with the logarithm of the corresponding exact value.} \label{t:comp}}

\subsection{The Canonical Free Energy}  \label{freeenergy}
An immediate consequence of \eref{dim} is a general form of the unrefined generating function:
\bea
g^{(N_f, SU(N_c))}(t) = \frac{P(t)}{\prod_i (1-t^{n_i})^{d_i}}~,  \label{genform}
\eea 
where $P(t)$ is a palindromic polynomial with $P(1) \neq 0$, and the order of the pole $t=1$ of $g^{(N_f, SU(N_c))}(t)$ is 
\bea
\sum_{i} d_i = \dim \CM_{(N_f, SU(N_c))} = 2N_f N_c~. \label{degree}
\eea

We can define the {\bf canonical free energy} of the system as
\bea
F(t) = - \log  g^{(N_f, SU(N_c))}(t) ~. \label{freeen}
\eea
It is easy to see from \eref{genform}, \eref{degree} and \eref{freeen} that in the large and $N_c$ limit
\bea
F(t) \sim f(t) N_f N_c~, 
\eea
where $f(t)$ is some function of order 1. In other words, in this limit, the canonical free energy scales linearly with the dimension of the moduli space, which in turn is linear in both the number of colours and the number of flavours.

\subsection{Complete Intersection Moduli Space} \label{secCI}
Having seen from a number of examples in preceding subsections that the moduli space of the theories with 1 flavour and 1 adjoint chiral multiplet is a complete intersection, we shall demonstrate that this statement is true for any $SU(N_c)$ gauge group. 

\paragraph{Generators and relations.} The generators of the theories with 1 flavour and 1 adjoint matter are
\beq
\ba{llll}
\mbox{\bf Casimir invariants}: & s^{k} &\rightarrow & u_k = {\rm Tr} (\phi^{k}),~k=2, \ldots,\: N_c  \nn \\
&&&  1~\text{operator for each $k$}~, \nn  \\
\mbox{\bf Meson}: & t \tilde{t}  &\rightarrow & M = Q_{a} \widetilde{Q}^{a}   \nn \\
&&& 1~\text{operator}~, \nn \\
\mbox{\bf Adjoint mesons}: & s^l t \tilde{t}  & \rightarrow & (A_l) =  Q_{a} \: (\phi^l)^a_b \: \widetilde{Q}^{b},~l = 1,
\ldots,\: N_c-1 \nn \\
&&& 1~\text{operator for each $l$}~, \nn \\
\mbox{\bf Adjoint baryon}: & s^{(N_c-1)N_c/2} t^{N_c}  & \rightarrow & \CB_{0,1,2,\ldots,N_c-1} = \epsilon^{ a b c \ldots d}   Q_{a} {(P_1)}_{b} {(P_2)}_{c} \ldots  (P_{N_c-1})_{d}\nn \\
&&& \text{1 operator}~, \nn \\
\mbox{\bf Adjoint antibaryon}: & s^{(N_c-1)N_c/2} \tilde{t}^{N_c}  & \rightarrow & \mbox{adjoint baryon with $Q \rightarrow\widetilde{Q}$}\nn \\
&&& \text{1 operator}~. \nn \\
\ea
\eeq
We see that there are altogether $2N_c+1$ generators.  Note that in all examples we checked the number of relations is 1. We therefore assume that there is precisely one basic relation at order $s^{(N_c-1)N_c} t^{N_c} \tilde{t}^{N_c}$. 

Since the dimension of the moduli space (which is $2N_c$ from \eref{dim}) is equal to the number of generators (which is $2N_c+1$) minus the number of basic relations (which is assumed to be 1), it gives a strong indication that the moduli space is a {\bf complete intersection}.

\paragraph{General formula.} As a consequence, we can write down a fully refined generating function for an arbitrary $N_c$ as
\bea \label{ci}
g^{(1,SU(N_c))}(s,t, \tilde{t}) = \frac{1- s^{(N_c-1)N_c} t^{N_c} \tilde{t}^{N_c}}{(1- s^{(N_c-1)N_c/2} t^{N_c}) (1- s^{(N_c-1)N_c/2} \tilde{t}^{N_c}) \prod_{k=2}^{N_c} (1-s^k)   \prod_{l=0}^{N_c-1} (1-s^l t \tilde{t})}~. \nn \\
\eea

\subsubsection{A General Expression for The Relation} \label{relci}
In \sref{su2gaugegroup}, the relation for the case of $N_c=2, N_f =1$ is written explicitly in \eref{relsu2}.  It is interesting to find a general expression of the relation for any $N_c$ (with $N_f =1$).\footnote{Special thanks to Nathan Seiberg, Kenneth Intriligator and Michael Douglas for discussions.}  In this and only this subsection, we include the factor of $1/k$ into the Casimir invariant $u_k$, namely $u_k =\frac{1}{k} {\rm Tr} (\phi^{k})$.

\paragraph{The case of $N_c=3$.} Let us introduce the operators $A_3$ and $A_4$ in the usual way, \emph{i.e.} $A_l =  Q_{a} \: \phi^l \: \widetilde{Q}^{a}$. Note that they can be written in terms of basic generators as 
\bea 
A_3 =u_3 A_0 + u_2 A_1~, \qquad
A_4 = u_3 A_1 + u_2 A_2~, \label{nc3a}
\eea
where $A_0$ denotes the meson $M$.
Then, the relation can be written as
\bea
\CB \widetilde{\CB} = A_0 A_2 A_4 - A_0 A_3^2 + 2 A_1 A_2 A_3 - A_1^2 A_4 - A_2^3~,  \label{3rel}
\eea
where $\CB$ and $\widetilde{\CB}$ respectively denote the baryon and antibaryon.
The moduli space is $\BC^9/\mathcal{I}_3$, where the ideal $\mathcal{I}_3$ is given by the 3 relations: \eref{nc3a} and \eref{3rel}.

\paragraph{The case of $N_c=4$.} We introduce the operators $A_4, \ldots, A_6$, which can be written in terms of basic generators as
\bea
A_4 &=&  u_4 A_0 - \frac{1}{2} u_2^2 A_0 + u_3 A_1 +  u_2 A_2~, \nn\\
A_5 &=&  u_4 A_1 -\frac{1}{2} u_2^2 A_1 +  u_3 A_2 + u_2 A_3~, \nn\\
A_6 &=& u_2 u_4 A_0 - \frac{1}{2} u_2^3 A_0 + u_2 u_3 A_1 + \frac{1}{2} u_2^2 A_2 +  u_4 A_2 +  u_3 A_3~. \label{nc4a}
\eea
Then, the relation can be written as
\bea
\CB \widetilde{\CB} &=&  A_0 A_2 A_4 A_6 - A_0 A_2 A_5^2 - A_0 A_3^2 A_6 + 2 A_0 A_3 A_4 A_5 - A_0 A_4^3 - A_1^2 A_4 A_6 + A_1^2 A_5^2 \nn\\
&& + 2 A_1 A_2 A_3 A_6 - 2 A_1 A_2 A_4 A_5 - 2 A_1 A_3^2 A_5 + 2 A_1 A_3 A_4^2 - A_2^3 A_6 + 2 A_2^2 A_3 A_5  \nn\\
&& + A_2^2 A_4^2- 3 A_2 A_3^2 A_4 + A_3^4~. \label{4rel}
\eea
The moduli space is $\BC^{12}/\mathcal{I}_4$, where the ideal $\mathcal{I}_4$ is given by the 4 relations: \eref{nc4a} and \eref{4rel}.

\paragraph{A general expression.}  We can generalise \eref{3rel} and \eref{4rel} to any number of colours.  The relation can be written compactly as\footnote{We thank Michael Douglas for pointing out this elegant expression.}
\bea
\CB \widetilde{\CB} &=& \det \mathcal{A}~, \label{magic}
\eea
where $\mathcal{A}_{ij} = A_{i+j}$ and $0 \leq i,j \leq N_c-1$.

It is interesting to examine this formula in the spacial case of $N_c = 2$.  The adjoint baryon is given by $\CB = \epsilon^{ab} Q^1_a (P_1)^1_b = A^{11}$, whereas the adjoint antibaryon is given by $\widetilde{\CB} = \epsilon_{ab} Q^{a2} (P_1)^{b2} = \epsilon_{ab} \epsilon^{ad} Q_{d}^2 \epsilon^{bc} (P_1)_{c}^2 = - \delta^d_b Q_{d}^2  \epsilon^{bc} (P_1)_{c}^2 = - \epsilon^{bc} Q_b ^2 (P_1)_c^2 = - A^{22}~\text{(note the minus sign)}$.  Similarly, it is easy to see that $\det \mathcal{A}= A_0 A_2 - A_1^2 = M^{12} \left( \frac{1}{2} u M^{12} \right) - (A^{12})^2$. We thus correctly recover the formula \eref{relsu2}.

\paragraph{The relations via Newton's formula and the Cayley-Hamilton theorem.} Consider the characteristic polynomial $p(x) = \det (x \mathbb{I} - \phi)$ of $\phi$, which can be written as
\bea
p(x) = \sum_{j=0}^{N_c} x^j s_{N_c - j}~,
\eea
with $s_0 = 1$ and $s_1 = - {\rm Tr}(\phi) = 0$.
Note that $s_k$ is a symmetric polynomial $s_k = (-)^k \sum_{i_1 < \ldots < i_k} \lambda_{i_1} \ldots \lambda_{i_k}$ and $u_k = \frac{1}{k} \sum_{i} \lambda_i^k$, where $\lambda$'s are the eigenvalues of $\phi$.
It follows that the $u$'s and $s$'s are related by Newton's formula (see \emph{e.g.}, \cite{Argyres:1994xh, Hanany:1995na}): 
\bea
n s_{n}+\sum_{j=1}^{n} j u_{j} s_{n - j} = 0 \qquad (2 \leq n \leq N_c)~. \label{us}
\eea
Note that the case of $n=1$ is trivial.
Using the Cayley--Hamilton theorem, one obtains the matrix relation\footnote{It should be noted that if we take a trace of \eref{ch}, we obtain a special case $n=N_c$ of \eref{us}.}:
\bea
p(\phi) =  \sum_{j=0}^{N_c} \phi^j s_{N_c - j} = 0~. \label{ch}
\eea
Multiplying \eref{ch} by $\phi^m$ (with $0 \leq m \leq N_c-2$) and then by $Q$ to the left and $\widetilde{Q}$ to the right, one obtains
\bea
\sum_{j=0}^{N_c} A_{j+m} s_{N_c-j} = 0 \qquad (0 \leq m \leq N_c-2)~. \label{As}
\eea
Note that \eref{us} gives $N_c-1$ relations between $u$'s and $s$'s and \eref{As} gives $N_c-1$ relations between $A$'s and $s$'s.

\paragraph{Counting generators and relations.}  
There are altogether $4N_c-1$ variables: $A_l$ (with $l=0, \ldots, 2N_c-2$); $s_m$, $u_m$ (with $m=2, \ldots ,N_c$); $\CB$, $\widetilde{\CB}$.
However, there are $2N_c-1$ relations: $N_c-1$ relations \eref{us} between $u$'s and $s$'s; $N_c-1$ relations \eref{As} between $A$'s and $s$'s; and 1 relation \eref{magic}.
Thus, by the assumption of the moduli space being a complete intersection, we find that the dimension of the moduli space is $2N_c$, in agreement with the earlier result.

\section{The $Sp(N_c)$ Gauge Groups} \label{secSp}
Let us turn to the $Sp(N_c)$ gauge theory\footnote{We shall use the notation where the rank of $Sp(n)$ is $n$ and $Sp(1)$ is isomorphic to $SU(2)$.} with $2N_{f}$ chiral superfields transforming in the fundamental representation ($N_f$ flavours)\footnote{Note that the number of fundamental chiral multiplets must be even due to the global $\mathbb{Z}_2$ anomaly.} and 1 chiral superfield transforming in the adjoint representation.   The anomaly-free global symmetry of this theory \cite{Luty:1996cg} is $SU(2N_f) \times U(1) \times U(1)_R$.  

\comment{The field content and global charges are summarised in Table \ref{TabSp}.
\TABULAR{|c||c|ccc|}
{\hline
& \textsc{gauge symmetry} & & \textsc{global symmetry} &   \\
& $Sp(N_c)$ & $SU(2N_f)$ & $U(1)$ & $U(1)_R$  \\
\hline \hline
$Q$ & $\Box$ & $\Box$ & $\frac{N_c + 1}{N_f}$ & $1$  \\
$\phi$ & {\bf Adj} & $\mathbf{1}$ & $-1$ & $0$  \\
\hline}{The field content and global charges of the $Sp(N_c)$ gauge theory.  We note that $\Box$ represents the fundamental representation, ${\bf Adj}$ represents the adjoint representation, and $\mathbf{1}$ represents the trivial representation. \label{TabSp}} }

\subsection{Examples of Hilbert Series}
Below we shall derive Hilbert series for various cases.

\subsubsection{The $Sp(2)$ Gauge Group}
Let us now examine the $Sp(2)$ gauge theory with $2N_f$ chiral multiplets in the fundamental representation and 1 chiral superfield in the adjoint representation. The Molien--Weyl formula for this theory is
\bea
g^{(N_{f}, Sp(2))} &=& \int_{Sp(2)} d\mu_{Sp(2)}(z_{1},z_{2}) PE\left[ 2N_f [1,0] t\;+ [2,0] s\right]  \nn \\
&=& \oint \limits_{|z_1|=1} \frac{d z_{1}}{2\pi i z_{1}} \oint \limits_{|z_2|=1} \frac{d z_{2}}{2\pi i z_{2}}\frac{(1-z^{2}_{1})(1-z_{2})(1-\frac{z^{2}_{1}}{z_{2}})(1-\frac{z^{2}_{2}}{z^{2}_{1}})}{\left((1-t z_{1})(1-t\frac{z_{2}}{z_{1}})(1-t\frac{z_{1}}{z_{2}})(1-t\frac{1}{z_{1}})\right)^{2N_{f}}}\times \nn \\
&& \frac{1}{(1-s)^{2}(1-s z^{2}_{1})(1-s z_{2})(1-s\frac{z^{2}_{1}}{z_{2}})(1-s\frac{z^{2}_{2}}{z^{2}_{1}})(1-s\frac{z^{2}_{1}}{z^{2}_{2}})(1-s\frac{z_{2}}{z^{2}_{1}})(1-s\frac{1}{z_{2}})(1-s\frac{1}{z^{2}_{1}})} \nn \\
\eea
Applying the residue theorem, we can compute Hilbert series for various $N_f$:
{\small
\bea
g^{(1,Sp(2))} (s,t) &=& \frac{(1-s^6 t^4)(1-s^4 t^4)}{(1-s^{2})(1-s^{4})(1-t^{2})(1-st^{2})^{3}(1-s^2 t^2)(1-s^{3}t^{2})^{3}}~, \nn \\
&=& 1+s^2+2 s^4+2 s^6+3 s^8+t^2+3 s t^2+2 s^2 t^2+6 s^3 t^2+3 s^4 t^2+9 s^5 t^2+4 s^6 t^2+\nn \\
&& 12 s^7 t^2+5 s^8 t^2+ t^4+3 s t^4+8 s^2 t^4+9 s^3 t^4+18 s^4 t^4+15 s^5 t^4+30 s^6 t^4+ \nn \\ 
&& 21 s^7 t^4+40 s^8 t^4+t^6+3 s t^6+8 s^2 t^6+ 19 s^3 t^6+24 s^4 t^6+43 s^5 t^6+45 s^6 t^6+ \nn \\
&& 74 s^7 t^6+66 s^8 t^6+t^8+3 s t^8+8 s^2 t^8+19 s^3 t^8+39 s^4 t^8+53 s^5 t^8+90 s^6 t^8+\nn \\
&& 102 s^7 t^8+156 s^8 t^8+ O(s^9)O(t^9)~, \nn\\
g^{(2,Sp(2))} (s,t) &=& 1+s^2+2 s^4+2 s^6+3 s^8+6 t^2+10 s t^2+12 s^2 t^2+20 s^3 t^3+18 s^4 t^2+30 s^5 t^2+\nn \\
&& 24 s^6 t^2+40 s^7 t^2+30 s^8 t^2+21 t^4+60 s t^4+111 s^2 t^4+165 s^3 t^4+232 s^4 t^4+270 s^5 t^4+ \nn \\
&& 357 s^6 t^4+375 s^7 t^4+478 s^8 t^8+ 56 t^6+210 s t^6+500 s^2 t^6+890 s^3 t^6+1330 s^4 t^6+\nn\\
&& 1836 s^5 t^6+ 2300 s^6 t^6 +2866 s^7 t^6+ 3270 s^8 t^6+126 t^8+560 s t^8+1560 s^2 t^8+ \nn \\
&& 3220 s^3 t^8+5405 s^4 t^8+7991 s^5 t^8+10955 s^6 t^8+13906 s^7 t^8+18190 s^8 t^8+ O(s^9)O(t^9)~, \nn \\
g^{(3,Sp(2))} (s,t)&=& 1+s^2+2 s^4+2 s^6+3 s^8+15 t^2+21 s t^2+30 s^2 t^2+42 s^3 t^2+45 s^4 t^2+63 s^5 t^2+ \nn \\
&& 60 s^6 t^2+84 s^7 t^2 +75 s^8 t^2+120 t^4+315 s t^4+561 s^2 t^4+840 s^3 t^4+1122 s^4 t^4+\nn \\
&& 1365 s^5 t^4+1689 s^6 t^4+1890 s^7 t^4+2250 s^8 t^4+679 t^6+2485 s t^6+5530 s^2 t^6+ \nn \\
&& 9436 s^3 t^6+13895 s^4 t^6+18571 s^5 t^6+23310 s^6 t^6+28168 s^7 t^6 +32725 s^8 t^6+ \nn \\
&& 3045 t^8+13770 s t^8+36120 s^2 t^8+70050 s^3 t^8+113190 s^4 t^8+162960 s^5 t^8+\nn \\
&& 216825 s^6 t^8+272160 s^7 t^8+328896 s^8 t^8+  O(s^9)O(t^9)~, \nn \\
g^{(4,Sp(2))} (s,t)&=& 1+s^2+2 s^4+2 s^6+3 s^8+28 t^2+36 s t^2+56 s^2 t^2+72 s^3 t^2+84 s^4 t^2+108 s^5 t^2+ \nn \\
&& 112 s^6 t^2+144 s^7 t^2+140 s^8 t^2+406 t^4+1008 s t^4+1786 s^2 t^4+2646 s^3 t^4+3488 s^4 t^4+ \nn \\
&& 4284 s^5 t^4+5198 s^6 t^4+5922 s^7 t^4+6900 s^8 t^4+4032 t^6+14196 s t^6+30576 s^2 t^6+ \nn \\
&& 51192 s^3 t^6+74424 s^4 t^6+98352 s^5 t^6+122892 s^6 t^6+147228 s^7 t^6+171360 s^8 t^6+\nn \\
&& 30744 t^8+ 135120 s t^8+340641 s^2 t^8+640800 s^3 t^8+1015092 s^4 t^8+1438332 s^5 t^8+ \nn \\
&& 1889421 s^6 t^8+2351496 s^7 t^8+2819127 s^8 t^8+O(s^9)(t^9)~.
\eea}
A general form of the generating function when we set $s = t =\tilde{t}$ is
\bea
g^{(N_{f}, Sp(2))} (t) &=& \frac{P_{24N_{f}-16}(t)}{(1+t)^{6N_{f}-7}(1-t^2)^{2N_{f}+1}(1+t^2)(1-t^3)^{4N_{f}-1}(1-t^5)^{2N_{f}}}~.
\eea

\paragraph{Plethystic Logarithms.}
We shall calculate the plethystic logarithms of the generating functions using formula \eref{PL}:
{\small
\bea
\PL \left[g^{(1, Sp(2))}(s,t) \right] &=& s^2+s^4+t^2+3 s t^2+s^2 t^2+3 s^3 t^2-s^4 t^4-s^6 t^4~, \nn \\
\PL \left[g^{(2, Sp(2))}(s,t) \right] &=& s^2+s^4+6 t^2+10 s t^2+6 s^2 t^2+10 s^3 t^2-s^2 t^4-15 s^3 t^4-21 s^4 t^4-15 s^5 t^4 \nn\\
&& -20 s^6 t^4-6 s^2 t^6-10 s^3 t^6+16 s^5 t^6+86 s^6 t^6+90 s^7 t^6+80 s^8 t^6+15 s^3 t^8  \nn \\
&& +51 s^4 t^8+140 s^5 t^8+120 s^6 t^8-31 s^7 t^8-236 s^8 t^8+O(s^9)O(t^9)~, \nn\\
\PL \left[g^{(3, Sp(2))}(s,t) \right] &=& s^2+s^4+15 t^2+21 s t^2+15 s^2 t^2+21 s^3 t^2-15 s^2 t^4-105 s^3 t^4-120 s^4 t^4 \nn\\
&& -105 s^5 t^4-105 s^6 t^4-t^6-35 s t^6-189 s^2 t^6-175 s^3 t^6+36 s^4 t^6+539 s^5 t^6 \nn \\
&& +1589 s^6 t^6+1575 s^7 t^6+1365 s^8 t^6+36 s t^8+540 s^2 t^8+2100 s^3 t^8+5019 s^4 t^8 \nn \\
&& +8118 s^5 t^8+4185 s^6 t^8-4515 s^7 t^8-16869 s^8 t^8+O(s^9)O(t^9)~, \nn \\
\PL \left[g^{(4, Sp(2))}(s,t) \right] &=& s^2+s^4+28 t^2+36 s t^2+28 s^2 t^2+36 s^3 t^2-70 s^2 t^4-378 s^3 t^4-406 s^4 t^4 \nn \\
&& -378 s^5 t^4-336s^6 t^4-28 t^6-420 s t^6-1512 s^2 t^6-1176 s^3 t^6+448 s^4 t^6 \nn \\
&& +4452 s^5 t^6+10920 s^6 t^6+10584 s^7 t^6 +8988 s^8 t^6+63 t^8+1728 s t^8 \nn \\
&& +12481 s^2 t^8+38220 s^3 t^8+80493 s^4 t^8+108432 s^5 t^8+41615 s^6 t^8 \nn\\
&& -81564 s^7 t^8-247521 s^8 t^8+O(s^9)O(t^9)~.
\eea}

\subsubsection{The $Sp(3)$ Gauge Group}\setall
We now move to examining the generating funtions and their plethystic logarithms for the $Sp(3)$ gauge group with $2N_f$ chiral fields transforming in the fundamental representation and one in the adjoint representation of the group. The Molien-Weyl formula for this theory is
\bea
g^{(N_{f}, Sp(3))} &=& \int_{Sp(3)} d\mu_{Sp(3)}(z_{1},z_{2},z_{3}) \PE \left[ 2N_f [1,0,0] t\;+ [2,0,0] s\right]  \nn \\
&=& \oint \limits_{|z_1|=1} \frac{d z_{1}}{2\pi i z_{1}} \oint \limits_{|z_2|=1} \frac{d z_{2}}{2\pi i z_{2}}\oint \limits_{|z_3|=1} \frac{d z_{3}}{2\pi i z_{3}}\frac{(1-z^{2}_{1})(1-\frac{z^{2}_{1}}{z_{2}})(1-z_{2})(1-\frac{z^{2}_{2}}{z^{2}_{1}})(1-\frac{z_{1}z_{2}}{z_{3}})}{(1-s\frac{1}{z^{2}_{1}})(1-s z^{2}_{1})(1-s\frac{z^{2}_{1}}{z^{2}_{2}})(1-s\frac{1}{z_{2}})(1-s\frac{z^{2}_{1}}{z_{2}})}\times\nn \\
&&\frac{(1-\frac{z^{2}_{2}}{z_{1}z_{3}})(1-\frac{z_{3}}{z_{1}})(1-\frac{z_{1}z_{3}}{z_{2}})(1-\frac{z^{2}_{3}}{z^{2}_{2}})
}{(1-s\frac{z^{2}_{2}}{z^{2}_{1}})(1-s\frac{z^{2}_{2}}{z^{2}_{3}})(1-s\frac{z_{1}}{z_{3}})(1-s\frac{z_{2}}{z_{1}z_{3}})(1-s\frac{z_{1}z_{2}}{z_{3}})(1-s\frac{z^{2}}{z_{1}z_{3}})(1-s\frac{z_{3}}{z_{1}})(1-s\frac{z_{1}z_{3}}{z^{2}_{2}})}
\times \nn \\
&&\frac{1}{(1-s\frac{z_{3}}{z_{1}z_{2}})(1-s\frac{z_{2}}{z^{2}_{1}})(1-s\frac{z_{1}z_{3}}{z_{2}})(1-s\frac{z^{2}_{3}}{z^{2}_{2}})(1-s z_{2})(1-s)^{3}((1-t\frac{1}{z_{1}})(1-t z_{1}))^{2N_{f}}}\times\nn \\
&&\frac{1}{((1-t\frac{z_{1}}{z_{2}})(1-t\frac{z_{2}}{z_{1}})(1-t \frac{z_{2}}{z_{3}})(1-t\frac{z_{3}}{z_{2}}))^{2N_{f}}}
\nn \\
\eea
Applying the residue theorem we can compute Hilbert series for various $N_{f}$:
{\small
\bea
g^{(1,Sp(3))} (s,t) &=& \frac{(1-s^6 t^4)(1-s^8 t^4)(1-s^{10}t^4)}{(1-s^{2})(1-s^{4})(1-s^{6})(1-t^{2})(1-st^{2})^{3}(1-s^2 t^2)(1-s^{3}t^{2})^{3}(1-s^4 t^2)(1-s^5 t^2)^{3}}~, \nn \\
&=& 1+s^2+2 s^4+3 s^6+4 s^8+t^2+3 s t^2+2 s^2 t^2+6 s^3 t^2+4 s^4 t^2+12 s^5 t^2+6 s^6 t^2+\nn \\
&& 18 s^7 t^2+9 s^8 t^2+t^4+3 s t^4+8 s^2 t^4+9 s^3 t^4+20 s^4 t^4+21 s^5 t^4+43 s^6 t^4+ \nn \\
&& 36 s^7 t^4+71 s^8 t^4+t^6+3 s t^6+8 s^2 t^6+19 s^3 t^6+26 s^4 t^6+52 s^5 t^6+65 s^6 t^6+ \nn \\
&& 116 s^7 t^6+123 s^8 t^6+t^8+3 s t^8+8 s^2 t^8+19 s^3 t^8+41 s^4 t^8+62 s^5 t^8+116 s^6 t^8+\nn \\
&&157 s^7 t^8+265 s^8 t^8+ O(s^9)O(t^9)~, \nn\\
g^{(2,Sp(3))} (s,t) &=& 1+s^2+2 s^4+3 s^6+4 s^8+6 t^2+10 s t^2+12 s^2 t^2+20 s^3 t^2+24 s^4 t^2+40 s^5 t^2+ \nn \\
&& 36 s^6 t^2+60s^7 t^2+54 s^8 t^2+21 t^4+60 s t^4+112 s^2 t^4+180 s^3 t^4+289 s^4 t^4+405 s^5 t^4+ \nn \\
&& 571 s^6 t^4+690 s^7 t^4+939 s^8 t^4+56 t^6+210 s t^6+512 s^2 t^6+1000 s^3 t^6+1738 s^4 t^6+ \nn \\
&& 2790 s^5 t^6+4094 s^6 t^6+5770 s^7 t^6+7600 s^8 t^6+126 t^8+560 s t^8+1617 s^2 t^8+3700 s^3 t^8+ \nn \\
&& 7257 s^4 t^8+12725 s^5 t^8+20552 s^6 t^8+30860 s^7 t^8+44162 s^8 t^8+ O(s^9)O(t^9)~, \nn\\
g^{(3,Sp(3))} (s,t)&=& 1 + s^2+2 s^4+3 s^6+4 s^8+15 t^2+21 s t^2+30 s^2 t^2+42 s^3 t^2+60 s^4 t^2+84 s^5 t^2+ \nn \\
&& 90 s^6 t^2+126 s^7 t^2+135 s^8 t^2+120 t^4+315 s t^4+576 s^2 t^4+945 s^3 t^4+1467 s^4 t^4+\nn \\
&& 2100 s^5 t^4+2820 s^6 t^4+3570 s^7 t^4+4608 s^8 t^4+680 t^6+2520 s t^6+5944 s^2 t^6+11501 s^3 t^6+ \nn \\
&& 19889 s^4 t^6+31262 s^5 t^6+45629 s^6 t^6+63000 s^7 t^6+83549 s^8 t^6+3060 t^8+14280 s t^8+ \nn \\
&& 40950 s^2 t^8+92274 s^3 t^8+178626 s^4 t^8+308589 s^5 t^8+488922 s^6 t^8+724509 s^7 t^8+\nn \\
&& 1019424 s^8 t^8+ O(s^9)O(t^9)~, \nn \\
g^{(4,Sp(3))} (s,t)&=& 1+s^2+2 s^4+3 s^6+4 s^8+28 t^2+36 s t^2+56 s^2 t^2+72 s^3 t^2+112 s^4 t^2+144 s^5 t^2+\nn \\
&& 168 s^6 t^2+216 s^7 t^2+252 s^8 t^2+406 t^4+1008 s t^4+1856 s^2 t^4+3024 s^3 t^4+4678 s^4 t^4+\nn \\
&& 6678 s^5 t^4+8874 s^6 t^4+11340 s^7 t^4+14442 s^8 t^4+4060 t^6+14616 s t^6+34048 s^2 t^6+\nn \\
&& 65472 s^3 t^6+112280 s^4 t^6+175044 s^5 t^6+253764 s^6 t^6+348276 s^7 t^6+460768 s^8 t^6+\nn \\
&& 31464 t^8+146097 s t^8+414036 s^2 t^8+922593 s^3 t^8+1765443 s^4 t^8+3014997 s^5 t^8+\nn \\
&& 4727424 s^6 t^8+6942129 s^7 t^8+9691512 s^8 t^8+O(s^9)(t^9)~.
\eea}

\paragraph{Plethystic Logarithms.}
We shall calculate the plethystic logarithms of the generating functions using formula \eref{PL}:
{\small
\bea
\PL \left[g^{(1, Sp(3))}(s,t) \right] &=& s^2+s^4+s^6+t^2+3 s t^2+s^2 t^2+3 s^3 t^2+s^4 t^2+3 s^5 t^2-s^6 t^4-s^8 t^4-s^{10}t^4~, \nn \\
\PL \left[g^{(2, Sp(3))}(s,t) \right] &=& s^2+s^4+s^6+6 t^2+10 s t^2+6 s^2 t^2+10 s^3 t^2+6 s^4 t^2+10 s^5 t^2-s^4 t^4-15 s^5 t^4-\nn \\ 
&& 21 s^6 t^4-15 s^7 t^4-21 s^8 t^4-6 s^4 t^6-10 s^5 t^6-6 s^6 t^6-10 s^7 t^6-s^4 t^8+15 s^7 t^8+\nn \\
&& 52 s^8 t^8+O(s^9)O(t^9)~, \nn\\
\PL \left[g^{(3, Sp(3))}(s,t) \right] &=& s^2+s^4+s^6+15 t^2+21 s t^2+15 s^2 t^2+21 s^3 t^2+15 s^4 t^2+21 s^5 t^2-15 s^4 t^4-\nn \\
&& 105 s^5 t^4-120 s^6 t^4-105 s^7 t^4-120 s^8 t^4-s^2 t^6-35 s^3 t^6-190 s^4 t^6-210 s^5 t^6-\nn \\
&& 189 s^6 t^6-175 s^7 t^6+36 s^8 t^6-15 s^2 t^8-105 s^3 t^8-105 s^4 t^8+36 s^5 t^8+555 s^6 t^8+\nn \\
&& 2241 s^7 t^8+5679 s^8 t^8+O(s^9)O(t^9)~, \nn \\
\PL \left[g^{(4, Sp(3))}(s,t) \right] &=& s^2+s^4+s^6+28 t^2+36 s t^2+28 s^2 t^2+36 s^3 t^2+28 s^4 t^2+36 s^5 t^2-70 s^4 t^4-\nn \\
&& 378 s^5 t^4-406 s^6 t^4-378 s^7 t^4-406 s^8 t^4-28 s^2 t^6-420 s^3 t^6-1540 s^4 t^6-1596 s^5 t^6-\nn \\
&& 1512 s^6 t^6-1176 s^7 t^6+448 s^8 t^6-t^8-63 s t^8-720 s^2 t^8-2352 s^3 t^8-1700 s^4 t^8+\nn \\
&& 1791 s^5 t^8+13265 s^6 t^8+42363 s^7 t^8+95521 s^8 t^8+O(s^9)O(t^9)~.
\eea}

\subsection{Generators of the Chiral Ring}
Below we summarise the generators of  $Sp(N_c)$ adjoint SQCD \cite{Leigh:1995qp, Luty:1996cg} and representations of $SU(2N_f) \times SU(2N_f)$ in which they transform.
{\small
\beq
\ba{llllll}
\mbox{\bf Casimir invariants}: & s^{2k} &\rightarrow & u_{2k} ={\rm Tr} (\phi^{2k}) \qquad (k=1,\: 2,\: \ldots, N_c)  \nn \\
&&& [0,\ldots,  0] \quad 1~\text{dimensional}~, \nn  \\
\mbox{\bf Mesons}: & t^2  &\rightarrow &  M^{ij}= J^{ab} Q^{i}_a {Q}^{j}_b  \qquad (a,b=1,\: 2,\: \ldots, 2N_c) \nn \\
&&& [0,1,0,\ldots,0]  \quad N_f(2N_f-1)~\text{dimensional}~, \nn \\
\mbox{\bf Even adjoint mesons}: & s^{2l} t^2 &\rightarrow &  (A_{2l})^{ij} = J^{a_1 b_1} J^{c_1 b_2} J^{c_2 b_3} \ldots J^{c_{2l-1} b_{2l}} J^{c_{2l} a_{2l}}  Q^i_{a_1} \phi_{b_1 c_1} \phi_{b_2 c_2} \ldots \phi_{b_{2l} c_{2l}}  Q^j_{a_{2l}} \nn \\ 
&&& [0,1,0, \ldots,0] \quad N_f(2N_f-1)~\text{dimensional} \quad (l=1, \ldots, N_c-1)~,\nn\\
\mbox{\bf Odd adjoint mesons}: & s^{2k-1} t^2 &\rightarrow &  (A_{2k-1})^{ij} = J^{a_1 b_1} J^{c_1 b_2} \ldots J^{c_{2k-1} a_{2k-1}}  Q^i_{a_1} \phi_{b_1 c_1} \phi_{b_2 c_2} \ldots \phi_{b_{2k-1} c_{2k-1}}  Q^j_{a_{2k-1}}  \nn \\ 
&&& [2,0,\ldots,0] \quad  N_f(2N_f+1)~\text{dimensional} \quad (k=1, \ldots, N_c)~,\nn \\
\ea
\eeq}
The total number of generators is 
\bea
N_c(1+4N_f^2)~ . \label{totgensp}
\eea
We note that for $N_c=1$, $Sp(1)$ is isomorphic to $SU(2)$.  In which case, we recover the $SU(2)$ adjoint SQCD and hence \eref{totgensp} reduces to \eref{totgensu2}.

\subsection{Complete Intersection Moduli Space}
We claim that the moduli space of the $Sp(N_c)$ adjoint SQCD with $2$ fundamental chiral multiplets ($N_f=1$) and 1 adjoint chiral multiplet is a complete intersection.  A general expression of the phethystic logarithm for the complete intersection case can be written as
\bea
\PL[g^{(1, Sp(N_c))}(s,t)] = \sum^{N_c}_{k=1}s^{2k} +\sum^{N_c}_{k=1} (3s^{2k-1}+s^{2(k-1)}) t^2 -\sum^{N_c}_{k=1}s^{2(N_c+k-1)}t^4~.
\eea
Note that the number of relations is equal to the rank of the gauge group and not to 1 as might naively be expected from the case of $SU(N_c)$ gauge group.

\section{The $SO(N_c)$ Gauge Groups} \setall \label{secSO}
Let us turn to the $SO(N_c)$ gauge theory with $N_{f}$ chiral superfields transforming in the fundamental (vector) representation ($N_f$ flavours) and 1 chiral superfield transforming in the adjoint representation.   The global symmetry of this theory \cite{Leigh:1995qp} is $SU(N_f) \times U(1)_R$.  

\comment{The field content and global charges are summarised in Table \ref{TabSO}.
\TABLE{
\begin{tabular}{|c||c|cc|}
\hline
& \textsc{gauge symmetry}  & \multicolumn{2}{c|}{\textsc{global symmetry}}    \\
& $SO(N_c)$ & $SU(N_f)$ &  $U(1)_R$  \\
\hline \hline
$Q$ & $\Box$ & $\Box$ &  $1$  \\
$\phi$ & {\bf Adj} & $\mathbf{1}$ & $0$  \\
\hline
\end{tabular}
\caption{The field content and global charges of the $SO(N_c)$ gauge theory.  We note that $\Box$ represents the fundamental representation, ${\bf Adj}$ represents the adjoint representation, and $\mathbf{1}$ represents the trivial representation.}  \label{TabSO}} }

\subsection{Examples of Hilbert Series}
We shall derive Hilbert series for various cases.  Note that the following subsections are \emph{not} merely a collection of results.  They will turn out to be essential for analysing the generators of the chiral ring. 

\subsubsection{The $SO(3)$ Gauge Group}
Let us now examine the $SO(3)$ gauge theory with $N_f$ chiral multiplets in the fundamental representation and 1 chiral superfield in the adjoint representation. The Molien--Weyl formula for this theory is
\bea
g^{(N_{f},SO(3))} &=& \frac{1}{2\pi i} \oint_{|z| =1} \frac{dz}{z}  (1-z)~\PE \: \left[ N_f [1] t + [1] s \right] \nn \\
&=& \frac{1}{2\pi i} \oint_{|z| =1} \frac{dz}{z} \frac{1-z}{(1-t)^{N_{f}}(1-tz)^{N_{f}}(1-\frac{t}{z})^{N_{f}}(1-s)(1-sz)(1-\frac{s}{z})}~. \quad
\eea
Applying the residue theorem, we find that 
\bea
g^{(1,SO(3))}(s,t) &=& \frac{1}{\left(1-s^2\right) (1-s t) \left(1-t^2\right)}~, \nn \\
g^{(2,SO(3))}(s,t) 
&=& \frac{1- s^2 t^4}{\left(1-s^2\right) (1-s t)^2(1-st^2) \left(1-t^2\right)^3}~, \nn \\
g^{(3,SO(3))}(s,t) &=& \frac{1 - t + t^2 + 3 s t^2 - 3 s t^3 - s^2 t^3 + s^2 t^4 - s^2 t^5}{(1 - s^2) (1 - t)^6 (1 + t)^5 (1 - s t)^3}~, \nn \\
g^{(4,SO(3))}(s,t) &=& 1+s^2+s^4+4 s t+4 s^3 t+10 t^2+6 s t^2+20 s^2 t^2+6 s^3 t^2+20 s^4 t^2+4 t^3 \nn \\
&& +40 s t^3+24 s^2 t^3+60 s^3 t^3+24 s^4 t^3+55 t^4+60 s t^4+155 s^2 t^4+105 s^3 t^4 \nn \\
&& +190 s^4 t^4+ O(s^5) O(t^5) ~.
\eea
\paragraph{Plethystic logarithms.} We shall calculate plethystic logarithms of generating functions  using formula \eref{PL}:
\bea
\PL \left[ g^{(1,SO(3))}(s,t) \right] &=& s^2 + s t + t^2~, \nn \\
\PL \left[ g^{(2,SO(3))}(s,t) \right] &=& s^2 + 2 s t + 3 t^2 + s t^2 - s^2 t^4~, \nn \\
\PL \left[ g^{(3,SO(3))}(s,t) \right] &=& s^2 + 3 s t + 6 t^2 + 3 s t^2 + t^3 - s^2 t^3 - 3 s t^4 - 6 s^2 t^4+ O(s^5) O(t^5)~, \nn \\
\PL \left[ g^{(4,SO(3))}(s,t) \right] &=& s^2+4 s t+10 t^2+6 s t^2+4 t^3-4 s^2 t^3-16 s t^4-21 s^2 t^4+s^3 t^4+ O(s^5) O(t^5)~.\qquad \quad
\eea
The $N_f =1$ moduli space is freely generated, \emph{i.e.} there is no relation between the generators. 
Since the plethystic logarithm for $N_f=2$ is a polynomial (not an infinite series), it follows that the $N_f =2$ moduli space is a complete intersection. 

\comment{
\paragraph{Generators.} Armed with plethystic logarithms, we can write down the generators of the chiral ring.  We remind the reader that $\phi_{ab}$ is an antisymmetric tensor, and so it has only 3 independent components: $\phi_{12}, \phi_{13}, \phi_{23}$.

The generators and $SU(N_f)$ representations in which they transform are listed below:
\beq
\ba{llllll}
\mbox{\bf Casimir invariant}: & s^{2} &\rightarrow & u = {\rm Tr} (\phi^2)  \nn \\
&&& [0,\ldots,  0] \quad 1~\text{dimensional}~, \nn  \\
\mbox{\bf Adjoint-quark invariant}: & s t &\rightarrow & \mathfrak{M}^{i} = \epsilon^{a b c} \phi_{ab} Q^i_{c}~,
\nn \\
&&& [1, 0, \ldots,  0] \quad N_f ~\text{dimensional}~, \nn  \\
\mbox{\bf Mesons}: & t^2  &\rightarrow &  M^{ij}=Q^{i}_a {Q}^{j}_b \delta^{ab}   \nn \\
&&& [2,0,\ldots,0]  \quad \frac{1}{2}N_f(N_f+1)~\text{dimensional}~, \nn \\
\mbox{\bf Adjoint mesons}: & st^2  &\rightarrow &  A^{ij}= Q^{i}_a \phi^{ab} {Q}^{j}_b  \nn \\
&&& [0,1,0,\ldots,0]  \quad \frac{1}{2}N_f(N_f-1)~\text{dimensional}~, \nn \\
\mbox{\bf Baryons}: & t^{3}  &\rightarrow & B^{i_1 i_2 i_3} = \epsilon^{a_1a_2 a_{3}}  Q^{i_1}_{a_1}  {Q}^{i_2}_{a_2} Q^{i_{3}}_{a_{3}} \nn \\
&&& [0,0,1,0, \ldots, 0] \quad  {N_f \choose 3}~\text{dimensional}~.\nn \\
\ea
\eeq

The total number of generators is
\beq
\frac{(6+2N_{f}+N_{f}^{2})(N_{f}+1)}{6} .
\eeq}

\subsubsection{The $SO(4)$ Gauge Group}
Let us now examine the $SO(4)$ gauge theory with $N_f$ chiral multiplets in the fundamental representation and 1 chiral superfield in the adjoint representation. The Molien--Weyl formula for this theory is
\bea
g^{(N_{f},SO(4))} &=& \frac{1}{(2\pi i)^2} \oint_{|z_1| =1}  \frac{dz_1}{z_1} \oint_{|z_2| =1}  \frac{dz_2}{z_2} \left(1-\frac{z_1}{z_2}\right)  (1-z_1 z_2) \nn \\
&& \times \PE \: \left[ N_f [1,0] t + [1,1] s \right]~. \qquad
\eea
Applying the residue theorem, we find that 
{\footnotesize
\bea
g^{(1,SO(4))}(s,t) &=& \frac{1}{\left(1-s^2\right)^2 \left(1-t^2\right) \left(1-s^2 t^2\right)}~, \nn \\
g^{(2,SO(4))}(s,t) &=& \frac{(1-s^2 t^4)(1-s^4 t^4)}{(1-s^2)^2(1-t^2)^3(1-s t^2)^2(1-s^2 t^2)^3}~, \nn \\
g^{(3,SO(4))}(s,t) &=& \frac{1+3 s t^2+3 s^2 t^2+3 s^3 t^4-2 s^2 t^6-8 s^3 t^6-8 s^4 t^6-2 s^5 t^6+3 s^4 t^8+3 s^5 t^{10}+3 s^6 t^{10}+s^7 t^{12}}{\left(1-s^2\right)^2 (1-s t)^3 (1+s t)^3 \left(1-t^2\right)^6 \left(1-s t^2\right)^3}~, \nn \\
g^{(4,SO(4))}(s,t) &=& 1+2 s^2+3 s^4+10 t^2+12 s t^2+30 s^2 t^2+24 s^3 t^2+50 s^4 t^2+56 t^4+120 s t^4+267 s^2 t^4+ \nn \\
&& 330 s^3 t^4+513 s^4 t^4+ O(s^5)O(t^5)~, \nn \\
g^{(5,SO(4))}(s,t) &=& 1+2 s^2+3 s^4+15 t^2+20 s t^2+45 s^2 t^2+40 s^3 t^2+75 s^4 t^2+125 t^4+300 s t^4+620 s^2 t^4+ \nn \\
&& 810 s^3 t^4+1185 s^4 t^4+ O(s^5)O(t^5)~.
\eea}
\paragraph{Plethystic logarithms.} We shall calculate plethystic logarithms of generating functions  using formula \eref{PL}:
\bea
\PL \left[ g^{(1,SO(4))}(s,t) \right] &=& 2s^2 + t^2 +s^2 t^2~, \nn \\
\PL \left[ g^{(2,SO(4))}(s,t) \right] &=&  2 s^2+3 t^2+2 s t^2+3 s^2 t^2-s^2t^4-s^4 t^4~, \nn \\
\PL \left[ g^{(3,SO(4))}(s,t) \right] &=& 2 s^2 + 6 t^2 + 6 s t^2 + 6 s^2 t^2 - 6 s^2 t^4 - 6 s^3 t^4 - 6 s^4 t^4 + O(s^5)O(t^5)~, \nn \\ 
\PL \left[ g^{(4,SO(4))}(s,t) \right] &=& 2 s^2 + 10 t^2 + 12 s t^2 + 10 s^2 t^2 + t^4 - 23 s^2 t^4 - 30 s^3 t^4 - 20 s^4 t^4 + O(s^5)O(t^5)~,\nn \\
\PL \left[ g^{(5,SO(4))}(s,t) \right] &=&  2 s^2 + 15 t^2 + 20 s t^2 + 15 s^2 t^2 + 5 t^4 - 65 s^2 t^4 - 90 s^3 t^4 - 50 s^4 t^4+ O(s^5)O(t^5)~, \nn \\
\PL \left[ g^{(6,SO(4))}(s,t) \right] &=&2 s^2 + 21 t^2 + 30 s t^2 + 21 s^2 t^2 + 15 t^4 - 150 s^2 t^4 -  210 s^3 t^4 - 105 s^4 t^4+ O(s^5)O(t^5)~. \nn \\
\eea
The $N_f =1$ moduli space is freely generated, \emph{i.e.} there is no relation between the generators. 
Since the plethystic logarithm for $N_f=2$ is a polynomial (not an infinite series), it follows that the $N_f =2$ moduli space is a complete intersection.

\subsubsection{The $SO(5)$ Gauge Group}
Let us now examine the $SO(5)$ gauge theory with $N_f$ chiral multiplets in the fundamental representation $[1,0]$ and 1 chiral superfield in the adjoint representation $[0,2]$. The Molien--Weyl formula for this theory is
\bea
g^{(N_{f},SO(5))} &=& \frac{1}{(2\pi i)^2} \oint_{|z_1| =1}  \frac{dz_1}{z_1} \oint_{|z_2| =1}  \frac{dz_2}{z_2}  
\left(1 - z_1 \right) \left(1 -\frac{ z_1^2}{z_2^2} \right) \left(1 - z_2^2 \right) \left(1 - \frac{z_2^2}{z_1} \right) \nn \\
&& \times \PE \: \left[ N_f [1,0] t + [0,2] s \right]~.
\eea
Applying the residue theorem, we find that 
\bea
g^{(1,SO(5))} (s,t) &=& \frac{1}{(1-s^2)(1-  s^4)(1 - s^2 t)(1 - t^2)(1 - s^2 t^2)}~, \nn\\
g^{(2,SO(5))} (s,t) &=&  \frac{(1- s^4 t^4)(1 - s^6 t^4)}{(1- s^2)( 1- s^4)( 1-  s^2 t)^2( 1-  t^2)^3( 1- s t^2)( 1-  s^2 t^2)^3( 1- s^3 t^2)}~, \nn \\
g^{(3,SO(5))} (s,t) &=& \frac{1}{(1 - s)^2 (1 + s)^2 (1 + s^2) (1 - t)^6 (1 + t)^6 (1 - s t)^3 (1 + s t)^3 (1 - s^2 t)^3 (1 - s t^2)^3} \times \nn\\
&& (1+3 s^2 t^2+3 s^3 t^2+s t^3-s^5 t^3-3 s^3 t^4+6 s^5 t^4+3 s^3 t^5-3 s^7 t^5-s^4 t^6-9 s^5 t^6-9 s^6 t^6 \nn \\
&& -s^7 t^6-3 s^4 t^7+3 s^8 t^7+6 s^6 t^8-3 s^8 t^8-s^6 t^9+s^{10} t^9+3 s^8 t^{10}+3 s^9 t^{10}+s^{11} t^{12} )~. \nn\\
\eea
\paragraph{Plethystic logarithms.} We shall calculate plethystic logarithms of generating functions  using formula \eref{PL}:
\bea
\PL \left[ g^{(1,SO(5))} (s,t) \right] &=&s^2 + s^4 + s^2 t + t^2 + s^2 t^2~, \nn \\
\PL \left[ g^{(2,SO(5))} (s,t) \right] &=& s^2 + s^4 + 2 s^2 t + 3 t^2 + s t^2 + 3 s^2 t^2 + s^3 t^2 - s^4 t^4 - s^6 t^4~, \nn \\
\PL \left[ g^{(3,SO(5))} (s,t) \right] &=& s^2 + s^4 + 3 s^2 t + 6 t^2 + 3 s t^2 + 6 s^2 t^2 + 3 s^3 t^2 + s t^3 - s^5 t^3 - 3 s^3 t^4 \nn \\
&& - 6 s^4 t^4 - 3 s^5 t^4 - 6 s^6 t^4 - 3 s^4 t^5 - s^2 t^6 - s^4 t^6 + 9 s^6 t^6 + O(s^7)O(t^7)~, \nn \\
\PL \left[ g^{(4,SO(5))} (s,t) \right] &=& s^2 + s^4 + 4 s^2 t + 10 t^2 + 6 s t^2 + 10 s^2 t^2 + 6 s^3 t^2 + 4 s t^3 - 4 s^5 t^3 - 16 s^3 t^4\nn \\
&&  - 21 s^4 t^4 - 15 s^5 t^4 - 21 s^6 t^4 - 4 s^3 t^5 - 24 s^4 t^5 - 10 s^2 t^6 - 6 s^3 t^6 - 10 s^4 t^6 \nn \\
&& + 10 s^5 t^6 + 102 s^6 t^6 + O(s^7)O(t^7)~, \nn\\
\PL \left[ g^{(5,SO(5))} (s,t) \right] &=& s^2 + s^4 + 5 s^2 t + 15 t^2 + 10 s t^2 + 15 s^2 t^2 + 10 s^3 t^2 + 10 s t^3 - 10 s^5 t^3 \nn \\
&& - 50 s^3 t^4 - 55 s^4 t^4 - 45 s^5 t^4 - 55 s^6 t^4 + t^5 - s^2 t^5 - 24 s^3 t^5 - 101 s^4 t^5 + s^5 t^5   \nn \\
&& + s^6 t^5  - 60 s^2 t^6 - 45 s^3 t^6 - 50 s^4 t^6 + 75 s^5 t^6 + 555 s^6 t^6 + O(s^7)O(t^7)~. \nn\\
 \eea
 The $N_f =1$ moduli space is freely generated, \emph{i.e.} there is no relation between the generators. 
Since the plethystic logarithm for $N_f=2$ is a polynomial (not an infinite series), it follows that the $N_f =2$ moduli space is a complete intersection.

\subsubsection{The $SO(6)$ Gauge Group}
Let us now examine the $SO(6)$ gauge theory with $N_f$ chiral multiplets in the fundamental representation and 1 chiral superfield in the adjoint representation. The Molien--Weyl formula for this theory is
\bea
g^{(N_{f},SO(6))} &=& \frac{1}{(2\pi i)^3} \oint_{|z_1| =1}  \frac{dz_1}{z_1} \oint_{|z_2| =1}  \frac{dz_2}{z_2}  \oint_{|z_3| =1} \frac{dz_3}{z_3}
 \left(1 - \frac{z_2^2}{z_1}\right) \left(1 - \frac{z_1^2}{z_2 z_3} \right) \left(1 - \frac{z_1 z_2}{z_3} \right) \times \nn \\
 &&  \left(1 - \frac{z_1 z_3}{z_2} \right) \left(1 - z_2 z_3 \right) \left(1 - \frac{z_3^2}{z_1} \right) \PE \: \left[ N_f [1,0,0] t + [0,1,1] s \right]~.
\eea
Applying the residue theorem, we find that 
\bea
g^{(1,SO(6))} (s,t) &=& \frac{1}{(1-s^2)(1-s^3)(1-s^4)(1-t^2)(1-s^2 t^2)(1-s^4 t^2)}~, \nn \\
g^{(2,SO(6))} (s,t) &=& \frac{(1-s^4 t^4)(1 - s^6 t^4)(1 - s^8 t^4)}{(1-s^2)(1- s^3)(1- s^4)(1-  t^2)^3(1-  s t^2)( 1-  s^2 t^2)^4(1- s^3 t^2)(1-  s^4 t^2)^3}~, \nn \\
g^{(3,SO(6))} (s,t) &=& 1+s^2+s^3+2 s^4+s^5+3 s^6+2 s^7+4 s^8+3 s^9+5 s^{10}+6 t^2+3 s t^2+15 s^2 t^2\nn\\ 
&& +12 s^3 t^2+30 s^4 t^2 +24 s^5 t^2+48 s^6 t^2+42 s^7 t^2+72 s^8 t^2+63 s^9 t^2+99 s^{10} t^2+21 t^4 \nn \\
&& +18 s t^4+81 s^2 t^4+84 s^3 t^4+204 s^4 t^4 +204 s^5 t^4+381 s^6 t^4+387 s^7 t^4+621 s^8 t^4\nn\\
&& +624 s^9 t^4+915 s^{10} t^4+56 t^6+63 s t^6+281 s^2 t^6+354 s^3 t^6 +867 s^4 t^6+1028 s^5 t^6 \nn \\
&& +1907 s^6 t^6+2182 s^7 t^6+3446 s^8 t^6+3825 s^9 t^6+5474 s^{10} t^6 + O(s^{11})O(t^7)~, \nn \\
g^{(4,SO(6))} (s,t) &=& 1 + s^2 + s^3 + 2 s^4 + s^5 + 3 s^6 + 10 t^2 + 6 s t^2 + 26 s^2 t^2 +  22 s^3 t^2 + 52 s^4 t^2 \nn \\
&&+ 44 s^5 t^2 + 84 s^6 t^2 + 55 t^4 + 61 s t^4 + 236 s^2 t^4  + 272 s^3 t^4 + 602 s^4 t^4 + 654 s^5 t^4 + \nn \\
&& +1139 s^6 t^4 + 220 t^6 + 340 s t^6 + 1316 s^2 t^6 + 1916 s^3 t^6 + 4252 s^4 t^6 + 5540 s^5 t^6  \nn \\
&&  + 9518 s^6 t^6+ O(s^{7})O(t^7)~, \nn \\
g^{(5,SO(6))} (s,t) &=& 1 + s^2 + s^3 + 2 s^4 + s^5 + 3 s^6 + 15 t^2 + 10 s t^2 + 40 s^2 t^2 + 35 s^3 t^2 + 80 s^4 t^2  \nn \\
&& + 70 s^5 t^2 + 130 s^6 t^2 + 120 t^4 + 155 s t^4 + 550 s^2 t^4 + 675 s^3 t^4  + 1415 s^4 t^4 + 1610 s^5 t^4 \nn \\
&&  + 2695 s^6 t^4 + 680 t^6 + 1275 s t^6 + 4555 s^2 t^6 + 7195 s^3 t^6 + 15105 s^4 t^6 + 20650 s^5 t^6  \nn \\
&& + 34055 s^6 t^6 + O(s^{7})O(t^7)~, \nn \\
g^{(5,SO(6))} (s,t) &=& 1 + s^2 + s^3 + 2 s^4 + s^5 + 3 s^6 + 21 t^2 + 15 s t^2 + 57 s^2 t^2 + 51 s^3 t^2 + 114 s^4 t^2   \nn \\
&& +102 s^5 t^2 + 186 s^6 t^2 + 231 t^4 + 330 s t^4 + 1107 s^2 t^4 + 1416 s^3 t^4 + 2865 s^4 t^4 + 3357 s^5 t^4   \nn \\
&& +5478 s^6 t^4 + 1772 t^6 + 3780 s t^6 + 12832 s^2 t^6 + 21316 s^3 t^6 + 43220 s^4 t^6 + 60768 s^5 t^6   \nn \\
&& +97744 s^6 t^6 + O(s^{7})O(t^7)~. 
\eea

\paragraph{Plethystic logarithms.} We shall calculate plethystic logarithms of generating functions  using formula \eref{PL}:
\bea
\PL \left[ g^{(1,SO(6))} (s,t) \right] &=&s^2 + s^3 + s^4 + t^2 + s^2 t^2 + s^4 t^2~, \nn \\
\PL \left[ g^{(2,SO(6))} (s,t) \right] &=& s^2 + s^3 + s^4 + 3 t^2 + s t^2 + 4 s^2 t^2 + s^3 t^2 + 3 s^4 t^2 - s^4 t^4 - s^6 t^4 - s^8 t^4~, \nn \\
\PL \left[ g^{(3,SO(6))} (s,t) \right] &=& s^2 + s^3 + s^4 + 6 t^2 + 3 s t^2 + 9 s^2 t^2 + 3 s^3 t^2 + 6 s^4 t^2 - 6 s^4 t^4 - 3 s^5 t^4  \nn \\
&& - 9 s^6 t^4 - 3 s^7 t^4 - 6 s^8 t^4 - s^4 t^6 - s^5 t^6 - s^6 t^6 + 9 s^8 t^6 + 9 s^9 t^6 \nn \\
&& + 17 s^{10} t^6 +O(s^{11}) O(t^7)~, \nn \\
\PL \left[ g^{(4,SO(6))} (s,t) \right] &=& s^2 + s^3 + s^4 + 10 t^2 + 6 s t^2 + 16 s^2 t^2 + 6 s^3 t^2 + 10 s^4 t^2 + s t^4 - 22 s^4 t^4 \nn \\
&& - 16 s^5 t^4 - 36 s^6 t^4 - 6 s^3 t^6 - 16 s^4 t^6 - 32 s^5 t^6 - 10 s^6 t^6 + O(s^{7}) O(t^7)~,\nn\\
\PL \left[ g^{(5,SO(6))} (s,t) \right] &=& s^2 + s^3 + s^4 + 15 t^2 + 10 s t^2 + 25 s^2 t^2 + 10 s^3 t^2 + 15 s^4 t^2 + 5 s t^4 - 60 s^4 t^4 \nn \\
&& - 50 s^5 t^4 - 100 s^6 t^4 - 55 s^3 t^6 - 105 s^4 t^6 - 220 s^5 t^6 - 45 s^6 t^6 + O(s^{7}) O(t^7)~,\nn\\
\PL \left[ g^{(6,SO(6))} (s,t) \right] &=& s^2 + s^3 + s^4 + 21 t^2 + 15 s t^2 + 36 s^2 t^2 + 15 s^3 t^2 + 21 s^4 t^2 + 15 s t^4 - 135 s^4 t^4 \nn \\
&&- 120 s^5 t^4  - 225 s^6 t^4 + t^6 - s^2 t^6 - 261 s^3 t^6 - 436 s^4 t^6 - 903 s^5 t^6 - 139 s^6 t^6 \nn \\
&& + O(s^{7}) O(t^7)~.
\eea
The $N_f =1$ moduli space is freely generated, \emph{i.e.} there is no relation between the generators. 
Since the plethystic logarithm for $N_f=2$ is a polynomial (not an infinite series), it follows that the $N_f =2$ moduli space is a complete intersection.

\subsection{Generators of the Chiral Ring}
Using plethystic logarithms computed in preceding subsections, we can write down the generators of $SO(N_c)$ adjoint SQCD and representations of $SU(N_f)$ in which they transform. We will make a distinction between $SO(2m)$ and $SO(2m+1)$ gauge groups.

\subsubsection{The $SO(2m)$ Gauge Groups}
The generators of the chiral ring in the case of $SO(2m)$ are as follows:
\beq
\ba{llllll}
\mbox{\bf Casimir invariants}: & s^{2k} &\rightarrow & u_{2k} = {\rm Tr}(\phi^{2k}) \;\;\;\;\;k=1,\ldots,m-1\nn \\
&&& [0,\ldots,  0] \quad 1~\text{dimensional}~, \nn  \\
\mbox{\bf Even adjoint mesons}: & s^{2k}t^2  &\rightarrow &  (A_{2k})^{ij}=Q^i_{a_1} (\phi^{2k})_{a_1 a_2}  Q^j_{a_2}   \;\;\;\;\;k=0,\ldots,m-1\nn \\
&&& [2,0,\ldots,0]  \quad \frac{1}{2}N_f(N_f+1)~\text{dimensional}~, \nn \\
\mbox{\bf Odd adjoint mesons}: & s^{2k+1}t^2  &\rightarrow &  (A_{2k+1})^{ij}=Q^i_{a_1} (\phi^{2k+1})_{a_1 a_2}  Q^j_{a_2}   \;\;\;\;\;k=0,\ldots,m-2\nn \\
&&& [0,1,0,\ldots,0]  \quad \frac{1}{2}N_f(N_f-1)~\text{dimensional}~, \nn \\
\mbox{\bf Adjoint baryons}: &  s^k t^{l}  &\rightarrow &  \CB_k^{i_1 \ldots i_l}= \epsilon^{a_1 \ldots a_{2k} a_{2k+1} \ldots a_{2k+l}} \phi_{a_1 a_2}\ldots \phi_{a_{2k-1} a_{2k}} Q^{i_1}_{a_{2k+1}} \ldots Q^{i_l}_{a_{2k+l}} \nn \\
&&&  \text{with} \; ~ 2k+l =2m, \; ~ k=0,\ldots,m  \nn \\
&&& [0,\ldots,0,1_{l;L},0, \ldots, 0]   \quad {N_f \choose {2m-2k}} ~\text{dimensional}~.\nn \\
\ea
\eeq
The total number of generators is
\beq
m ( 1 + N_{f}^{2} ) - \frac{N_{f}(N_{f}-1)}{2} - 1 + \sum_{k=0}^{m} {N_f \choose 2m - 2 k} ~,
\eeq
which behaves as $N_f^{N_c}/N_c!$ for large values of $N_c$.

\subsubsection{The $SO(2m+1)$ Gauge Groups}
The generators of the chiral ring in the case of $SO(2m+1)$ are as follows:
\beq
\ba{llllll}
\mbox{\bf Casimir invariants}: & s^{2k} &\rightarrow & u_{2k} = {\rm Tr}(\phi^{2k}) \;\;\;\;\;k=1,\ldots,m\nn \\
&&& [0,\ldots,  0] \quad 1~\text{dimensional}~, \nn  \\
\mbox{\bf Even adjoint mesons}: & s^{2k}t^2  &\rightarrow &  (A_{2k})^{ij}=Q^i_{a_1} (\phi^{2k})_{a_1 a_2}  Q^j_{a_2}    \;\;\;\;\;k=0,\ldots,m-1\nn \\
&&& [2,0,\ldots,0]  \quad \frac{1}{2}N_f(N_f+1)~\text{dimensional}~, \nn \\
\mbox{\bf Odd adjoint mesons}: & s^{2k+1}t^2  &\rightarrow &  (A_{2k+1})^{ij}=Q^i_{a_1} (\phi^{2k+1})_{a_1 a_2}  Q^j_{a_2}    \;\;\;\;\;k=0,\ldots,m-1\nn \\
&&& [0,1,0,\ldots,0]  \quad \frac{1}{2}N_f(N_f-1)~\text{dimensional}~, \nn \\
\mbox{\bf Adjoint baryons}: &  s^k t^{l}  &\rightarrow &  \CB_k^{i_1 \ldots i_l}= \epsilon^{a_1 \ldots a_{2k} a_{2k+1} \ldots a_{2k+l}} \phi_{a_1 a_2}\ldots \phi_{a_{2k-1} a_{2k}} Q^{i_1}_{a_{2k+1}} \ldots Q^{i_l}_{a_{2k+l}} \nn \\
&&&  \text{with} \; ~ 2k+l =2m+1, \; ~ k=0,\ldots,m  \nn \\
&&& [0,\ldots,0,1_{l;L},0, \ldots, 0]   \quad {N_f \choose {2m+1-2k}} ~\text{dimensional}

\ea
\eeq
The total number of generators is
\beq
m \left ( 1 + N_{f}^{2} \right ) + \sum_{k=0}^{m}{N_f \choose 2m+1 - 2 k} ,
\eeq
which behaves as $N_f^{N_c}/N_c!$ for large values of $N_c$.
It is to be noted that, among the adjoint mesons of $SO(N_c)$ gauge theories we have just listed, the generator $A_0$ is what we referred to  as meson in the preceding sections. We have listed it among the adjoint mesons only for simplicity. 

\subsection{Complete Intersection Moduli Space}
We have seen from several examples in the preceding sections that 
\begin{itemize}
\item The moduli space of the $SO(N_c)$ gauge theories with 1 fundamental chiral superfield and 1 adjoint chiral superfield is freely generated,
\item The moduli space of the $SO(N_c)$ gauge theories with 2 fundamental chiral superfields and 1 adjoint chiral superfield is a complete intersection.
\end{itemize}
Generalising these examples, we write down general expressions for the fully refined plethystic logarithms in the case of 2 fundamental chiral superfields as 
\bea
\PL[g^{(2,SO(2m+1))}(s,t_1,t_2)] &=& \sum^{m}_{k=1} s^{2k}+ s^m t_1 + s^m t_2
+\sum^{m}_{k=1} \left[ s^{2(k-1)} ( t_1^2 + t_1t_2 + t_2^2) + s^{2k-1}t_1t_2 \right] \nn \\
&& -\sum^{m}_{k=1} s^{2(m+k-1)}(t_1t_2)^2~, \nn \\
\PL[g^{(2,SO(2m))}(s,t_1,t_2)] &=& \sum^{m-1}_{k=1} s^{2k}+s^{m}+s^{m-1}t_1t_2
+\sum^{m}_{k=1}s^{2(k-1)}(t_1^2+t_1t_2+t_2^2)+ \sum^{m-1}_{k=1}s^{2k-1}t_1t_2 \nn \\
&& -\sum^{m}_{k=1} s^{2(m+k-2)}(t_1t_2)^2~.
\eea
Note that the number of relations is equal to the rank of the gauge group and not to 1 as might naively be expected from the case of $SU(N_c)$ gauge group.
The plethystic logarithms in the case of 1 fundamental chiral superfields can be easily obtained  by setting $t_1 = t,~ t_2 =0$:
\bea
\PL[g^{(1,SO(2m+1))}(s,t)] &=& \sum^{m}_{k=1} s^{2k}+ s^m t+\sum^{m}_{k=1}s^{2(k-1)} t^{2}~, \nn \\
\PL[g^{(1,SO(2m))}(s,t)] &=& \sum^{m-1}_{k=1} s^{2k}+s^{m}+\sum^{m}_{k=1}s^{2(k-1)}t^{2}~.
\eea

\section{The $G_2$ Gauge Group} \label{secg2}
Let us now examine the $G_2$ gauge theory with $N_f$ chiral multiplets in the fundamental representation and 1 chiral superfield in the adjoint representation. 

\subsection{Examples of Hilbert Series}
The Molien--Weyl formula for this theory is
\bea
g^{(N_{f},G_2)} &=& \frac{1}{(2 \pi i)^2 } \oint_{|z_1|=1} \frac{ \ud z_1}{z_1} \oint_{|z_2|=1} \frac{ \ud z_2}{z_2} \left(1-z_1\right) \left(1-\frac{z_1^2}{z_2}\right) \left(1-\frac{z_1^3}{z_2}\right)\left(1-z_2\right) \times \nn \\
&&   \left(1-\frac{z_2}{z_1}\right) \left(1-\frac{z_2^2}{z_1^3}\right) \PE \left[ N_f [1,0] t + [0,1] s \right]~. 
\eea
Applying the residue theorem, we find that 
\bea
g^{(1,G_{2})}(s,t) &=& \frac{1-s^{12}t^6}{(1-s^2)(1-s^6)(1-s^3 t)(1-t^2)(1-s^2 t^2)(1-s^4 t^2)(1-s^3 t^3)(1-s^6 t^3)}~ , \nn \\
g^{(2,G_{2})}(s,t) &=& 1 + s^2 + s^4 + 2 s^6 + 2 s^8 + 2 s^{10} + 3 s^{12} + 2 s^3 t + 2 s^5 t + 2 s^7 t + 4 s^9 t + 4 s^{11} t + \nn\\
 &&   3 t^2 + s t^2 + 6 s^2 t^2 + 2 s^3 t^2 + 9 s^4 t^2 + 3 s^5 t^2 + 15 s^6 t^2 + 4 s^7 t^2 + 18 s^8 t^2 + 5 s^9 t^2 + \nn\\
 && 21 s^{10} t^2 + 6 s^{11} t^2 + 27 s^{12} t^2 + 2 s t^3 + 2 s^2 t^3 + 12 s^3 t^3 + 6 s^4 t^3 + 20 s^5 t^3 + 12 s^6 t^3 + \nn\\
 && 28 s^7 t^3 + 16 s^8 t^3 + 42 s^9 t^3 + 20 s^{10} t^3 + 50 s^{11} t^3 + 26 s^{12} t^3 + 6 t^4 + 3 s t^4 + 17 s^2 t^4 +  \nn\\
&& 12 s^3 t^4 + 37 s^4 t^4 + 25 s^5 t^4 + 70 s^6 t^4 + 41 s^7 t^4 + 99 s^8 t^4 + 61 s^9 t^4 + 128 s^{10} t^4 +  \nn\\
&& 77 s^{11} t^4 +166 s^{12} t^4 + 6 s t^5 + 8 s^2 t^5 + 38 s^3 t^5 + 32 s^4 t^5 + 84 s^5 t^5 + 74 s^6 t^5 + 146 s^7 t^5 +\nn\\
&&  122 s^8 t^5 + 226 s^9 t^5 + 174 s^{10} t^5 + 300 s^{11} t^5 + 232 s^{12} t^5 + 10 t^6 + 6 s t^6 + 37 s^2 t^6 + \nn\\
&&  36 s^3 t^6 + 105 s^4 t^6 +100 s^5 t^6 + 233 s^6 t^6 + 198 s^7 t^6 + 385 s^8 t^6 + 330 s^9 t^6 + 560 s^{10} t^6 + \nn\\
&& 466 s^{11} t^6 + 764 s^{12} t^6 + O(s^{13}) O(t^7)~, \nn \\
g^{(3,G_{2})}(s,t) &=& 1+s^2+s^4+2 s^6+2 s^8+2 s^{10}+3 s^{12}+3 s^3 t+3 s^5 t+3 s^7 t+6 s^9 t+6 s^{11} t+6 t^2+ \nn \\
&& 3 s t^2+ 12 s^2 t^2+6 s^3 t^2+18 s^4 t^2+9 s^5 t^2+30 s^6 t^2+12 s^7 t^2+36 s^8 t^2+15 s^9 t^2+42 s^{10} t^2+\nn \\
&& 18 s^{11} t^2+54 s^{12} t^2+t^3+8 s t^3+10 s^2 t^3+37 s^3 t^3+27 s^4 t^3+63 s^5 t^3+47 s^6 t^3+89 s^7 t^3+\nn \\ 
&& 64 s^8 t^3+  128 s^9 t^3+81 s^{10} t^3+154 s^{11} t^3+101 s^{12} t^3+21 t^4+21 s t^4+72 s^2 t^4+75 s^3 t^4+ \nn \\
&& 162 s^4 t^4+153 s^5 t^4+297 s^6 t^4+246 s^7 t^4+423 s^8 t^4+348 s^9 t^4+549 s^{10} t^4+441 s^{11} t^4 \nn \\
&& +699 s^{12} t^4+6 t^5+51 s t^5+90 s^2 t^5+255 s^3 t^5+300 s^4 t^5+570 s^5 t^5+621 s^6 t^5+978 s^7 t^5 \nn \\
&& +990 s^8 t^5+1464 s^9 t^5+1383 s^{10} t^5+1932 s^{11} t^5+1794 s^{12} t^5+57 t^6+89 s t^6+317 s^2 t^6 \nn \\
&& +471 s^3 t^6+980 s^4 t^6+1235 s^5 t^6+2104 s^6 t^6+2339 s^7 t^6+3487 s^8 t^6+3687 s^9 t^6 \nn \\
&& +5040 s^{10} t^6+5106 s^{11} t^6+6725 s^{12} t^6+ O(s^{13}) O(t^7)~, \nn\\
g^{(4,G_{2})}(s,t) &=& 1+s^2+s^4+2 s^6+2 s^8+2 s^{10}+3 s^{12}+4 s^3 t+4 s^5 t+4 s^7 t+8 s^9 t+8 s^{11} t +\nn \\
&& 10 t^2+6 s t^2+20 s^2 t^2+12 s^3 t^2+30 s^4 t^2+18 s^5 t^2+50 s^6 t^2+24 s^7 t^2+60 s^8 t^2+30 s^9 t^2+ \nn\\
&& 70 s^{10} t^2+36 s^{11} t^2+90 s^{12} t^2+4 t^3+20 s t^3+28 s^2 t^3+84 s^3 t^3+72 s^4 t^3+144 s^5 t^3+ \nn \\
&& 120 s^6 t^3+204 s^7 t^3+164 s^8 t^3+288 s^9 t^3+208 s^{10} t^3+348 s^{11} t^3+256 s^{12} t^3+56 t^4+ \nn \\
&& 75 s t^4+212 s^2 t^4+256 s^3 t^4+483 s^4 t^4+516 s^5 t^4+870 s^6 t^4+821 s^7 t^4+1241 s^8 t^4\nn\\
&& 1612 s^{10} t^4+1447 s^{11} t^4+2034 s^{12} t^4+40 t^5+224 s t^5+436 s^2 t^5+1028 s^3 t^5+1356 s^4 t^5+ \nn \\
&& 2292 s^5 t^5+2696 s^6 t^5+3896 s^7 t^5+4232 s^8 t^5+5740 s^9 t^5+5852 s^{10} t^5+7544 s^{11} t^5+ \nn \\
&& 7512 s^{12} t^5+240 t^6+550 s t^6+1640 s^2 t^6+2756 s^3 t^6+5140 s^4 t^6+7010 s^5 t^6+10820 s^6 t^6+ \nn \\
&& 12980 s^7 t^6+17810 s^8 t^6+20036 s^9 t^6+25550 s^{10} t^6+27466 s^{11} t^6+33740 s^{12} t^6 +\nn\\
&& O(s^{13}) O(t^7)~.
\eea
\newline
\paragraph{Plethystic logarithms.} We shall calculate plethystic logarithms of generating functions  using formula \eref{PL}:
\bea
\PL\left[g^{(1,G_{2})}(s,t) \right] &=& s^2+s^6+s^3 t+t^2+s^2 t^2+s^4 t^2+s^3 t^3+s^6 t^3-s^{12}t^6~, \nn\\
\PL\left[g^{(2,G_{2})}(s,t) \right] &=& s^2 + s^6 + 2 s^3 t + 3 t^2 + s t^2 + 3 s^2 t^2 + s^3 t^2 + 3 s^4 t^2 + s^5 t^2 + 2 s t^3 + 2 s^2 t^3\nn\\
&&   + 4 s^3 t^3 + 2 s^4 t^3 + 2 s^5 t^3 + 4 s^6 t^3 + s^2 t^4 + 3 s^3 t^4 - s^6 t^4 - 2 s^8 t^4 - 3 s^9 t^4 - s^{10} t^4 \nn \\
&& - 4 s^5 t^5 - 8 s^6 t^5 - 8 s^7 t^5 - 8 s^8 t^5 - 8 s^9 t^5 - 8 s^{10} t^5 - 4 s^{11} t^5 - 3 s^4 t^6 - 5 s^5 t^6  \nn \\
&& - 10 s^6 t^6 - 17 s^7 t^6 - 17 s^8 t^6 - 13 s^9 t^6 - 11 s^{10} t^6 - 7 s^{11} t^6 - 7 s^{12} t^6 + O(s^{13}) O(t^7)~. \nn \\
\PL\left[g^{(3,G_{2})}(s,t) \right] &=& s^2+s^6+3 s^3 t+6 t^2+3 s t^2+6 s^2 t^2+3 s^3 t^2+6 s^4 t^2+3 s^5 t^2+t^3+8 s t^3+9 s^2 t^3 \nn \\
&& +11 s^3 t^3+8 s^4 t^3+8 s^5 t^3+10 s^6 t^3-s^8 t^3+3 s t^4+9 s^2 t^4+15 s^3 t^4-3 s^5 t^4-6 s^6 t^4 \nn \\
&& -6 s^7 t^4-15 s^8 t^4-18 s^9 t^4-6 s^{10} t^4-3 s^3 t^5-18 s^4 t^5-54 s^5 t^5-87 s^6 t^5-84 s^7 t^5 \nn \\
&& -81 s^8 t^5-78 s^9 t^5-63 s^{10} t^5-27 s^{11} t^5+6 s^{12} t^5-10 s^2 t^6-35 s^3 t^6-81 s^4 t^6 \nn \\
&& -136 s^5 t^6-209 s^6 t^6-260 s^7 t^6-224 s^8 t^6-138 s^9 t^6-89 s^{10} t^6-26 s^{11} t^6 \nn \\
&&  +33 s^{12} t^6+ O(s^{13}) O(t^7)~, \nn\\
\PL\left[g^{(4,G_{2})}(s,t) \right] &=& s^2+s^6+4 s^3 t+10 t^2+6 s t^2+10 s^2 t^2+6 s^3 t^2+10 s^4 t^2+6 s^5 t^2+4 t^3+20 s t^3\nn \\
&& +24 s^2 t^3+24 s^3 t^3+20 s^4 t^3+20 s^5 t^3+20 s^6 t^3-4 s^8 t^3+t^4+15 s t^4+35 s^2 t^4+45 s^3 t^4\nn\\
&& -16 s^5 t^4-22 s^6 t^4-30 s^7 t^4-56 s^8 t^4-60 s^9 t^4-21 s^{10} t^4+s^{11} t^4-4 s^2 t^5-28 s^3 t^5 \nn \\
&& -124 s^4 t^5-288 s^5 t^5-424 s^6 t^5-408 s^7 t^5-380 s^8 t^5-356 s^9 t^5-260 s^{10} t^5-96 s^{11} t^5 \nn\\
&& +40 s^{12} t^5-16 s t^6-116 s^2 t^6-332 s^3 t^6-646 s^4 t^6-1028 s^5 t^6-1454 s^6 t^6-1620 s^7 t^6 \nn \\
&& -1288 s^8 t^6-684 s^9 t^6-312 s^{10} t^6+78 s^{11} t^6+474 s^{12} t^6+ O(s^{13}) O(t^7)~.
\eea

\subsection{Generators of the Chiral Ring} 
We note that $G_2$ has 3 independent invariant tensors \cite{Pesando:1995bq, Giddings:1995ns, Distler:1996ub}: $\delta^{ab}$, $\epsilon^{a_1 \ldots a_7}$ and $f^{a_1 a_2 a_3}$, where the last two tensors are totally antisymmetric and the indices run over 1 to 7. 
According to \eref{G2adj}, we shall denote the adjoint field by an antisymmetric tensor $\phi^{ab}$ with the property
\bea
f_{abc}\phi^{ab} = 0~.
\eea
Note that $\phi^{ab}$ has $\frac{1}{2}(7\times 6) - 7 = 14$ independent components which is equal to the dimension of the adjoint representation of $G_2$.  
Equipped with plethystic logarithms, we can write down the generators of the chiral ring as follows:
\small{
\beq
\ba{llllll}
\mbox{\bf Casimir invariants}:  & s^{2}  &\rightarrow & u_2 = {\rm Tr}(\phi^2) \nn \\
&&&  [0,\ldots,  0]  
\nn  \\
\quad & s^6 &\rightarrow&  u_6 = {\rm Tr} (\phi^6) \nn \\
&&&   [0, \ldots,  0] 
\nn  \\
\mbox{\bf Adjoint-quark invariants}: & s^{3}t &\rightarrow &  A^i = \epsilon^{a_1 \ldots a_7} \phi_{a_1 a_2} \phi_{a_3 a_4} \phi_{a_5 a_6} Q^i_{a_7} \nn \\
&&&  [1,0,\ldots,  0]  
\nn  \\
\mbox{\bf Even adjoint mesons}: & s^{2k} t^2  &\rightarrow &  (A_{2k})^{ij}=Q^i_{a_1} (\phi^{2k})_{a_1 a_2}  Q^j_{a_2} \quad k=0,1,2 \nn \\
&&&  [2,0,\ldots,0]  
\nn \\
\mbox{\bf Odd adjoint mesons}: & s^{2k+1} t^2  &\rightarrow &  (A_{2k+1})^{ij}=  Q^i_{a_1} (\phi^{2k+1})_{a_1 a_2}  Q^j_{a_2} \quad k=0,1,2  \nn \\
&&&  [0,1,0,\ldots,0]  
\nn \\
\mbox{\bf 3-baryons}: &  t^3  &\rightarrow &  B^{ijk}= f^{abc} Q^i_a Q^j_b Q^k_c  \nn \\
&&&  [0,0,1,0,\ldots,0]  
\nn \\

\mbox{\bf Adjoint 3-baryons}: & s t^3  &\rightarrow & \CB^{i_1 i_2 j_1}_{0,0,1} = f^{a_1 a_2 b_1} Q^{i_1}_{a_1} Q^{i_2}_{a_2} (P_1)^{j_1}_{b_1}    \nn \\
&&& [1,1,0,\ldots,0] 
\nn \\
\quad & s^2 t^3  &\rightarrow &  \CB^{i_1 i_2 j_1}_{0,0,2} = f^{a_1 a_2 b_1} Q^{i_1}_{a_1} Q^{i_2}_{a_2} (P_2)^{j_1}_{b_1} ~,~ \CB^{i_1 j_1 j_2}_{0,1,1} = f^{a_1 b_1 b_2} Q^{i_1}_{a_1} (P_1)^{j_1}_{b_1} (P_1)^{j_2}_{b_2}~,  \nn \\
&&& \CB^{i_1 i_1 i_3} =  \epsilon^{a_1 \ldots a_7} Q^{i_1}_{a_1} Q^{i_2}_{a_2} Q^{i_3}_{a_3} \phi_{a_4 a_5} \phi_{a_6 a_7}  \nn \\
&&& [1,1,0, \ldots,0]+ [0,0,1,0,\ldots,0] 
\nn \\
\quad & s^3 t^3  &\rightarrow & \CB^{i_1 i_2 j_1}_{0,0,3} = f^{a_1 a_2 b_1} Q^{i_1}_{a_1} Q^{i_2}_{a_2} (P_3)^{j_1}_{b_1}~,~ \CB^{i_1 j_1  k_1}_{0,1,2} = f^{a_1 a_2 a_3} Q^{i_1}_{a_1} (P_1)^{j_1}_{b_1} (P_2)^{k_1}_{c_1}~,  \nn \\
&&& \CB^{i_1 i_2 i_3}_{1,1,1} = f^{a_1 a_2 a_3} (P_1)^{i_1}_{a_1} (P_1)^{i_2}_{a_2} (P_1)^{i_3}_{a_3}  \nn \\
&&& [3,0,\ldots,0]+[0,0,1,0,\ldots,0] 
\nn \\
\quad & s^4 t^3  &\rightarrow &   \CB_{0,0,4} = f Q Q P_4~,~\CB_{0,1,3} = f Q P_1 P_3~,~\CB_{0,2,2} = f Q P_2 P_2~,\nn \\
&&& \CB_{1,1,2} = f P_1 P_1 P_2   \nn \\
&&& [1,1,0,\ldots,0] 
 \nn \\
\quad & s^5 t^3  &\rightarrow & \CB_{0,0,5} = f Q Q P_5~,~\CB_{0,1,4} = f Q P_1 P_4~,~\CB_{0,2,3} = f Q P_2 P_3~, \nn \\
&&& \CB_{1,2,2} = f P_1 P_2 P_2~,~\CB_{1,1,3} = f P_1 P_1 P_3   \nn \\
&&& [1,1,0,\ldots,0]  
\nn \\
\quad & s^6 t^3  &\rightarrow &   \CB_{0,0,6} = f Q Q P_6~,~\CB_{0,1,5} = f Q P_1 P_5~,~\CB_{0,2,4} = f Q P_2 P_4~, \nn\\ 
&&& \CB_{0,3,3} = f Q P_3 P_3~,~ \CB_{1,2,3} = f P_1 P_2 P_3~,~ \CB_{2,2,2} = f P_2 P_2 P_2~, \nn \\
&&& \CB_{1,1,4} = f P_1 P_1 P_4  \nn \\
&&& [3,0,\ldots,0] 
\nn \\

\mbox{\bf 4-baryons}: & t^4  &\rightarrow &  b^{i_1 \ldots i_4}= \epsilon^{a_1 \ldots a_7} Q^{i_1}_{a_1} \ldots  Q^{i_4}_{a_4} f_{a_5 a_6 a_7}  \nn \\
&&& [0,0,0,1,0\ldots,0] 
\nn \\
\mbox{\bf Adjoint 4-baryons}:  & st^4  &\rightarrow &  \mathfrak{b}^{i_1 \ldots i_4}_{0,0,0,1}= \epsilon^{a_1 \ldots a_7} Q^{i_1}_{a_1} Q^{i_2}_{a_2} Q^{i_3}_{a_3}  (P_1)^{i_4}_{a_4} f_{a_5 a_6 a_7}    \nn \\
&&& [1,0,1,0\ldots,0]  
\nn \\
\quad & s^2t^4  &\rightarrow & \mathfrak{b}^{i_1 \ldots i_4}_{0,0,0,2}= \epsilon^{a_1 \ldots a_7} Q^{i_1}_{a_1} Q^{i_2}_{a_2} Q^{i_3}_{a_3}  (P_2)^{i_4}_{a_4} f_{a_5 a_6 a_7}~, \nn \\
& & &  \mathfrak{b}^{i_1 \ldots i_4}_{0,0,1,1}= \epsilon^{a_1 \ldots a_7} Q^{i_1}_{a_1} Q^{i_2}_{a_2} (P_1)^{i_3}_{a_3}  (P_1)^{i_4}_{a_4} f_{a_5 a_6 a_7}   \nn \\
&&& [1,0,1,0\ldots,0] + [0,2,0,\ldots,0]  
\nn \\
\quad & s^3t^4  &\rightarrow & \mathfrak{b}^{i_1 \ldots i_4}_{0,0,1,2}= \epsilon^{a_1 \ldots a_7} Q^{i_1}_{a_1} Q^{i_2}_{a_2} (P_1)^{i_3}_{a_3}  (P_2)^{i_4}_{a_4} f_{a_5 a_6 a_7} ~,\nn\\
&&&  \mathfrak{b}^{i_1 \ldots i_4}_{0,1,1,1}= \epsilon^{a_1 \ldots a_7} Q^{i_1}_{a_1} (P_1)^{i_2}_{a_2} (P_1)^{i_3}_{a_3}  (P_1)^{i_4}_{a_4} f_{a_5 a_6 a_7}    \nn \\
&&& [2,1,0\ldots,0] 
\nn \\
\ea \eeq}

\noindent  Note that the Casimir invariant $u_4 \equiv {\rm Tr} (\phi^4)$ can be obtained from $u_2$ as follows:
\bea
u_4 = \frac{1}{4} u_2^2~.
\eea
Therefore, we do not include $u_4$ in the above list.  

As for the case of the $SU(N_c)$ gauge groups, we emphasise that the representations written above are \emph{not} the ones in which  the generators transform; however, they are the ones in which the relations have already been taken into account. 

\comment{
\paragraph{${}^{\dagger}$The generator at order $s t^4$.}  Note that the generator {\footnotesize $\CB_{0,0,0,1}^{i_1 i_2 i_3 j_1} = \epsilon^{a_1a_2 a_{3} b_{1}}  Q^{i_1}_{a_1} {Q}^{i_2}_{a_2} Q^{i_{3}}_{a_{3}} (P_1)^{j_{1}}_{b_{1}}$} satisfies a relation:
\bea
 \CB_{0,0,0,1}^{[i_1 i_2 i_3 j_1]} = 0~, \label{rel14}
\eea 
where the square bracket denotes an antisymmetrisation without a normalisation factor. This means that the completely antisymmetric part, which transforms in the $SU(N_f)$ representation [0,0,0,1,0,\ldots,0], vanishes. Note that we can construct the generator by considering the following $SU(N_f)$ tensor product:
\bea
[0,0,1,0,\ldots,0] \times [1,0,\ldots,0]= [1,0,1,0, \ldots, 0] + [0,0,0,1,0, \ldots, 0]~. \nn
\eea
Therefore, after taking \eref{rel14} into account, we conclude that $\CB_{0,0,0,1}^{i_1 i_2 i_{3} j_1}$ transforms in the $SU(N_f) \times SU(N_f)$ representation [1,0,1,0, \ldots, 0; 0,\ldots, 0], as stated in the list above. }

\paragraph{Total number of generators.} Using the trick mentioned in Section \ref{trick}, we find that the total number of generators is $\frac{1}{12} \left(24-524 N_f+957 N_f^2-430 N_f^3+69 N_f^4\right)$.

\section{A Geometric Aper\c{c}u} \label{apercu} \setall
In the preceding sections, we computed Hilbert series and their plethystic logarithms which count generators and relations in the chiral rings of adjoint SQCD.  In the following, we shall extract a number of useful geometrical properties of moduli spaces from Hilbert series.  

\subsection{Palindromic Numerator}  We have observed in many case studies before that the numerator of the Hilbert series for adjoint SQCD is palindromic, {\em i.e.}\ it can be written in the form of a degree $k_1,~k_2 $ polynomial in $s,~t$:
\begin{equation}
P_{k_1, k_2} (s,t) = \sum_{m=0} ^{k_1} \sum_{n=0} ^{k_2} a_{m,n} s^m t^n ~,
\end{equation}
with symmetric coefficients $a_{(k_1 -m) ,(k_2-n)} = a_{m,n}$.
We establish the following theorem:
\begin{theorem} \label{palin}
Let $P(s,t)$ be a numerator of the Hilbert series $g(s,t)$ such that $P(1,1) \ne 0$.  Then, $P(s, t)$ is palindromic.
\end{theorem}

\noindent We shall use the following lemma to prove the above theorem.
\begin{lemma} \label{prop}
Let $d$ be the dimension of the moduli space and let $d_G$ be the dimension of the gauge group $G$.
Then, the Hilbert series obey:
\bea \label{palindrome}
g^{(N_f, G)} \left(1/s, 1/t \right) &=& (-1)^{d} s^{d_G} t^{d} g^{(N_f, G)}(s,t) 
\eea
\end{lemma}
\noindent {\bf Proof.} \: For definiteness, we shall prove only the case of $G=SU(N_c)$, \emph{i.e.,}
\bea
g^{(N_f,SU(N_c))} \left(1/s, 1/t \right) &=& (-1)^d s^{N_c^2-1} t^{d} g^{(N_f,SU(N_c))}(s,t)~, \label{supalin}
\eea
where $d= 2N_f N_c$. The others can be proven in a similar fashion.  

We write down the Molien--Weyl formula as  
\bea
g^{(N_f, SU(N_c))}(s,t) &=& \int_{SU(N_c)} \ud \mu_{SU(N_c)} \: \mathrm{PE} \left[ N_f \left( [1,0, \ldots,0] +[0, \ldots, 0 , 1]\right) t + [1,0, \ldots, 0 , 1] s \right] \nn \\
&=&  \int_{SU(N_c)} \frac{ \ud \mu_{SU(N_c)} }{ \prod_w (1- t  \prod_{l=1}^{N_c-1} z_l^{w_l})^{N_f} (1- t  \prod_{l=1}^{N_c-1} z_l^{-w_l})^{N_f}} 
\times \frac{1}{(1-s)^{N_c-1}}  \times \nn \\
&& \frac{1}{\prod_{\alpha^+} \left(1- s \prod_{l=1}^{N_c-1} z_l^{\alpha^+_l} \right)\left(1- s \prod_{l=1}^{N_c-1} z_l^{-\alpha^+_l} \right) }~, \label{molienint}
\eea
where $w$ denotes the collection of the weights of the fundamental representation, $\alpha^+$ denotes the collection of the positive roots, and the subscript $l$ denotes the $l$-th component of the weight or root.

Let us turn to $g^{(N_f, SU(N_c))}(1/s,1/t)$.
Under the transformations $s \rightarrow 1/s$ and $t \rightarrow 1/t$, the integrand in (\ref{molienint}) changes to (up to a minus sign)
{\footnotesize
\bea \label{t2nNf}
 \frac{t^{2N_fN_c}}{ \prod_w (1- t  \prod_{l=1}^{N_c-1} z_l^{w_l})^{N_f} (1- t  \prod_{l=1}^{N_c-1} z_l^{-w_l})^{N_f} } \cdot \frac{s^{N_c^2-1}}{(1-s)^{N_c-1}\prod_{\alpha^+} \left(1- s \prod_{l=1}^{N_c-1} z_l^{\alpha^+_l} \right)\left(1- s \prod_{l=1}^{N_c-1} z_l^{-\alpha^+_l} \right) }~. \nn
\eea}
Now let us obtain the correct overall sign.
An easy way to do so is to consider the expansion of $g^{(N_f, SU(N_c))}(s,t)$ as a Laurent series around $s,\:t = 1$:
\begin{equation}
g^{(N_f, N_c)}(s,t) = \sum_{m_1 = -p}^ \infty  \sum_{m_2 = -q} ^ \infty c_{m_1,m_2} (s-1)^{m_1} (t-1)^{m_2} \sim \frac{c_{-p,-q}}{(s-1)^p(t-1)^q} ~,
\end{equation}
(as $s,\: t \rightarrow 1$) where $p+q=d$. 
(Recall that $d$ is the dimension of the moduli space, which is equal to the order of the pole at $t=1$ of the unrefined Hilbert series $g^{(N_f, N_c)}(t,t)$ .) Therefore, we see that as $s,\: t \rightarrow 1$, the signs of $g^{(N_f, SU(N_c))}(1/s, 1/t)$ and $g^{(N_f, SU(N_c))}(s, t)$ differ by $(-1)^d$. Combining this with (\ref{t2nNf}), we prove the first formula in (\ref{supalin}). $\Box$ \\

\noindent We are now ready for our claim.\\
\noindent{\bf Proof of Theorem \ref{palin}.}\: We note that the denominator of the Hilbert series $g(s,t)$ is in the form $ \prod_{k} (1-s^{a_k})^{b_{k}} \prod_{l} (1-t^{c_l})^{d_{l}}$, where $a_k,\: b_k,\: c_l,\: d_l$ are non-negative integers. 
Observe that upon the transformation $s \rightarrow 1/s$ and $t \rightarrow 1/t$, the denominator picks up the sign $(-1)^{\sum_{k} b_k + \sum_{l} d_l}$.
Now if the numerator $P(t)$ does not vanish at $s=t=1$, then $\sum_{k} b_k + \sum_{l} d_l$ is exactly the order of the pole at $t=1$ of the unrefined Hilbert series $g(t,t)$, which is equal to the dimension $d$ of the moduli space.
Since $P(s,t) = g(s,t)  \prod_{k} (1-s^{a_k})^{b_{k}} \prod_{l} (1-t^{c_l})^{d_{l}}$, it follows from (\ref{palindrome}) that $P(s, t)$ is palindromic.
$\Box$

\subsection{The Adjoint SQCD vacuum Is Calabi-Yau} 
Similar situations were encountered in \cite{Gray, Hanany:2008kn, Forcella:2008bb}.  Due to a well-known theorem in commutative algebra called the Hochster--Roberts theorem\footnote{This theorem states that the invariant ring of a linearly reductive group acting on a regular ring is Cohen--Macaulay.}${}^,$\footnote{We are grateful to Richard Thomas for drawing our attention to this important theorem.}  \cite{hochster}, our coordinate rings of the moduli space are Cohen--Macaulay.
Therefore, as an immediate consequence of Theorem \ref{palin} and the Stanley theorem\footnote{This theorem states that the numerator to the Hilbert series of a graded Cohen--Macaulay domain $R$ is palindromic if and only if $R$ is Gorenstein.} \cite{stanley}, the chiral rings are also algebraically Gorenstein.   Since affine Gorenstein varieties means Calabi--Yau, we reach an important conclusion:
\begin{theorem}
The moduli space of the adjoint SQCD is an affine Calabi--Yau cone over a weighted projective variety.
\end{theorem}


\subsection{The Adjoint SQCD Moduli Space Is Irreducible}
In this subsection, we demonstrate that the classical moduli space of adjoint SQCD is irreducible.  Similar situations were encountered in \cite{Gray, Hanany:2008kn}.

As in \cite{Gray}, the moduli space (in the absense of a superpotential) of adjoint SQCD can be described by a symplectic quotient:
\beq 
\BC^n // G = \BC^n / G^c~,
\eeq
where $G^c$ denotes the complexification of gauge group and $n$ is the total number of the chiral superfields (both transforming in fundamental and adjoint representations):  
\begin{equation*}
n = \left \{
\begin{array}{rl}
2N_cN_f + (N_c^2 -1) \quad & \text{for} \quad  G= SU(N_c) \\
4N_cN_f + \frac{1}{2}N_c(N_c+1) \quad &  \text{for} \quad G= Sp(N_c) \\
N_cN_f + \frac{1}{2}N_c(N_c-1) \quad & \text{for} \quad G= SO(N_c) \\
7N_f + 14  \quad & \text{for} \quad G= G_2.
\end{array} \right. 
\end{equation*}
Since $\BC^n$ is irreducible and $G^c$ is a continuous group, we expect the resulting quotient to be also irreducible\footnote{We are grateful to Alberto Zaffaroni for this point.}.

\section*{Acknowledgements}
We would like to thank Scott Thomas for asking a question that initiated this project, as well as Nathan Seiberg, Kenneth Intriligator and Michael Douglas for discussions in \sref{relci}.
We are indebted to James Gray, Vishnu Jejjala and Yang-Hui He for discussions.
N.~M.~ would like to express his deep gratitude towards his family for the warm encouragement and support, as well as Carl M. Bender, Alexander Shannon, Fabian Spill and Sam Kitchen for their help and dicussions.  He is also grateful to the DPST project and the Royal Thai Government for funding his research.
G.~T.~ wants to offer his most heartfelt thanks to his family for the precious support during the preparation of this work, as well as to Elisa Rebessi for being such a wonderful source of understanding and encouragment for anything. He is grateful to Collegio di Milano, especially to Luis Quagliata, for the privilege he has been given of spending there three marvellous years.

\end{document}